\documentclass[a4paper,preprint,10pt]{elsarticle}
\usepackage{amsfonts}
\usepackage{epsfig,amsmath,latexsym,amssymb}
\usepackage{amscd}
\usepackage{multirow}
\usepackage{graphicx}
\usepackage{geometry}
\usepackage{lscape}
\usepackage{amsmath}
\usepackage{amssymb}
\usepackage{bm}
\usepackage{subfigure}
\usepackage{url}
\usepackage{pifont}
\usepackage{bbding}
\usepackage{color}
\usepackage{natbib}
\usepackage{float}
\usepackage{placeins}
\usepackage{bm}
\usepackage{lineno}
\definecolor {darkblue} {rgb} {0,0.08,0.45}

\providecommand{\U}[1]{\protect\rule{.1in}{.1in}}

\newcommand{\Remm}[1]{}
\newtheorem{theo}{Theorem}[section]
\newtheorem{lemma}[theo]{Lemma}
\newtheorem{prop}[theo]{Proposition}

\newtheorem{defi}[theo]{Definition}
\newtheorem{model ass}[theo]{Model Assumptions}

\newtheorem{remark}[theo]{Remarks}

\def\EndProof{{\begin{flushright}\vspace{-2mm}$\Box$\end{flushright}}}
\numberwithin{equation}{section}

\begin{document}

\begin{frontmatter}

\title{A Copula Based Bayesian Approach for Paid-Incurred Claims Models for Non-Life Insurance Reserving}
\author{Gareth W.~Peters$^{1,2}$ \quad Alice X.D.~Dong$^{3}$ \quad Robert ~Kohn$^{4}$}
\date{{\footnotesize {Working paper, version from \today }}}
\maketitle

\begin{abstract}

\noindent Our article considers the class of recently developed stochastic models that combine claims payments and incurred losses information into a coherent reserving methodology. In particular, we develop a family of hierarchical Bayesian paid-incurred-claims models, combining the claims reserving models of \citet{hertig85} and \citet{gogol1993using}. In the process we extend the independent log-normal model of \citet{merz2010paid} by incorporating different dependence structures using a Data-Augmented mixture Copula paid-incurred claims model.

The usefulness of incorporating both payment and incurred losses into estimating of the full predictive distribution of the outstanding loss liabilities and the resulting reserves is demonstrated in the following cases: (i) an independent payment data model; (ii) the independent payment and incurred claims data model of \citet{merz2010paid}; (iii) a novel dependent lag-year telescoping block diagonal Gaussian copula payment and incurred claim data model incorporating conjugacy via transformation; (iv) a novel data-augmented mixture Archimedean copula dependent payment and incurred claim data model.

Inference in such models is developed by adaptive Markov chain Monte Carlo sampling algorithms. These incorporate a data-augmentation framework utilised to efficiently evaluate the likelihood for the copula based payment and incurred claim model in the loss reserving triangles. The adaptation strategy of the Markov chain Monte Carlo is based on two components. The first component uses an adaptive strategy for learning the posterior structures for the parameters defined over a Euclidean space and the second component deals with an adaptive learning of the posterior for the covariance matrices restricted to the Riemann manifold corresponding to the space of positive definite matrices for the linear dependence structure specified for the payment and incurred claim model. 

\vspace{5mm}
\end{abstract}


\begin{center}
{\footnotesize {\ \textit{$^{1}$ Department of Statistical Science,
University College London UCL, London, UK; \\[0pt]
email: gareth.peters@ucl.ac.uk \\[0pt]
(Corresponding Author) \\[0pt]
$^{2}$ CSIRO Mathematics, Informatics and Statistics, Sydney, Australia \\[0pt]
$^{3}$ University of Sydney, School of Mathematics and Statistics \\[0pt]
$^{4}$ UNSW Australian School of Business \\[0pt] } } }
\end{center}

\end{frontmatter}

\section{Introduction}

As discussed in \citet{merz2010paid} the main task of reserving actuaries is to predict ultimate loss ratios and outstanding loss liabilities. In general such predictions are based on past information that comes from a variety of sources. Under a credibility based framework, the weighting of different data sources and their relative contribution to the estimated reserve is difficult to determine. Therefore, it is important to consider developing a unified prediction framework for the outstanding loss liabilities, known as the paid-incurred-claims (PIC) class of models. However, to date only simple dependence structures have been considered, with three parameters for the correlations which were not incorporated into the formal Bayesian estimation approach, and instead fixed deterministically \textit{a priori}. There are two technical difficulties in extending the current restrictive assumptions within a Bayesian framework. The first is being able to generate the positive definite matrices; the second is evaluating the joint likelihood of the mixture copula defined over the observed payments and incurred losses in each accident year row of the reserving matrix. Our article significantly extends the dependence structure of current PIC models by solving these two problems. The first problem is resolved through utilisation of a class of matrix-variate Inverse-Wishart priors coupled with an adaptive Markov chain sampler that restricts the proposed Markov chain states to remain on the manifold of such matrices. The second problem is solved by using a data augmentation strategy which treats the unobserved parts of the loss triangle as missing data so that one can perform evaluation of the copula based likelihood required for inference on the model parameters. 

In order to ensure the financial security of an insurance company, it is important to predict future claims liabilities and obtain the corresponding prediction intervals which take into account parameter uncertainty. The PIC model is a claims reserving method which statistically combines information about claims payments and incurred losses. It allows actuaries to best utilise the available information for loss reserves. The Munich chain ladder method introduced by \citet{quarg2004munich} is one of the first claims reserving approaches in the actuarial literature to unify outstanding loss liability prediction based on both sources of information. This method aims to reduce the gap between the two chain ladder predictions that are based on claims payments and incurred losses data, respectively. It is achieved by adjusting the chain ladder factors with paid-incurred ratios to reduce the gap between the two predictions. The main drawback with the Munich chain ladder method is that it involves several parameter estimates whose precisions are difficult to quantify within a stochastic model framework.

\citet{merz2010paid} recently introduced a log-normal PIC chain model and used Bayesian methods to estimate the missing (future) part of the claims reserving triangles based on both payment and loss incurred information. Its major advantage is that the full predictive distribution of the outstanding loss liabilities can be quantified. One important limitation of the model of \citet{merz2010paid} is that it does not develop the dependence properties of the PIC model that will be applicable to loss reserving data observed in practice. Our article extends the proposed Bayesian PIC models of \citet{merz2010paid} to capture additional dependence structures.

\subsection{Brief Background}

Dependence within payment data, within incurred loss data, and even between payment and incurred loss data commonly exists due to the nature of the loss process. Payment and incurred loss ratios in the previous development period are likely to impact that of the next development period. Hence, correlation between development periods is practically appealing in claims reserving practice. Moreover, incurred loss is essentially payment data plus case estimates which are projections foreseen by case managers to estimate the remaining payments. Correlation between payment and incurred loss data is also found. \citet{happ2011paid} propose a fixed covariance structure to describe the correlation between payment and incurred loss, assuming that the correlations between different development periods are identical. In reality, correlations differ across development periods for various reasons, such as different stages of the life cycle for a claim and internal policy changes. In order to fully incorporate the actual correlations, we introduce a block covariance structure to allow for the variation between different development periods within payment and incurred losses. We also develop a second class of hierarchical mixture of copulas models.

The estimation challenge involves constructing and sampling from the resulting Bayesian models for PIC with flexible dependence structures. To specify the model, we vectorize the triangular random structures for payments and incurred loss and, applying appropriate permutations, we then assume a copula dependence structure on the vectorized data. We use a Gaussian copula with an unknown correlation matrix, which is restricted to be block diagonal for parsimony, or a mixture Archimedian copulas across development periods.

We estimate the Bayesian models by MCMC methods, using data augmentation to generate missing data values in the loss triangle and use an adaptive Metropolis algorithm to generate the unknown parameters. Bayesian simulation methodology is used to carry out inference on all aspects of the models considered and to obtain predictive distributions for reserves.

\subsection{Contributions}
We design a novel class of PIC models and illustrate it with two examples. The first involves a mixture of Clayton and Gumbel copulas for upper and lower tail dependence features in the development years for payments and incurred losses. The second example involves a Gaussian copula model in which the covariance structure is a telescoping block diagonal form representation which captures dependence between development lag years in the payments and incurred losses. By a telescoping block diagonal matrix we mean one in which the main diagonal is comprised of sub-blocks for which each incremental sub-block contains one less row and column compared to the previous. In constructing these models we consider hierarchical Bayesian models with hyperparameters on the priors for development factors and specially developed matrix-variate priors on the covariance structures which preserves the conjugacy properties of the independence models developed in \citet{merz2010paid} and \citet{merz2010estimation}.

For the independent and Gaussian copula based PIC models  we develop a class of conjugate posterior models that can be efficiently estimated via an MCMC sampler known as a block Gibbs sampler. However, the extension to general copula dependence structures requires non-conjugate priors, making it necessary to develop adaptive MCMC algorithms. Adaptive sampling uses previous iterates of the Markov chain to form more efficient Metropolis proposals for the parameters, this class of MCMC algorithm has received growing attention in the statistics literature since it was recently developed and is now recognized as an important tool for Bayesian inference. There is an increasing interest in utilizing adaptive MCMC to facilitate more efficient sampling (\citet{andrieu2008tutorial}, \citet{atchade2005adaptive}). The adaptive techniques that we adopt in this paper fall within the general framework of  adaptive Metropolis, and employ the optimal scale factors (\citet{roberts2009examples}) from the Single Component Adaptive Metropolis (SCAM) algorithm (\citet{haario2005componentwise}). There have been some initial utilisations of adaptive MCMC specifically in financial modelling such as \cite{peters2011calibration} and the references therein. In addition the adaption strategies we consider in this paper involve extensions of Euclidean space Adaptive Metropolis to the space of positive definite matrices, creating a class of matrix variate Markov chain adaptive proposals. 

In the mixture copula based PIC models, we design data augmentation strategies which are a class of auxiliary variable methods. We modify these approaches to the PIC copula based models in order to circumvent the challenge of intractable likelihood evaluations which arise form the structure of the PIC reserving triangle. In particular we argue that the tail dependence features of the model should be consistent accross all development years for both payment and incurred loss data. This poses an evaluation challenge for the likelihood as it involves evaluation of marginal likelihood quantities given the observed data in accident year $i$, given by payment and incurred losses. The integral required when utilising mixture copula structures over the accident years is intractable, therefore we introduce auxiliary variables into the Bayesian model in a data-augmentation structure to overcome this dificulty. 

\section{Review of the Merz-Wuethrich Independence Copula Paid-Incurred Claims Model}
\begin{figure}[ht]
\includegraphics[height=6cm, width=\textwidth]{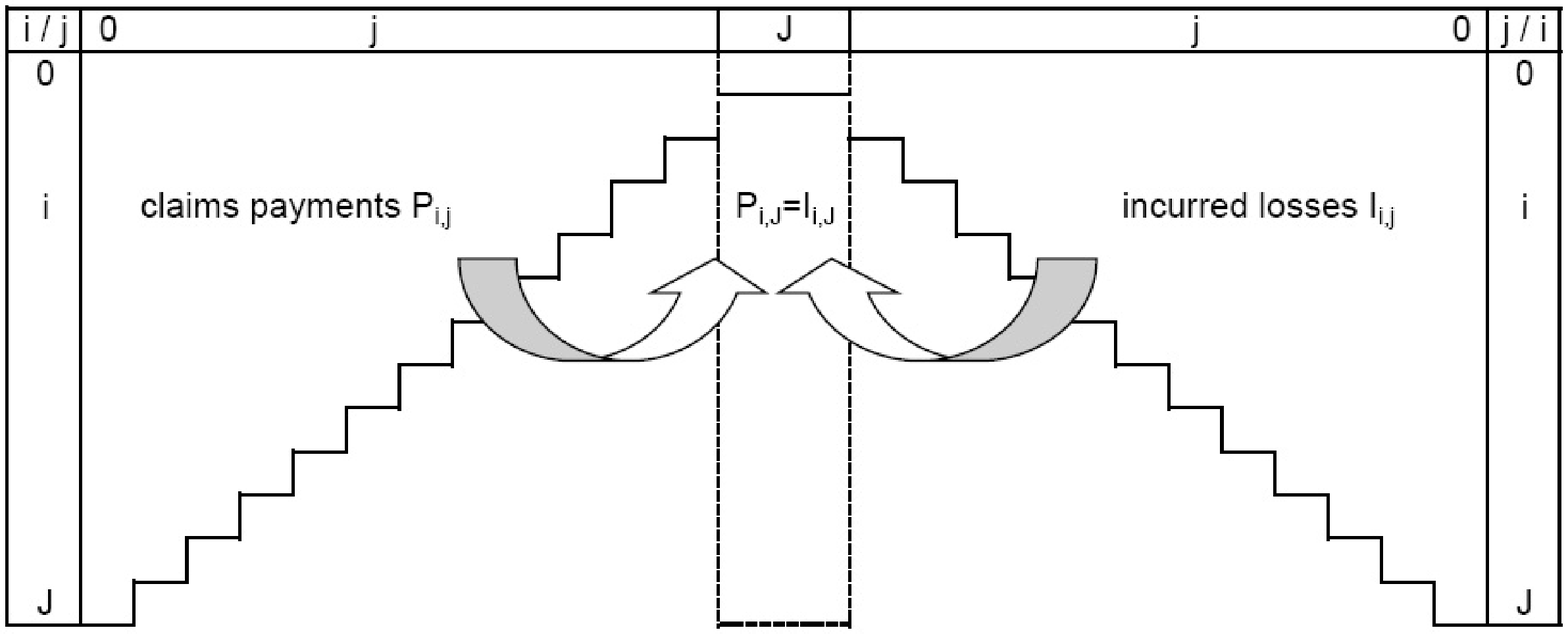} \par \noindent
\caption{Claims triangle for payment data and incurred data, including constraint on zero case estimates at development period $J$ (source \citet{merz2010paid}).}
\label{PICTriangle}
\end{figure}

This section introduces the PIC model which involves two sources of information. The first is the claims payment data, which involves payments made for reported claims. The second source of data incorporated into the statistical estimation are the incurred losses corresponding to the reported claim amounts. The differences between the incurred losses and the claim payments are known as the case estimates for the reported claims which should be equal once a claim is settled. This imposes a set of constraints on any statistical model developed to incorporate each of these sources of data into the parameter estimation. We use the constraints proposed in \citet{merz2010paid} which are used to specify a model based on a claims triangle constructed from vertical columns corresponding to development years of claims and rows corresponding to accident years. This structure for the observed data is summarized in triangular form which is utlised for both the claims payments and the incurred losses, as presented in Figure \ref{PICTriangle}.

Without loss of generality, we assume an equivalent number $J$ of accident years and development years.   Furthermore, we assume that all claims are settled after the $J$-th development year. Let $P_{i,j}$ be the cumulative claims payments in accident year $i$ after $j$ development periods and $I_{i,j}$ the corresponding incurred losses. Moreover, for the ultimate loss we assume the constraint discussed on the case estimates corresponds to the observation that predicted claims should satisfy $P_{i,J}=I_{i,J}$ with probability 1, which means that ultimately (at time $J$) the claims reach the same value and therefore satisfy the required constraint. 

We define (i) $P_{0:J,0:j} = \left\{P_{k,l}:0 \leq k \leq J, 0 \leq l \leq j\right\}$. (ii) Let $A$ and $B$ be square matrices. Then $diag(A,B)$ is the diagonal matrix, with the diagonal elements of A appearing topmost, then the diagonal elements of B. Let the matrices $A$ and $B$ be as in (ii). Then the direct sum of $A$ and $B$, written as $A \oplus B$ is the block diagonal matrix with $A$ in the top left corner and $B$ in the bottom right corner. It is clear that the definitions in (ii) and (iii) can be iterated. That is $diag(A,B,C) = diag(diag(A,B),C)$ and $A \oplus B \oplus C = (A \oplus B)\oplus C$. (iv) Define the $d \times d$ diagonal square identity matrix according to $\mathbb{I}_d$. (v) Define the indicator of an event by the dirac-delta function $\delta_{i}$. (vi) Define the vectorization operator on a $p \times n$ matrix $A$, denoted by $Vec(A)$, as the stacking of the columns to create a vector. 

As in \citet{merz2010paid}, we consider a Log-Normal PIC model as this facilitates comparison between existing results and the results we derive based on different dependence frameworks in extensions to this model. 

We now introduce the PIC model and the statistical assumptions for the independent case, followed by remarks on the resulting marginal posterior models for the payment and incurred losses.

\begin{model ass}[Independent PIC Log-Normal (\textsl{Model I})]\label{model ass}
The model assumptions for the independent model of \citet{merz2010paid} are:
\begin{itemize}
\item{The cumulative payments $P_{i,j}$ are given by the forward recursion
\begin{equation*}
P_{i,0} =  \exp\left(\xi_{i,0}\right) \;\; \mathrm{and} \;\; \frac{P_{i,j}}{P_{i,j-1}} = \exp\left(\xi_{i,j}\right)\;\; \mathrm{for} \; j=1,\ldots,J
\end{equation*}}
\item{The incurred losses $I_{i,j}$ are given by the backward recursion
\begin{equation*}
I_{i,J} =  P_{i,J}  \;\; \mathrm{and} \;\; \frac{I_{i,j-1}}{I_{i,j}} = \exp\left(-\zeta_{i,j-1}\right).
\end{equation*}}
\item{The random vector
$(\xi_{0,0}, \ldots, \xi_{J,J}, \zeta_{0,0},\ldots, \zeta_{J,J-1})$
has independent components with
\begin{eqnarray*}
\xi_{i,j} & \sim & N \left( \Phi_j~,~ \sigma_j^{2} \right)
\qquad \text{ for $i\in \{0,\ldots, J\}$ and
$j\in \{0,\ldots, J\}$},\\
\zeta_{k,l} & \sim & N \left(\Psi_l~,~ \tau_l^{2} \right)
\qquad \text{ for $k\in \{0,\ldots, J\}$ and
$l\in \{0,\ldots, J-1\}$};
\end{eqnarray*}}
\item{ The parameter vector for the model is $\mathbf{\Theta}=\left(
\Phi_0,\ldots, \Phi_{J},\Psi_0,\ldots, \Psi_{J-1},
\sigma_0,\ldots, \sigma_{J},\tau_0,\ldots, \tau_{J-1}\right)$. It is assumed that the components of $\mathbf{\Theta}$ are independent apriori. The prior density for $\mathbf{\Theta}$ has independent components, with
$\sigma_j,\tau_j$ both positive for all $j$.
\item It follows that 
\begin{equation}
\log\left(\frac{P_{i,j}}{P_{i,j-1}}\right) \sim N \left( \Phi_j~,~ \sigma_j^{2} \right) \;\; \mathrm{and} \;\; \log\left(\frac{I_{i,j}}{I_{i,j+1}}\right) \sim N \left(-\Psi_l~,~ \tau_l^{2} \right)
\end{equation}}
\end{itemize}
\end{model ass}

Let $\{\mathbf{P},\mathbf{I}\}=\{P_{i,j},I_{k,l};~ 0\le i,j,k, \le J,~0\le l \le J-1 \}$. Then, based on Model Assumptions 2.1 and the observed matrices $P$ and $I$, the likelihood for $\bm{\Theta}$ is given by three components, see derivation in \citet[Section 3.3, Equation 3.5]{merz2010paid}. The first and third components correspond to the payment and incurred data and the second component corresponds to the imposition of the restriction that ultimate claims for payments $P_{i,J}$ match incurred $I_{i,J}$ for all accident years, giving:
\begin{eqnarray} \label{LH_IndependentPIC}
f\left(
\left.\mathbf{P},\mathbf{I}\right|\boldsymbol{\Theta}\right)&=&
\underbrace{\prod^{J}_{j=0}\prod^{J-j}_{i=0}
\frac{1}{\sqrt{2\pi}\sigma_{j}P_{i,j}}~
\exp \left\{ -\frac{1}{2\sigma_j^{2}}(\Phi_{j}-\log(\frac{P_{i,j}}{P_{i,j-1}}))^2
\right\} \nonumber}_{\text{Component1: payment}}
\\&&\times ~
\underbrace{\prod^{J}_{i=1}
\frac{1}{\sqrt{2\pi(\upsilon_{J-i}^{2}-\omega_{J-i}^{2})}I_{i,J-i}}~
\exp \left\{ -\frac{1}{2(\upsilon_{J-i}^{2}-\omega_{J-i}^{2})}(\mu_{J-i}-\eta_{J-i}-\log(\frac{P_{i,J-i}}{P_{i,J-i}}))^2
\right\} \nonumber}_{\text{Component2: Discounted final development year restricted payment and incurred}}
\\&&\times ~
\underbrace{\prod^{J-1}_{j=0}\prod^{J-j-1}_{i=0}
\frac{1}{\sqrt{2\pi}\tau_{j}I_{i,j}}~
\exp \left\{ -\frac{1}{2\tau_j^{2}}(-\Psi_{j}+\log(\frac{I_{i,j}}{I_{i,j+1}}))^2
\right\}.}_{\text{Component3: incurred}}
\end{eqnarray}
where $\upsilon_{j}^{2} = \sum_{m=0}^{J}\sigma_{m}^{2}+\sum_{n=j}^{J-1}\tau_{n}^{2};$  \hspace{4mm}
$\omega_{j}^{2} = \sum_{m=0}^{j}\sigma_{m}^{2};$ \hspace{4mm} $\eta_{j} = \sum_{m=0}^{j}\Phi_{m}; 
$ \hspace{1mm} and  $\mu_{j} = \sum_{m=0}^{J}\Phi_{m}-\Sigma_{n=j}^{J-1}\varphi_{n}.$

As noted in \citet{merz2010paid}, there are several consequences of the model assumptions made regarding the restriction $I_{i,J} = P_{i,J}$ which applies for all $i \in \left\{1,2\ldots,J\right\}$. The first is that this condition is sufficient to guarantee that the ultimate loss will coincide for both claims payments and incurred loss data. The second is that this model assumes that there is no tail development factor beyond the ultimate year $J$. However this could be incorporated into such models, see \citet{merz2010estimation}.

\citet{merz2010paid} discuss the relationship between the proposed Independent Log-Normal PIC model and existing models in the literature for payment loss based reserving and incurred loss based reserving. In particular, \citet{merz2010paid} [Section 2.1 and 2.2] show that the resulting cumulative payments $P_{i,j}$, conditional on model parameters $\boldsymbol{\Theta}$, will satisfy the model proposed in \citet{hertig85} and the incurred losses $I_{i,j}$, conditional on model parameters $\boldsymbol{\Theta}$, will satisfy the model proposed in \citet{gogol1993using}. Lemma \ref{LemmaHertigGogol} summarizes their results.

\begin{lemma} \label{LemmaHertigGogol} The relationships between consecutive payment development year losses in a given accident year is given conditionally according to
\begin{equation}
\left[\left.\log\left(\frac{P_{i,j}}{P_{i,j-1}}\right)\right|P_{0:J,0:j-1},\boldsymbol{\Theta}\right] \sim \mathcal{N}\left(\Phi_j,\sigma_j^2\right), \; \forall j \geq 0
\end{equation}
in agreement with Hertig's model. With conditional moments given according to the Chain Ladder property as in \citet[Lemma 5.2]{merz2010paid} by,
\begin{equation}
\mathbb{E}\left[P_{i,j}|P_{0:J,0:j-1},\boldsymbol{\Theta}\right] = P_{i,j-1}\exp\left(\Phi_j + \sigma_j^2/2\right).
\end{equation}
Furthermore, conditional upon the model parameters $\boldsymbol{\Theta}$, for all $0 \leq j < j+l \leq J$ the relationships between consecutive incurred losses in a given accident year are given in \citet{merz2010paid} [Proposition 2.2] according to
\begin{equation}
\left[\left.\log\left(I_{i,j+l}\right)\right|I_{0:J,0:j-1},I_{i,J},\boldsymbol{\Theta}\right] \sim \mathcal{N}\left( \mu_{j+1} + \frac{\nu_{j+1}^2}{\nu_j^2}\left(\log(I_{i,j}) - \mu_j \right), \nu_{j+1}^2(1 - \nu_{j+1}^2/\nu_j^2) \right),
\end{equation}
These results are consistent with the model assumptions of Gogol, and are derived using properties of multivariate normal distribution, see Lemma 2.1 in \citet{merz2010paid}.
\end{lemma}

Furthermore, for all accident years $i \in \left\{1,2,\ldots,J\right\}$, the resulting conditional expected ultimate payment loss equals the expected ultimate incurred loss, given the model parameters $\boldsymbol{\Theta}$, and is expanded in terms of the model parameters according to Equation (\ref{EqnCondExpectUltimateLoss}), which are given by \citet[Equation 1.1]{merz2010paid} as,
\begin{equation} \label{EqnCondExpectUltimateLoss}
E\left.\left[ P_{i,J} \right| \mathbf{\Theta} \right]
=E\left.\left[ I_{i,J} \right| \mathbf{\Theta} \right]
=\exp\left( \sum_{m=0}^J \Phi_m + \sigma_m^2/2 \right).
\end{equation}

\section{Incorporating the Gaussian Copula into Paid-Incurred-Claims Models}
This section discusses an important aspect of extending the original Log-Normal PIC model of \citet{merz2010paid}. In particular, when this model was developed in the independent setting it was observed by the authors that the assumption of conditional independence between $\xi_{i,j}$ and $\zeta_{k,l}$ for all $i,j,k,l \in \left\{1,2,\ldots,J\right\}$ was not necessarily consistent with observations. In particular, they note that \citet{quarg2004munich} discovered evidence for strong linear correlation between incurred and paid ratios. In Section 3.1 we explore in detail a different approach to incorporate dependence structures into the Log-Normal PIC model. Some aspects of the new approach have subsequently been proposed in the literature, while others are novel developments proposed in our article. We note that in developing the extended models, the convenient properties of conjugacy in the Bayesian framework, which aids estimation, is often lost. Hence, after presenting the models we develop efficient state of the art statistical estimation strategies based on adaptive MCMC.

\subsection{Dependence via Payment Loss Ratios and Incurred Loss Ratios (Model II)}
This section generalizes the model by \citet{happ2011paid}, which has a static covariance structure, see \citet[Figure 1.1]{happ2011paid}. We use a Bayesian approach, based on results in Lemma \ref{lemmaWishartDist} and Model Assumptions \ref{modelass2}, to estimate the extended models. We use properties of the matrix-variate Wishart and Inverse Wishart distributions to develop a Gaussian copula based statistical model.  The relevant matrix-variate distributional assumptions and properties are provided in Lemma \ref{lemmaWishartDist} and Lemma \ref{lemmaInvWishartDist}.

\begin{model ass}[Dependent Payment-Incured Ratios: PIC Log-Normal (\textsl{Model II})]\label{modelass2}
The model assumptions for the Gaussian copula PIC Log-Normal model involve:
\begin{itemize}
\item The random matrix $\Sigma_i \in \mathbb{R}^{(2J+1)\times(2J+1)}$ representing the covariance structure for the random vector constructed from log payment ratios $\left(\xi_{i,j} = \log\left(\frac{P_{i,j}}{P_{i,j-1}}\right) \right)$ and log incurred loss ratios $\left(\zeta_{i,j} = \log\left(\frac{I_{i,j}}{I_{i,j+1}}\right) \right)$ in the $i$-th development year, denoted by $\Xi_i = \left(\xi_{i,0},\xi_{i,1},\zeta_{i,1},\xi_{i,2},\zeta_{i,2},\ldots,\xi_{i,J},\zeta_{i,J}\right)$, is assumed distributed according to an inverse Wishart distribution prior (see definition and properties in Lemma \ref{lemmaWishartDist} and Lemma \ref{lemmaInvWishartDist}),
\begin{equation}
\Sigma_i \sim \mathcal{IW}\left(\Lambda_i,k_i\right)
\end{equation}
where $\Lambda_i$ is a $\left((2J+1)\times(2J+1)\right)$ positive definite matrix and $k_i > 2J$.
\item Conditionally, given $\mathbf{\Theta}=\left(
\Phi_0,\ldots, \Phi_{J},\Psi_0,\ldots, \Psi_{J}\right)$ and the $(2J + 1) \times (2J + 1)$-dimensional covariance matrix $\Sigma$, we have:
\begin{itemize}
\item{The random matrix, constructed from log payment ratios $\left(\xi_{i,j} = \log\left(\frac{P_{i,j}}{P_{i,j-1}}\right) \right)$ and log incurred loss ratios $\left(\zeta_{i,j} = \log\left(\frac{I_{i,j}}{I_{i,j+1}}\right) \right)$, denoted by $\Xi$ and comprised of columns  $\Xi_i = \left(\xi_{i,0},\xi_{i,1},\zeta_{i,1},\xi_{i,2},\zeta_{i,2},\ldots,\xi_{i,J},\zeta_{i,J}\right)$, is assumed distributed according to a matrix-variate Gaussian distribution $f_{\Xi}^{MVN}\left(\Xi|M,\Sigma,\Omega\right)$, see the definition and properties in Lemma \ref{lemmaMatrixGaussianDist}. The sufficient matrices are then the $\left((2J + 1) \times (J+1)\right)$ mean matrix $M = \left[\bm{\Theta}',\ldots,\bm{\Theta}'\right]$, column dependence given by $\left((2J + 1) \times (2J+1)\right)$ dimensional covariance matrix $\Sigma$ and row dependence given by $\left((J+1) \times (J+1)\right)$ dimensional matrix $\Omega$. If $\Omega = \mathbb{I}_{J+1}$, the covariance of the vectorization of $\widetilde{\Xi} = Vec(\Xi)$ is
\begin{align}
{\footnotesize{
\widetilde{\Sigma}=\mathbb{C}\text{ov}\left(\widetilde{\Xi}\right) = \bigoplus_{i=0}^J \Sigma_{i} = \left(
\begin{array}{cccc}
\Sigma_0 	& \bm{0} 		& \ldots 	& \bm{0} \\
\bm{0} 		& \Sigma_1 	& \ldots 	& \bm{0} \\
\vdots 		& \ddots 		& \ddots 	& \vdots \\
\bm{0} 		& \bm{0}  	&	\bm{0} & \Sigma_J \end{array}
\right)}},
\end{align}
where it is assumed in the model in \citet{happ2011paid} that $\Sigma_{i} = \mathbb{C}ov(\Xi_i) = \Sigma$. However, this need not be the case and it is possible to consider two extensions, the first in which $\mathbb{C}ov(\Xi_i)$ varied as a function of $i \in \left\{0,1,\ldots,J\right\}$ and the second being the most general of these model structures given by the assumption
\begin{align}
{\footnotesize{
\mathbb{C}\text{ov}\left(\widetilde{\Xi}\right) = \Sigma \otimes \Omega.
}}
\end{align}
}
\item{For all accident years, $i \in \left\{0,1,\ldots,J\right\}$, the ultimate payment losses and incurred losses are equal a.s., $P_{i,J} = I_{i,J}.$}
\end{itemize}
\item{The matrix $\widetilde{\Sigma}$ is positive definite and the components of $\bm{\Theta}$ are independent with prior distributions
\begin{equation}
\Phi_i \sim \mathcal{N}\left(\phi_i,s_i^2\right) \; \text{ and } \; \Psi_j \sim \mathcal{N}\left(\psi_j,t^2_j\right),
\end{equation}
and hyper-prior distributions
\begin{equation}
s^2_i \sim \mathcal{IG}\left(\alpha_i,\beta_i\right) \; \text{ and } \; t^2_j \sim \mathcal{IG}\left(a_j,b_j\right),
\end{equation}
for all $i \in \left\{1,\ldots,J\right\}$ and $j \in \left\{0,\ldots,J\right\}$.
}
\end{itemize}
\end{model ass}

This model extends the model developed in \citet{happ2011paid} which assumes that $\Sigma$ is fixed and known with a tri-diagonal structure. The extension we consider generalizes the dependence structure to be unknown \textit{a priori} and given an inverse Wishart prior for matrix $\widetilde{\Sigma}$, so it forms part of the inference given the data, in the Bayesian inference. In addition, unlike in \citet{happ2011paid} where they assume $\Sigma = \Sigma_i, \forall i \in \left\{0,1,\ldots,J\right\}$, we also allow for variation in $\Sigma_i$ across development years.
 
Given these model assumptions, we now consider two consequences of the proposed model structures for the dependence between the log payment ratios and the log incurred loss ratios given in Lemma \ref{lemmaIndependentDevYr} and Lemma \ref{lemmaMultiVarGaussian}.

\begin{lemma} \label{lemmaIndependentDevYr} Conditional upon $\Lambda_i$ and $k_i$, for all i in $\left\{0,1,\ldots,J\right\}$, and given the marginal distributions for $\Sigma_i$ follow $\Sigma_i \sim \mathcal{IW}\left(\Lambda_i,k_i\right)$ with $\Lambda_i$ a $\left((2J+1)\times(2J+1)\right)$ positive definite matrix and $k_i > 2J$, the joint distribution for the $\left((2J^2 + 3J + 1) \times (2J^2 + 3J + 1)\right)$ covariance matrix $\widetilde{\Sigma}$ for the vectorized matrix for $\Xi$, given by $\widetilde{\Xi} = Vec(\Xi)$, under the assumption of independendence between development years,
\begin{align}
\widetilde{\Sigma}=\mathbb{C}\text{ov}\left(\widetilde{\Xi}\right) = \bigoplus_{i=0}^J \Sigma_{i} = (\Sigma_0 \oplus \cdots \oplus \Sigma_J),
\end{align}
results in a joint distribution given by:
\begin{equation}
\widetilde{\Sigma} \sim \mathcal{IW}\left(\widetilde{\Lambda},\widetilde{k}\right),
\end{equation}
with degrees of freedom $\widetilde{k} = \sum_{i=0}^J k_i > 2J^2 + 3J$ and scale  matrix
\begin{equation}
\widetilde{\Lambda} = \bigoplus_{i=0}^J \Lambda_{i}.
\end{equation}
Furthermore, the joint prior mean and mode for the distribution of the random matrix $\widetilde{\Lambda}$ are
\begin{equation}
\begin{split}
\mathbb{E}\left[\widetilde{\Sigma}|\widetilde{\Lambda},\widetilde{k}\right] &= \frac{1}{\left(\sum_{i=0}^J k_i\right)-\left(2J^2 + 3J\right)}\widetilde{\Lambda}, \; \mathrm{and}\\
m\left(\widetilde{\Sigma}\right)&= \frac{1}{2J^2 + 3J + 1+\sum_{i=0}^J k_i}\widetilde{\Lambda}.
\end{split}
\end{equation}
\end{lemma}
{\textsl{The proof of this result is a consequence of the results in Lemma \ref{lemmaWishartDist}, the model assumptions and the properties of an inverse Wishart distributions; see  \citet{gupta2000matrix}[Chapter 3].}}
\EndProof

\begin{remark} We can demonstrate that under the proposed model assumptions the selection of the factorized covariance structure in Lemma \ref{lemmaIndependentDevYr} produces Bayesian conjugacy in the joint posterior of the model parameters given observed payment losses and incurred losses.
\end{remark}

\begin{remark} It is noted in \citet{happ2011paid} and Lemma \ref{lemmaIndependentDevYr} that in formulating the likelihood structure for this dependent model it is more convenient to work with the one-to-one (invertible) transformation for the observed data defined marginally for the $i$-th development year according to
\begin{equation}
\left[\bm{X}_i|\bm{\Theta}\right] = \left[B_i\Xi_i|\bm{\Theta}\right] \sim \mathcal{N}\left(B_iM_i,B_i\Sigma_iB_i'\right),
\end{equation}
where $M_i$ is the $i$-th column of matrix $M$ and $\bm{X}_i \in \mathbb{R}^{2J+1}$ defined by \\$\bm{X}_i = \left[\log I_{i,0},\log P_{i,0},\log I_{i,1},\log P_{i,1}, \ldots,\log I_{i,J-1},\log P_{i,J-1},\log I_{i,J}\right]$. This results in the joint matrix variate Normal distribution for random matrix $X = \left[\bm{X}_0',\bm{X}_1',\ldots,\bm{X}_J'\right]$ of all observed losses for payment and incurred data given after vectorisation $\widetilde{\bm{X}} = Vec\left(\bm{X}\right)$ by
\begin{equation} \label{EqnTransformedGaussianVect}
\left[\widetilde{\bm{X}}|\bm{\theta}\right] = \left[B\widetilde{\Xi}|\bm{\Theta}\right] \sim \mathcal{N}\left(BVec(M),B\left(\Sigma \otimes \Omega\right)B^T\right).
\end{equation}
\end{remark}

Furthermore, if we consider the property of multivariate Gaussian distributions given in Lemma \ref{lemmaMultiVarGaussian} we can find for the $i$-th accident year the required conditional distribution of the unobserved claims for payment and incurred loss data under the specified model. Furthermore, we can find the conditional distribution for unobserved claims for payment and incurred losses in the $i$-th accident year, given all observed claims triangles for payments and incurred losses data, see Lemma \ref{lemmaMultiVarGaussian} below. This is directly relevant for specifying the resulting likelihood model.

\begin{lemma} \label{lemmaMultiVarGaussian} Consider a $(n \times 1)$ random vector $\bm{Y}$ with multivariate Gaussian distribution, $\bm{Y} \sim \mathcal{N}\left(\bm{\mu},\Sigma\right)$, where $\bm{\mu} = \left[ \mu_1, \ldots, \mu_n\right]$ and $\mathbb{C}\text{ov}\left(\bm{Y}\right) = \Sigma$, and partition $Y = \left[Y^{(1)'},Y^{(2)'}\right]'$. Then the conditional distribution of $Y^{(1)}$ given $Y^{(2)}$ and the marginal distribution of $\bm{Y}^{(1)}$ is
\begin{equation}
\left[\bm{Y}^{(1)} | \bm{Y}^{(2)}\right] \sim \mathcal{N}\left(\bar{\bm{\mu}},\bar{\Sigma}\right),
\end{equation}
with $\bar{\bm{\mu}} = \bm{\mu}_1 + \Sigma_{1,2}\Sigma_{2,2}^{-1}\left(\bm{Y}^{(2)} - \bm{\mu}^{(2)}\right)$ and the Schur complement $\bar{\Sigma} = \Sigma_{1,1} - \Sigma_{1,2}\Sigma_{2,2}^{-1}\Sigma_{2,1}$ under the partitioning of the mean and covariance given by
\begin{equation}
\mu = \left[
\begin{array}{c}
\bm{\mu}_{1} \\
\bm{\mu}_{2}
\end{array}
\right] \; \text{ and } \;
\Sigma =
\left(
\begin{array}{cc}
\Sigma_{1,1} 	& \Sigma_{2,1} \\
\Sigma_{1,2} 		&  \Sigma_{2,2}
\end{array}
\right).
\end{equation}
\end{lemma}

Definition \ref{DefnPermMatrices} below defines a family of permutation matrix operators. This permutation family allows the representation of the vectorization of the two loss triangles under different permutations that facilitate dependence specifications in the proposed models that admit conjugacy.

\begin{defi} \label{DefnPermMatrices} Let $Y$ be an $n \times n$ matrix, with $\widetilde{Y}= \left[Y_{1,1},Y_{1,2},\ldots \right]'$ and with $Vec(Y)$ defined as\\ $Vec(Y) = \left[Y_{1,1},Y_{1,2},\ldots,Y_{1,n},Y_{2,1},\ldots,Y_{2,n},\ldots,Y_{n,n}\right]'$. Define the family of permutation matrix operators, denoted by $\mathcal{P}^*_{\bm{i_{~}}}$ and indexed by $p \times 2, \; p\leq n^2$, indices matrix (vector of tuple elements) $\bm{i_{~}}$ with $j$-th element $\left[\bm{i_{~}}\right]_j = \left\{ \left(k,l\right); k,l \in \left\{1,2,\ldots,n\right\} \right\}$, and defined according to the mapping $\mathcal{P}^*_{\bm{i_{~}}}: Vec(Y) \mapsto Vec(Y)^*$ given by
\begin{equation}
\begin{split}
\mathcal{P}_{\bm{i_{~}}}^*\left(Vec(Y)\right) &= P_{\bm{i_{~}}}^*Vec(Y)\\
&= \left[Y_{\left[\bm{i_{~}}\right]_1},Y_{\left[\bm{i_{~}}\right]_2},\ldots,Y_{\left[\bm{i_{~}}\right]_p},Vec(Y)_{\setminus \bm{i_{~}}}'\right]',
\end{split}
\end{equation}
where we define $Y_{\left[\bm{i_{~}}\right]_j}$ as the element of matrix $Y$ corresponding to the resulting tuple index location in the $j$-th element (column) of (tuple vector) $\bm{i_{~}}$, $P_i^*$ an $n^2 \times n^2$ permutation matrix defined by
\begin{equation}
P_{\bm{i_{~}}}^* = P_{\bm{i_{~}}} \oplus \mathbb{I}_{n^2 - p}
=\left[
\begin{array}{cc}
P_{\bm{i_{~}}}  & 0_{n^2-p,n^2-p}\\
0_{n^2-p,n^2-p} & \mathbb{I}_{n^2-p}
\end{array}
\right],
\end{equation}
and $P_{\bm{i_{~}}}$ is a matrix with only non-zero identity elements at the $p$ locations in the indices matrix tuples in $\bm{i_{~}}$ corresponding to index elements.
\end{defi}

Using the property of the multivariate Gaussian distribution in Lemma \ref{lemmaMultiVarGaussian}, one can state the result in Proposition \ref{Propa} which is based on a generalization of the result in \citet{happ2011paid}[Lemma 2.1] to the model developed above. We consider two cases for the dependence structures in Proposition \ref{Propa} and Proposition \ref{Propb}.

\begin{prop} \label{Propa} Consider the $i$-th accident year. Conditional on the model parameters $\bm{\Theta}$ and the covariance matrix of the $i$-th accident year
\begin{equation}
\Sigma_i =
\left(
\begin{array}{cc}
\left[\Sigma_{i}\right]_{1,1} 	& \left[\Sigma_{i}\right]_{2,1} \\
\left[\Sigma_{i}\right]_{1,2} 		&  \left[\Sigma_{i}\right]_{2,2}
\end{array}
\right),
\end{equation}
the dependence structure $\Omega = \mathbb{I}_{J+1}$ and the observed payment losses and incurred losses in the $i$-th accident year, denoted by $\bm{X}_i^{(1)} = \left[\log I_{i,0},\log P_{i,0},\log I_{i,1},\log P_{i,1}, \ldots,\log I_{i,J-i},\log P_{i,J-i}\right]$ with $\bm{X}_i \in \mathbb{R}^{q}$, the conditional distribution for the log of the unobserved payment losses and incurred losses \\
$(\bm{X}_i^{(2)} = \left[\log I_{i,J-i+1},\log P_{i,J-i+1},\ldots,\log I_{i,J-1},\log P_{i,J-1},\log I_{i,J}\right])$ is given by
\begin{equation}
\left[\bm{X}_i^{(2)}|\bm{X}_i^{(1)}, \bm{\Theta}\right] \sim \mathcal{N}\left(\bar{\bm{\mu}}^{(2)},\bar{\Sigma_i}^{(2)}\right)
\end{equation}
where $\bar{\bm{\mu}}_i^{(2)} = \bm{\mu}_i^{(2)} + \left[\Sigma_{i}\right]_{2,1}\left[\Sigma_{i}\right]_{1,1}^{-1}  \left(\bm{X}_i^{(1)} - \bm{\mu}_i^{(1)}\right)$ and $\bar{\Sigma_i}^{(2)} = \left[\Sigma_{i}\right]_{22}$.
\end{prop}

\begin{prop}[Conditional Distribution of Unobserved Payment and Incurred Losses]{\label{Propb} Consider the $i$-th accident year and define indices for this year (vector of tuples), given by matrix \\
$\bm{i_{~}} = \left\{\left(k,j\right):\forall j \in \left\{J-k+1,\ldots,J\right\}\right\}\cup\left\{\left(k,j\right):\forall k \in \left\{0,1,\ldots,J\right\}, j \in \left\{0,\ldots,J-k\right\}\right\}$. Then consider the transformed vector of log payment and incurred losses $\mathcal{P}_{\bm{i_{~}}}^*\left(\widetilde{X}\right)$ defined by
\begin{equation}
\mathcal{P}_{\bm{i_{~}}}^*\left(\widetilde{X}\right) \sim \mathcal{N}\left(P_{\bm{i_{~}}}^* Vec(M), P_{\bm{i_{~}}}^*\left(\Sigma \otimes \Omega\right)(P_{\bm{i_{~}}}^*)'\right),
\end{equation}
for which the first $J-i$ elements of the permuted random vector $\left[\widetilde{X}^*\right]^{(1)} = \left[\mathcal{P}_{\bm{i_{~}}}^*\left(\widetilde{X}\right)\right]_{1:J-i-1}$ correspond to all un-observed payment and incurred loss random variables, and the remaining $J-i$ to $J-i+\left(\sum_{n=-1}^{J}(J-n)\right)$ elements are the observed payment and incurred data, denoted  $\left[\widetilde{X}^*\right]^{(2)} = \left[\mathcal{P}_{\bm{i_{~}}}^*\left(\widetilde{X}\right)\right]_{J-i:J-i+\left(\sum_{n=-1}^{J}(J-n)\right)}$. Then, conditional on the model parameters $\bm{\Theta}$, the general dependence structre $\widetilde{\Sigma} = \Sigma\otimes\Omega$ with matrices $\Sigma$ and $\Omega$, and $\left[\widetilde{X}^*\right]^{(2)}$ the following results hold:
\begin{itemize}
\item{The conditional distribution for the log of the \underline{unobserved payment losses and incurred losses} in the $i$-th year, corresponding to the first $J-i$ elements of the permuted random vector $\left[\widetilde{X}^*\right]^{(1)} = \left[\mathcal{P}_{\bm{i_{~}}}^*\left(\widetilde{X}\right)\right]_{1:J-i-1}$ is given by
\begin{equation}
\left[\left[\widetilde{X}^*\right]^{(1)}|\left[\widetilde{X}^*\right]^{(2)}, \bm{\Theta}\right] \sim \mathcal{N}\left(\bar{\bm{\mu}}^{(1)},\bar{\Sigma_i}^{(1)}\right).
\end{equation}}
\item{The covariance matrix $\bar{\Sigma_i}^{(1)}$ is the postive definite \\ $\left(J-i+\left(\sum_{n=-1}^{J}(J-n)\right)\right)\times\left(J-i+\left(\sum_{n=-1}^{J}(J-n)\right)\right)$ sub-matrix denoted below by $\Gamma$ and defined by the top sublock of the permuted covariance matrix
\begin{equation}
P_{\bm{i_{~}}}^*\left(\Sigma \otimes \Omega\right)(P_{\bm{i_{~}}}^*)'
 = \left[
\begin{array}{cc}
\Gamma &  \left[P_{\bm{i_{~}}}^*\left(\Sigma \otimes \Omega\right)(P_{\bm{i_{~}}}^*)'\right]_{2,1} \\
\left[P_{\bm{i_{~}}}^*\left(\Sigma \otimes \Omega\right)(P_{\bm{i_{~}}}^*)'\right]_{1,2} & \left[P_{\bm{i_{~}}}^*\left(\Sigma \otimes \Omega\right)(P_{\bm{i_{~}}}^*)'\right]_{2,2} 	
\end{array}
\right].
\end{equation}
}
\item{Given, this covariance matrix one specifies the conditional mean vector, denoted by\\ $\bar{\bm{\mu}}^{(1)} = \bm{\mu}^{(1)} + \Gamma_{2,1}\Gamma_{1,1}^{-1}  \left(\left[\widetilde{X}^*\right]^{(2)} - \bm{\mu}^{(2)}\right)$, according to the subblocks of the $\Gamma$ covariance matrix defined with respect to the first $J-i$ elements $\left[\widetilde{X}^*\right]^{(1)}$ and remaining elements of $\left[\widetilde{X}^*\right]^{(2)}$ as well as $\bm{\mu}^{(1)} = \left[P_{\bm{i_{~}}}^* Vec(M)\right]_{1:J-i}$ and the second $J-i$ to $J-i+\left(\sum_{n=-1}^{J}(J-n)\right)$ elements are given by $\bm{\mu}^{(2)} = \left[P_{\bm{i_{~}}}^* Vec(M)\right]_{J-i:J-i+\left(\sum_{n=-1}^{J}(J-n)\right)}$.}
\end{itemize}
}
\end{prop}

Having specified these statistical assumptions, we can formulate the joint likelihood from the observed data for both payments and incurred claims conditional upon the model parameters according to Equation (\ref{LH_DependentPICModI}). The joint data likelihood function in the dependent Log-Normal PIC Model I for the random vector of observations corresponding to the first $\sum_{n=-1}^{J}(J-n)$ elements of the permuted random vector, given by $\left[\widetilde{X}^*\right]^{(1)} = \left[\mathcal{P}_{\bm{i_{~}}}^*\left(\widetilde{X}\right)\right]_{1:\left(\sum_{n=-1}^{J}(J-n)\right)}$, where we define indices in this case by $\bm{i_{~}} = \left\{\left(i,j\right):\forall i \in \left\{0,1,\ldots,J\right\}, j \in \left\{0,\ldots,J-i\right\}\right\}$. The resulting likelihood is given by the matrix-variate Gaussian distribution in Equation (\ref{LH_DependentPICModI}).
{\small{
\begin{equation} \label{LH_DependentPICModI}
\begin{split}
&f\left(
\left.\left[\widetilde{X}^*\right]^{(1)} \right|\boldsymbol{\Theta},\Sigma, \Omega\right)=\\
&\frac{\exp\left[ \left(\left[\widetilde{X}^*\right]^{(1)}-\left[\mathcal{P}_{\bm{i_{~}}}^*\left(Vec(M)\right)\right]^{(1)}\right) \left[ \left[P_{\bm{i_{~}}}^*\left(\Sigma \otimes \Omega\right)(P_{\bm{i_{~}}}^*)'\right]^{(1)}  \right]^{-1} \left(\left[\widetilde{X}^*\right]^{(1)}-\left[\mathcal{P}_{\bm{i_{~}}}^*\left(Vec(M)\right)\right]^{(1)}\right) \right]}{ \left(2\pi\right)^{\left(\sum_{n=-1}^{J}(J-n)\right)/2}\left|\left[P_{\bm{i_{~}}}^*\left(\Sigma \otimes \Omega\right)(P_{\bm{i_{~}}}^*)'\right]^{(1)} \right|^{\left(\sum_{n=-1}^{J}(J-n)\right)/2} }
\end{split}
\end{equation}
}}
We note that our proposed models also allow one to consider the dependence structures of \citet{happ2011paid} who assume that $\Sigma_i = \Sigma, \forall i \in \left\{0,1,\ldots,J\right\}$ and $\Omega = \mathbb{I}_{J+1}$, with the specific setting of $\Sigma$ via a tri-diagonal correlation matrix with three correlation parameters which are assumed either known \textit{a priori} or estimated prior to inference in the PIC model. Such an approach was motivated by the belief that a positive change in incurred loss results in an immediate payment in the same development period, and the remaining increased expectation is paid with some settlement delay.  Therefore, the incurred losses increments $\zeta_{i}^{j}$ are assumed to be positively correlated to the claims payments increments $\xi_{i,j}$, $\xi_{i,j+1}$ and $\xi_{i,j+2}$ with positive correlations $\rho_{0},\rho_{1},\rho_{2}$, respectively. However, the argument for more general dependence structure that were introduced as extensions to the model of \citet{happ2011paid} are developed to account for the fact that these assumption may not be true, especially in long tail portfolios, such as compulsory third party. If the status of a claimant changes and requires long term medical treatment and rehabilitation, it might result in substantially high loss in the subsequent lengthy lag periods. The paper also assumes that the dependence is the same across different lag years, which is not always a realistic feature of such data. Our article aims to fill this gap and enhance the correlation structure in PIC models whilst maintaining a parsimonious model specification.

\subsection{Dependence Between Development Lag Years for Payment Losses and Incurred Losses (Model III)} \label{DevLagYrs}
This section considers an alternative dependence structure motivated by the fact that dependence between lag years is practically appealing in claims reserving practice. It affects the estimation of outstanding claims the most, and is widely recognized by actuaries in claims reserving. Lag is the measure of the difference between incurred month and paid month. Depending on the nature of the portfolio,  many insurance claims often have lengthy settlement periods due to various reasons such as late reported claims, judicial proceedings, or schedules of benefits for employer's liability claims. During the lengthy lag periods, large payments in the previous lag period normally follow by small payments in the subsequent lag period.  There may in fact be positive correlation if all periods are equally impacted by a change in claims status, e.g. if a claim becomes litigated, resulting in a huge increase in claims cost. There may also be negative correlation if a large settlement in one period replaces a stream of payments in later periods. The model developed in this section mainly focuses on capturing this feature of dependence between lag years. To achieve this we propose a block covariance structure for the covariance matrix, which will respect the dependence between each lag point. The model we propose is summarised in Model Assumptions \ref{modelass3} below.

\begin{model ass}[Dependent Development Lag Years: PIC Log-Normal (\textsl{Model III})]\label{modelass3}
The following statistical model assumptions are developed:
\begin{itemize}
\item Let $\Sigma^{P}_i \in \mathbb{S}\mathbb{D}^+(J-i)$ be the $(J-i)\times(J-i)$ random covariance matrix on the space $\mathbb{S}\mathbb{D}^+(J-i)$ of positive definite covariance matrices of dimension $(J-i)\times(J-i)$ corresponding to the observed payment data $\left[\log P_{i,0},\log P_{i,1},\ldots,\log P_{i,J-i}\right]$ in the $i$-th accident year and analogously for incurred loss data $\Sigma^{I}_i \in \mathbb{S}\mathbb{D}^+(J-i)$. When $i = 0$ we consider $\Sigma^{P}_0 \in \mathbb{S}\mathbb{D}^+(J+1)$  and for incurred loss data $\log I_{0,0:J-1}$ with $\Sigma^{I}_0 \in \mathbb{S}\mathbb{D}^+(J)$.
Assume an inverse Wishart distribution (see Lemma \ref{lemmaInvWishartDist} and Lemma \ref{lemmaWishartDist}) for each matrix defined according to
\begin{equation}
\Sigma^P_i \sim \mathcal{IW}\left(\Lambda^P_i,k^P_i\right) \; \text{ and } \; \Sigma^I_i \sim \mathcal{IW}\left(\Lambda^I_i,k^I_i\right),
\end{equation}
where $\Lambda^P_i$ and $\Lambda^I_i$ are the inverse scale matrices for the prior for the payment and incurred loss data covariance priors respectively. Hence, the joint covariance between all observed payment and incurred loss data satisfies the telescoping diagonal block size covariance structure:
\begin{equation} \label{EqnTelescopingDependence}
\begin{split}
&\widetilde{\Sigma} = \mathbb{C}\text{ov}\left(\left[\log P_{0,0},\ldots,\log P_{0,J} ,\log P_{1,0},\log P_{1,J-1},\ldots,\log P_{J,0},\log I_{0,0},\ldots,\log I_{0,J-1},\ldots,\log I_{J,0}\right]\right)\\
&\;\;\;\; =\left(\bigoplus_{i=0}^J \Sigma^{P}_0\right)\oplus\left(\bigoplus_{i=0}^J \Sigma^{I}_0\right) \sim \mathcal{IW}\left(\left(\bigoplus_{i=0}^J \Lambda^{P}_0\right)\oplus\left(\bigoplus_{i=0}^J \Lambda^{I}_0\right), \sum_{i=0}^J \left(k_i^{P}+ k_i^I\right) \right).
\end{split}
\end{equation}
\item Conditionally, given $\mathbf{\Theta}=\left(
\Phi_0,\ldots, \Phi_{J},\Psi_0,\ldots, \Psi_{J}\right)$ and the covariance matrix $\widetilde{\Sigma}$, we have the following results
\begin{itemize}
\item{Consider the marginal distribution of the first $\left(\sum_{n=-1}^{J}(J-n)\right)$ elements of the vectorized random matrix of observed payment and incurred losses, with $i$-th column $\bm{X}_i \in \mathbb{R}^{2J+1}$ given by $$\bm{X}_i = \left[\log I_{i,0},\log P_{i,0},\log I_{i,1},\log P_{i,1}, \ldots,\log I_{i,J-1},\log P_{i,J-1},\log I_{i,J}\right].$$ 

Then given the matrix of permutation indices $\bm{i_{~}} = \left[(1,2),(1,4),\ldots,(1,2(J-1))\right.$\\ $\left.,(2,2),(2,4),\ldots,(2,2J-4),\ldots,(J,1),(1,1),(1,3),\ldots(J-1,1),(J-1,2)\right]$ characterizing the elements of the marginal distribution for the observations, the transform $\mathcal{P}_{\bm{i_{~}}}^*\left(Vec(\bm{X})\right)$ has multivariate Gaussian distribution with covariance structure $\widetilde{\Sigma}$. Note, $\mathcal{P}_{\bm{i_{~}}}^*\left(Vec(\bm{X})\right) = \left[\log P_{0,0},\log P_{0,1},\ldots,\log P_{0,J}\right.$\\
$\left.,\ldots,\log P_{J,0},\log I_{0,0},\ldots,\log I_{0,J-1},\log I_{1,0},\ldots,\log I_{J-1,0},\log I_{J-1,1}\right]$.}
\item{For all accident years, $i \in \left\{0,1,\ldots,J\right\}$, the ultimate payment losses and incurred losses are equal almost surely, $P_{i,J} = I_{i,J}.$}
\end{itemize}
\item{The matrix $\widetilde{\Sigma}$ is positive definite and the components of $\bm{\Theta}$ are independent with prior distributions
\begin{equation}
\Phi_i \sim \mathcal{N}\left(\phi_i,s_i^2\right) \; \text{ and } \; \Psi_j \sim \mathcal{N}\left(\psi_j,t_j^2\right)
\end{equation}
and hyper-prior distributions
\begin{equation}
s^2_i \sim \mathcal{IG}\left(\alpha_i,\beta_i\right) \; \text{ and } \; t^2_j \sim \mathcal{IG}\left(a_j,b_j\right)
\end{equation}
for all $i \in \left\{1,\ldots,J\right\}$ and $j \in \left\{0,\ldots,J\right\}$.
}
\end{itemize}
\end{model ass}

This proposed model is therefore another generalization of the dependence structure of the model structure proposed in \citet{happ2011paid}. As such, the likelihood structure is given by the multivariate Gaussian given in Equation (\ref{LH_DependentPICModI}) with the covariance matrix given by the telescoping diagonal block size covariance matrix structure in Equation (\ref{EqnTelescopingDependence}).

\subsection{Hierarchical Bayesian Conjugacy Under Gausian Copula Dependent PIC: Models I, II, III}
Under the Gaussian copula based dependence models, the ability to obtain the observed data likelihood in the form of a multivariate Gaussian distribution means that we obtain conjugacy properties. This makes the estimation of such models by MCMC more efficient because we can us Gibbs sampling in blocks. This section presents a generic set of such conjugate models for any of the dependence structures specified in Models I, II and III. 

\begin{lemma} \label{LemmaTransformData} Conditional upon the parameters $\bm{\Theta}$ and the covariance matrix $\Sigma$, the permuted data $\mathcal{P}_{\bm{i_{~}}}^*\left(Vec(\bm{X})\right)$ can be transformed to produce the independent likelihood in Equation (\ref{LH_IndependentPIC}). This is achieved by considering the class of vector transformations $\mathcal{T}: \mathbb{R}^{(d \times 1)} \mapsto \mathbb{R}^{(d \times 1)}$, such that if the initial covariance structure of random vector $\bm{X}$ was given by $\Sigma = \mathbb{C}\text{ov}\left(\bm{X}\right)$, then the resulting covariance structure $\mathbb{C}\text{ov}\left(\mathcal{T}(\bm{X})\right) = \mathbb{I}_{d}$. The required rotation-dilation transformation is obtained by the spectral decomposition of the covariance according to a spectral decomposition (see \citet{stoica1997introduction})
$\Sigma = U\Lambda^{\frac{1}{2}}U'$ where $U$ is a $(d \times d)$ matrix of eigenvectors of $\Sigma$ and $\Lambda$ is a diagonal $d \times d$ matrix of the eigenvalues of $\Sigma$. Therefore the following holds for each of the models under a transform of the vector of permuted observations $\mathcal{T}\left(\mathcal{P}_{\bm{i_{~}}}^*\left(Vec(\bm{X})\right)\right)$:
\begin{enumerate}
\item{Model II - When $\tilde{\Sigma} = \Sigma \otimes \Omega$, with $\Omega = \mathbb{I}_{J+1}$ then, $\mathcal{T}\left(\mathcal{P}_{\bm{i_{~}}}^*\left(Vec(\bm{X})\right)\right) = \left(U\Lambda^{\frac{1}{2}} \otimes \mathbb{I}_{J+1}\right) \mathcal{P}_{\bm{i_{~}}}^*\left(Vec(\bm{X})\right),$ where the $\left((2J+1)\times(2J+1)\right)$ covariance $\Sigma$ is decomposed as $U\Lambda^{\frac{1}{2}}U'$.}
\item{Model II - When $\tilde{\Sigma} = \bigoplus_{i=0}^J \Sigma_i$, $\mathcal{T}\left(\mathcal{P}_{\bm{i_{~}}}^*\left(Vec(\bm{X})\right)\right) = \left(\bigoplus_{i=0}^J U_i\Lambda_i^{\frac{1}{2}}\right) \mathcal{P}_{\bm{i_{~}}}^*\left(Vec(\bm{X})\right),$ where each accident year's dependence between payments and incurred losses is given by the $(2J+1)\times(2J+1)$ matrix $\Sigma_i$ which is decomposed as $U_i\Lambda_i^{\frac{1}{2}}U_i'$.}
\item{Model III - When $\tilde{\Sigma} = \left(\bigoplus_{i=0}^J \Sigma^{P}_0\right)\oplus\left(\bigoplus_{i=0}^J \Sigma^{I}_0\right)$, $$\mathcal{T}\left(\mathcal{P}_{\bm{i_{~}}}^*\left(Vec(\bm{X})\right)\right) = \left(\bigoplus_{i=0}^J U^{P}_i\left(\Lambda^{P}_i \right)^{\frac{1}{2}}\right)\oplus\left(\bigoplus_{i=0}^J U^{I}_i\left(\Lambda^{I}_i \right)^{\frac{1}{2}}\right) \mathcal{P}_{\bm{i_{~}}}^*\left(Vec(\bm{X})\right)$$ where each of the covariance matrices $\Sigma^P_i$ and $\Sigma^I_i$ decomposed to $U^P_i\left(\Lambda^P_i\right)^{\frac{1}{2}}(U^P_i)'$ and $U^I_i\left(\Lambda^I_i\right)^{\frac{1}{2}}(U^I_i)'$.}
\end{enumerate}
In each case, the resulting transformed random vector $\mathcal{T}\left(\mathcal{P}_{\bm{i_{~}}}^*\left(Vec(\bm{X})\right)\right)$, with elements $\widetilde{P}_{i,j}$ and $\widetilde{I}_{i,j}$, will produce a likelihood model given for the transformed data according to the independent Model I of \citet{merz2010paid} as defined in Equation (\ref{LH_IndependentPIC}). Of course this is defined now with respect to components in the likelihood corresponding to the transformed components, as detailed in Equation (\ref{EqnTransformedGaussianVect}).
\end{lemma}

\begin{remark}
The consequence is that results in Lemma \ref{LemmaTransformData} are that the conjugacy properties derived for the independent model in \citet{merz2010paid} can be directly applied post-transformation. This is of direct interest for MCMC based sampling schemes.
\end{remark}

In the models described so far, the following full conditional posterior distributions are now of relevance to the Bayesian MCMC estimation procedures developed for Models I, II and III.

\begin{lemma} The full conditional posterior distributions for sub-blocks of the model parameters can be decomposed under Model I, II and III into a conjugate model.
\begin{itemize}
\item{\textbf{Conjugate Posterior Distribution for Development Factors:} under the transformations $\mathcal{T}\left(\mathcal{P}_{\bm{i_{~}}}^*\left(Vec(\bm{X})\right)\right)$ on the data, described in Lemma \ref{LemmaTransformData}, the full conditional posterior distributions for sub-blocks of the transformed model parameters $\left(\widetilde{\Phi}_{0:J},\widetilde{\Psi}_{0:J}\right)$ are given by (see \citet{merz2010paid} [Theorem 3.4] for the independent case):
\begin{equation}
\left[\widetilde{\Phi}_{0:J},\widetilde{\Psi}_{0:J}|\Sigma,\Omega,\mathcal{T}\left(\mathcal{P}_{\bm{i_{~}}}^*\left(Vec(\bm{X})\right)\right)\right] \sim \mathcal{N}\left(\Pi,\Delta\right)
\end{equation}
with posterior mean $\Pi$ and posterior covariance $\Delta$, where the components of $\Delta^{-1} = \left(a_{n,m}\right)_{0 \geq n,m \leq 2J}$ are each given by
\begin{equation}
\begin{split}
a_{n,m} &= \left(s_n^{-2}+(J-n+1)\sigma_n^{-2}\right)\delta_{n=m} + \sum_{i=0}^{(n-1)\wedge(m-1)}\left(\nu^2_i - \omega^2_i\right)^{-1}, \; \text{ for } 0\leq n, m \leq J,\\
a_{J+1+n,J+1+m} &= \left(t_n^{-2} + (J-n)\tau_n^{-2} \right)\delta_{n=m} + \sum_{i=0}^{n\wedge m}\left(\nu^2_i - \omega^2_i\right)^{-1}, \; \text{ for } 0\leq n, m \leq J-1,\\
a_{n,J+1+m} &= \Delta_{n,J+1+m} = - \sum_{i=0}^{(n-1)\wedge m} \left(\nu^2_i - \omega^2_i\right)^{-1}, \; \text{ for } 0\leq n \leq J, 0 \leq m \leq J-1;\\
\end{split}
\end{equation}
where $\delta_{n=m}$ is the indicator of the event that index $m$ matches $n$, $m \wedge n$ is the minimum of $m$ and $n$ and the posterior mean is given on the transformed scale by,
\begin{equation}
\begin{split}
\left[\widetilde{\Phi}_{0:J},\widetilde{\Psi}_{0:J}\right] &= \Delta\left(\widetilde{c}_0,\widetilde{c}_1,\ldots,\widetilde{c}_J,\widetilde{b}_0,\ldots,\widetilde{b}_J\right),
\end{split}
\end{equation}
with
\begin{equation}
\begin{split}
\widetilde{c}_j &= s_j^{-2}\phi_j + \sigma^2_j \sum_{i=0}^{J-j} \log\left(\frac{\widetilde{P}_{i,j}}{\widetilde{P}_{i,j-1}}\right) + \sum_{i=J-j+1}^J \left(\nu_{J-i}^2 - \omega_{J-i}^2\right)^{-1}\log\left(\frac{\widetilde{I}_{i,J-i}}{\widetilde{P}_{i,J-i}}\right), \\
\widetilde{b}_j &= t_j^{-2}\psi_j + \tau^2_j \sum_{i=0}^{J-j-1} \log\left(\frac{\widetilde{I}_{i,j}}{\widetilde{I}_{i,j+1}}\right) - \sum_{i=J-j}^J \left(\nu_{J-i}^2 - \omega_{J-i}^2\right)^{-1}\log\left(\frac{\widetilde{I}_{i,J-i}}{\widetilde{P}_{i,J-i}}\right).
\end{split}
\end{equation}
Given the transform vector $\left[\widetilde{\Phi}_{0:J},\widetilde{\Psi}_{0:J}\right]$, the parameters on the orginal scale can be expressed according to the unique solution to the system of linear equations:
\begin{enumerate}
\item{Model II - On the untransformed scale, the solution is given by the following system of equations
\begin{equation}
\begin{split}
\left[\Phi_{0:J},\Psi_{0:J}\right]' = U^{-1}\Lambda^{-\frac{1}{2}} \left[\widetilde{\Phi}_{0:J},\widetilde{\Psi}_{0:J}\right].
\end{split}
\end{equation}
}
\item{Model II - On the untransformed scale, the solution is given by the following system of equations for each $i \in \left\{0,1,\ldots,J\right\}$, where we can randomly select $i$ or deterministically scan through $i$ for the results,
\begin{equation}
\begin{split}
\left[\Phi_{0:J},\Psi_{0:J}\right]' = U_i^{-1}\Lambda_i^{-\frac{1}{2}} \left[\widetilde{\Phi}_{0:J},\widetilde{\Psi}_{0:J}\right].
\end{split}
\end{equation}
}
\item{Model III - On the untransformed scale, the solution is given by the following system of equations,
\begin{equation*}
\begin{split}
\left[\Phi_{0:J},\Phi_{0:J-1},\Phi_{0:J-2},\ldots,\Phi_{J}\right]' &= \bigoplus_{i=0}^J \left(U^{P}_i\right)^{-1}\left(\Lambda^{P}_i \right)^{-\frac{1}{2}}\left[\widetilde{\Phi}_{0:J},\widetilde{\Phi}_{0:J-1},\widetilde{\Phi}_{0:J-2},\ldots,\widetilde{\Phi}_{J}\right],\\
\left[\Psi_{0:J},\Psi_{0:J-1},\Psi_{0:J-2},\ldots,\Psi_{J}\right]' &= \bigoplus_{i=0}^J \left(U^{I}_i\right)^{-1}\left(\Lambda^{I}_i \right)^{-\frac{1}{2}}\left[\widetilde{\Psi}_{0:J},\widetilde{\Psi}_{0:J-1},\widetilde{\Psi}_{0:J-2},\ldots,\widetilde{\Psi}_{J}\right].
\end{split}
\end{equation*}
}
\end{enumerate}
}
\item{\textbf{Conjugate Posterior Distribution for the Covariance Matrix:} Given the transformed observed payment and incurred losses have a multivariate Gaussian likelihood, as specified in Equation (\ref{LH_DependentPICModI}), with covaraince matrix $\widetilde{\Sigma} = \Sigma \otimes \Omega$ and mean vector $Vec\left(M\right)$. Then the posterior for the covariance matrix is the Inverse-Wishart-Gaussian distribution detailed in \citet{PetersBA1} [Section 3] and \citet{PetersBA2}
\begin{equation*} \label{EqnConjugateIW}
\begin{split}
\left[\widetilde{\Sigma}|\Phi_{0:J},\Psi_{0:J},\mathcal{T}\left(\mathcal{P}_{\bm{i_{~}}}^*\left(Vec(\bm{X})\right)\right)\right] \sim \mathcal{IW}\left(\Lambda + \mathcal{T}\left(\mathcal{P}_{\bm{i_{~}}}^*\left(Vec(\bm{X})\right)\right) \mathcal{T}\left(\mathcal{P}_{\bm{i_{~}}}^*\left(Vec(\bm{X})\right)\right)',dim\left(Vec(\bm{X})\right) + \widetilde{k}\right)
\end{split}
\end{equation*}
In cases in which the covariance matrix $\widetilde{\Sigma}$ takes any of the block diagonal forms presented in Models II and III, we may utilise Lemma \ref{lemmaWishartDist} and the result in Equation (\ref{EqnConjugateIW}) to further decompose the posterior covariance into blockwise components.
}
\item{\textbf{Conjugate Posterior Distribution for the Hyper-Parameters on Development Factors:} For all $i$ we have the following Inverse Gamma-Gaussian conjugacy for the hyper parameters in Models II and III,
\begin{equation*}
\begin{split}
\left[s^2_i|\Phi_{i}\right] \sim \mathcal{IG}\left(\alpha_i + \frac{1}{2}, \beta_i + \frac{\left(\Phi_i - \phi_i \right)^2}{2}\right)\; \; \text{ and } \; \; \left[t^2_i|\Psi_{i}\right] \sim \mathcal{IG}\left(a_i + \frac{1}{2}, b_i + \frac{\left(\Psi_i - \psi_i \right)^2}{2}\right).
\end{split}
\end{equation*}}
\end{itemize}

\end{lemma}

We next present alternative tail dependence structures for the PIC model. Previous studies on claims reserving that have incorporated copula based models, such as \citet{zhangpredicting} have done so through regression based frameworks. \citet{zhangpredicting} develop a parametric copula model to account for dependence between various lines of insurance claims. Their paper considers a bivariate Gaussian copula model with marginal generalized linear models to capture the positive correlation between the two insurance lines. Our article significantly extends the dependence modelling capability of the PIC model structure remaining in the frameworks presented above. However, to do so requires the introduction of auxiliary variables to enable computation. The approach developed involves modifying the posterior distribution by embeding the target posterior distribution for the model parameters into a much higher dimensional support comprised of the original model parameters and the additional auxiliary variables. The reason for this expansion of the posterior dimensions will be come clear below and is in general known in Bayesian statistics as an auxiliary variable framework.

\section{Incorporating Mixture-Archimedean Copula Dependence Structures into Paid-Incurred-Claims Models: Model IV}
This section presents an alternative parameteric approach to modelling and capturing dependence and tail dependence in the PIC model structure which involves considering copula based models within the PIC reserving framework. The dependence can be considered over the following combinations such as:
\begin{enumerate}
\item{Independent accident years and dependence between payment losses over the development years;}
\item{Independent accident years and dependence between incurred losses over the development years;}
\item{Independent accident years and dependence jointly between payment and incurred losses over the development years via a mixture copula, hierarchical copula (HAC) as in \citet{kurowicka2010dependence}, or a vine copula (d-vine, canononical vine) e.g. \citet{aas2009pair};}
\item{Dependent accident years and independent development years for payment, incurred or both sets of losses.}
\end{enumerate}

Our article concentrates on the mixture copula model which allows for combinations of upper and lower tail dependence of different strengths. We detail the class of auxiliary variable methods known in statistics as Data Augmentation and demonstrate how this class of models can be combined into our modelling framework to allow for consistent use of copula models in the PIC framework. There are many variations that can be explored in this approach. We give one such approach for Model IV, Assumptions \ref{modelass4}, that is directly comparable to that used for Model II in Assumption \ref{modelass2}.

We present fundamental properties of members of the Archimedean family of copula that we consider when constructing mixture copula models in the PIC framework in the Appendix, see Lemma \ref{LemmaArchCop} for the characteristics of the Archimedean family of copulas and Lemma \ref{lemmaArchCop} for the required distribution and densities for three members of this family. In addition references  \citet{denuit2005actuarial},\citet{aas2009pair},\citet{embrechts2009copulas}, \citet{min2010bayesian} and \citet{patton2009copula} provide more detail.

%

In Lemma \ref{LemmaArchCop} the property of associativity of Archimedean copula models is particularly useful in the PIC model framework as it allows us to obtain analytic expressions for the likelihood structure of the matrix-variate PIC model. This is particularly useful if one specifies the model as a hierarchical Archimedean Copula (HAC) construction.

We consider the following popular members of the Archimedean family of copula models, due to their analytic tractability, their non-zero tail dependence properties and their parsimonious parameterizations. In addition, generating random variates from these class of models is trivial given the generator for the member of the Archimedean family of interest. Lemma \ref{lemmaArchCop} in the appendix presents the three Archimedean copulas for Clayton, Gumbel and Frank copulas that we consider and their properties. We use the following notation for copula densities we consider on $[0,1]^d$, see \citet[Section 4.3, Table 4.1]{nelsen2006introduction} and Lemma \ref{lemmaArchCop}: 
the Clayton copula density is denoted by $c^C(u_1,...,u_n;\rho^C)$ with $\rho^C \in [0,\infty)$ the dependence parameter; the Gumbel copula density is denoted by $c^G(u_1,...,u_n;\rho^G)$ with $\rho^G \in [1,\infty)$ the dependence parameter; and the Frank copula density is denoted by $c^F(u_1,...,u_n;\rho^F)$ with $\rho^F \in \mathbb{R}/\{0\}$ the dependence parameter. 

In addition, we also note that the properties of these copulas of interest include that the Clayton copula does not have upper tail dependence, however its lower tail dependence can be expressed as $\lambda_L = 2^{-1/\rho^C}$. The Gumbel copula does not have lower tail dependence, however its upper tail dependence of the Gumbel copula can be expressed as $\lambda_U = 2 - 2^{1/\rho^G}$. The Frank copula does not have upper or lower tail dependence.

In this class of copula dependence models we consider the marginal distribution of each log payment or log incurred loss as distributed according to a Gaussian distribution and the joint distribution vector is modelled via a mixture copula comprised of the above three components from the Archimedean family. Such a copula construction will still produce a copula as shown in Lemma \ref{lemmaCopMix}.

\begin{lemma}{ \label{lemmaCopMix} Consider copula distributional members $C_i\left(u_1,u_2,\ldots,u_n\right) \in \mathcal{A}^n$, where $\mathcal{A}^n$ defines the space of all possible n-variate distributional members of the Archimedean family of copula models, specified in Lemma \ref{lemmaArchCop}. Any finite mixture distribution constructed from such copula components that admit tractable density functions $c_i\left(u_1,u_2,\ldots,u_n\right)$, denoted by $\tilde{c}\left(u_1,u_2,\ldots,u_n\right) = \sum_{i=1}^m w_i c_i\left(u_1,u_2,\ldots,u_n\right)$, such that $\sum_{i=1}^m w_i = 1$, is also the density of a copula distribution.}
\end{lemma}
The proof of Lemma \ref{lemmaCopMix} is provided in Appendix \ref{AppendixProofCopMix}.

\subsection{Understanding Bayesian Data Augmentation}
The modeling framework of Data Augmentation in the Bayesian framework is typically invoked to deal with situations in which the likelihood evaluation is intractable to perform point-wise. This would make Bayesian inference in such a model also generally intractable. For example if one considers the generic likelihood $p\left(\bm{y}_{1:n}|\bm{\theta}\right)$ with observation random vectors $\bm{Y}_{1:n}$, which can be evaluate point-wise as a function of parameter vector $\bm{\theta}$ with respect to a realization of the observation process $\bm{y}_{1:n}$.

In the setting we encounter in the PIC models, we can generically consider the data random vector observation is partitioned into two vector sub-components $\bm{Y} = \left[\bm{Y}^{(1)},\bm{Y}^{(2)}\right]$, of which only one component, say $\bm{Y}^{(1)}$, is actually observed. Then evaluation of the likelihood pointwise for $\bm{\theta}$ given a realization of $\bm{Y}_{1:n}^{(1)}$ would require solving the integral in Equation \ref{EqnLHGeneric}
\begin{equation} \label{EqnLHGeneric}
p\left(\bm{Y}_{1:n}^{(1)}|\bm{\theta}\right) = \int p\left(\bm{Y}_{1:n}^{(1)}|\bm{\theta},\bm{Y}_{1:n}^{(2)}\right) p\left(\bm{Y}_{1:n}^{(2)}|\bm{\theta}\right) d\bm{Y}_{1:n}^{(2)}.
\end{equation}
Generally, this integral will not admit a closed form solution. Therefore, the Bayesian Data Augmentation approach involves extending the target posterior $p\left(\bm{\theta}|\bm{Y}_{1:n}^{(1)}\right)$ which is intractable due to the intractability of the likelihood to a new posterior model on a higher dimensional space, in which the target distribution is a marginal as given in Equation \ref{EqnDAGeneric} 
\begin{equation} \label{EqnDAGeneric}
p\left(\bm{\theta},\bm{Y}_{1:n}^{(2)*}|\bm{Y}_{1:n}^{(1)}\right) = \frac{p\left(\bm{Y}_{1:n}^{(1)}|\bm{\theta},\bm{Y}_{1:n}^{(2)*}\right)p\left(\bm{Y}_{1:n}^{(2)*}|\bm{\theta}\right)p\left(\bm{\theta}\right)}{p\left(\bm{Y}_{1:n}^{(1)}\right)}
\end{equation}
where $\bm{Y}_{1:n}^{(2)*}$ are auxiliary random vectors with prior distribution $p\left(\bm{Y}_{1:n}^{(2)*}|\bm{\theta}\right)$, 'augmented' to the posterior parameter space to allow tractability of the posterior inference. This will be explained in detail for the PIC copula models below.

\subsection{Data Augmentation in the Bayesian PIC Copula Models}
Definition \ref{DefnLogAuxMod} gives some useful notation for the results that follow.

\begin{defi}[Auxiliary Data for Data Augmentation] \label{DefnLogAuxMod}
Consider the defined loss data under the one-to-one (invertible) transformation for the observed data
given by the joint matrix for all observations and auxiliary variables given by $X = \left[\bm{X}_0',\bm{X}_1',\ldots,\bm{X}_J'\right]$. In this framework, the $i$-th accident year is defined according to,
$\bm{X}_i = \left[\log I_{i,0},\log P_{i,0},\log I_{i,1},\log P_{i,1}, \ldots,\log I_{i,J-1},\log P_{i,J-1},\log I_{i,J}\right]$. Consider the permutation of each vector of log payments and log incurred losses given by\\
$\widetilde{\bm{X}}_i = \mathcal{P}^*_{\bm{i_{~}}}\left(\bm{X}_i\right) = \left[\log P_{i,0},\log P_{i,1},\ldots,\log P_{i,J},\log I_{i,0},\log I_{i,1},\ldots,\log I_{i,J-1}\right].$
Now consider the further partition by the decomposition of observed log payment losses and unobserved log payment losses as well as these quantities for the incurred losses defined for the $i$-th accident year by,
\begin{equation}
\begin{split}
\widetilde{\bm{X}}_i &= \left[\widetilde{\bm{X}}^P_{i,obs},\widetilde{\bm{X}}^P_{i,aux},\widetilde{\bm{X}}^I_{i,obs},\widetilde{\bm{X}}^I_{i,aux}\right] \\
&=\left[\widetilde{\bm{X}}^P_{0,i,obs},\ldots,\widetilde{\bm{X}}^P_{J-i,i,obs},\widetilde{\bm{X}}^P_{J-i+1,i,aux},\ldots,\widetilde{\bm{X}}^P_{J,i,aux},\widetilde{\bm{X}}^I_{0,i,obs},\ldots,\widetilde{\bm{X}}^I_{J-i,i,obs},\widetilde{\bm{X}}^I_{J-i+1,i,aux},\ldots,\widetilde{\bm{X}}^I_{J-1,i,aux}\right] \\
&=\left[ \underbrace{\log P_{i,0},\ldots,\log P_{i,J-i}}_{\text{observed Payments}},\underbrace{\log P_{i,J-i+1},\ldots,\log P_{i,J}}_{\text{unobserved Payments}},\underbrace{\log I_{i,0},\ldots,\log I_{i,J-i}}_{\text{observed Incurred}},\underbrace{\log I_{i,J-i+1},\ldots,\log I_{i,J-1}}_{\text{unobserved Incurred}} \right]'.\\
\end{split}
\end{equation}
Therefore the total data matrix of losses is given by $\widetilde{X} = \left[\widetilde{\bm{X}}_0,\ldots,\widetilde{\bm{X}}_J\right].$  Note, the introduction in this section of the notation subscripts $obs$ and $aux$ allows us to make explicit the fact that the upper triangle of log payment losses and the upper triangle of log incurred losses are un-observed quantities for these random variables, while the lower triangular regions for such losses are observed. We denote these random variables as auxiliary variables (augmented) to the observed data random variables to create a complete data set of all losses.
\end{defi}

By considering the unobserved data in the lower payment and incurred loss triangles as auxiliary variables to be jointly estimated along with the model parameters, we will demonstrate below that only under this approach is consistency ensured in the copula structure of the PIC model. However, we first make the following model assumptions about the statistical features of the PIC model.

The following assumptions illustrate a choice of copula models for the mixture from the Archimedean family. However, there are many related specifications and frameworks that can be explored in this context, be we leave that to future research.

\begin{model ass}[Data-Augmented Mixture Copula PIC (\textsl{Model IV})]\label{modelass4}
The model assumptions and specifications for the copula model we develop involve:
\begin{itemize}
\item Let the random matrix $\Sigma_i \in \mathbb{R}^{(2J+1)\times(2J+1)}$ be the covariance for $\widetilde{\bm{X}}_i = \left[\widetilde{\bm{X}}^P_{i,obs},\widetilde{\bm{X}}^P_{i,aux},\widetilde{\bm{X}}^I_{i,obs},\widetilde{\bm{X}}^I_{i,aux}\right]$ with $\widetilde{\bm{X}}_i \in \mathbb{R}^{2J+1}$ for all $i=0,\ldots,J$. We assume that $\Sigma$ is diagonal where
\begin{equation}
\Sigma_{i,i} \sim \mathcal{IG}\left(\alpha_i,\beta_i\right), \; \forall i \in \left\{0,\ldots,J\right\},
\end{equation}
where $\alpha_i$ and $\beta_i$ are the known hyper-parameters for shape and scale.
\item{\textbf{MARGINAL DISTRIBUTION:} given $\mathbf{\Theta}=\left(\Phi_0,\ldots, \Phi_{J},\Psi_0,\ldots, \Psi_{J}\right)$ and covariance matrices $\Sigma, \Omega \in \mathbb{R}^{(2J + 1) \times (2J + 1)}$ and $\bm{\rho}$, we assume the marginal distribution of the random matrix, of all log payments and log incurred losses $\widetilde{X}$, comprised of columns $\widetilde{\bm{X}}_i$ for the $i$-th accident year is matrix-variate Gaussian with density, defined as in Lemma \ref{lemmaMatrixGaussianDist}, with the $(2J + 1) \times (J+1)$  mean matrix $\widetilde{M} = \left[\bm{\Theta}',\ldots,\bm{\Theta}'\right]$, column dependence given by $(2J + 1) \times (2J+1)$  covariance matrix $\Sigma$ and row dependence given by $(J+1) \times (J+1)$ matrix $\Omega$. Here we only consider the case of $\Omega = \mathbb{I}_{J+1}$ for the marginal independent case.
}
\item{\textbf{DATA AUGMENTED PIC MIXTURE COPULA LIKELIHOOD:} Given $\widetilde{\bm{X}}^P_{0,aux},\widetilde{\bm{X}}^P_{1,aux},\ldots,\widetilde{\bm{X}}^P_{J-1,aux}$,
$\widetilde{\bm{X}}^I_{0,aux},\widetilde{\bm{X}}^I_{1,aux},\ldots,\widetilde{\bm{X}}^I_{J-1,aux}$, 
$\mathbf{\Theta}=\left(\Phi_0,\ldots, \Phi_{J},\Psi_0,\ldots, \Psi_{J}\right)$, covariance matrices $\Sigma, \Omega \in \mathbb{R}^{(2J + 1) \times (2J + 1)}$ and $\bm{\rho}$, the joint distribution of the random matrix ($\widetilde{X}$) of all log permuted payment and incurred losses is assumed (in this example) to be independent between accident years. For the $i$-th column (corresponding to $i$-th accident year), the joint distribution of all losses $(\widetilde{\bm{X}}_i)$ is assumed to be hierarchical Archimedean Copula (HAC) mixture copula specified by distribution,
\begin{equation}
\begin{split}
\left[\widetilde{X}\right]_{\bullet, i}  &\sim \tilde{C}_{\bm{\rho}_{i}}\left(F\left(\widetilde{\bm{X}}^P_{i,obs},\widetilde{\bm{X}}^P_{i,aux},\widetilde{\bm{X}}^I_{i,obs},\widetilde{\bm{X}}^I_{i,aux};\bm{[M]}_{\bullet, i},\Sigma\right)\right)\\
&=\tilde{C}^{P}_{\bm{\rho}^P_{i}}\left(F\left(\widetilde{\bm{X}}^P_{i,obs},\widetilde{\bm{X}}^P_{i,aux};\bm{[M]}^P_{\bullet i},\Sigma\right)\right) \tilde{C}^{I}_{\bm{\rho}^I_{i}}\left(F\left(\widetilde{\bm{X}}^I_{i,obs},\widetilde{\bm{X}}^I_{i,aux};\bm{[M]}^I_{\bullet i},\Sigma\right)\right),
\end{split}
\end{equation}
with supper script $P$ and $I$ denote the components for the log payments and log incurred losses in the $i$-th development year respectively and the density is given by
\begin{equation}
\begin{split}
&f\left(\widetilde{\bm{X}}^P_{i,obs},\widetilde{\bm{X}}^P_{i,aux},\widetilde{\bm{X}}^I_{i,obs},\widetilde{\bm{X}}^I_{i,aux}|\bm{[M]}_{\bullet i},\Sigma,\bm{\rho}^P_{i},\bm{\rho}^I_{i}\right) \\
&= \tilde{c}^P_{\bm{\rho}^P_{i}}\left(F\left(\widetilde{\bm{X}}^P_{i,obs},\widetilde{\bm{X}}^P_{i,aux};\bm{[M]}^P_{\bullet i},\Sigma\right)\right) \tilde{c}^I_{\bm{\rho}_{i}}\left(F\left(\widetilde{\bm{X}}^I_{i,obs},\widetilde{\bm{X}}^I_{i,aux};\bm{[M]}^I_{\bullet i},\Sigma\right)\right)\prod_{j=1}^{2J+1} \phi(\widetilde{X}_{j,i};M_{j,i},\Sigma_{i,i}),
\end{split}
\end{equation}
where
\begin{equation*}
\begin{split}
& \tilde{c}^{S}_{\bm{\rho}_{i}}\left(F\left(\widetilde{\bm{X}}^S_{i,obs},\widetilde{\bm{X}}^S_{i,aux};\bm{[M]}^{S}_{\bullet i},\Sigma\right)\right)
= w_1 c^{G}_{\rho^{(G,S)}_i}\left( F_{1,i}\left(\widetilde{\bm{X}}^S_{1,i,obs};\tilde{M}^S_{1,i},\Sigma_{1,1}\right),\ldots, F_{J,i}\left(\widetilde{\bm{X}}^S_{J,i,aux};\tilde{M}^S_{J,i},\Sigma_{J,J}\right) \right) \\
&\; \; + w_2 c^{F}_{\rho^{(F,S)}_i}\left( F_{1,i}\left(\widetilde{\bm{X}}^S_{1,i,obs};\tilde{M}^S_{1,i},\Sigma_{1,1}\right),\ldots, F_{J,i}\left(\widetilde{\bm{X}}^S_{J,i,aux};\tilde{M}^S_{J,i},\Sigma_{J,J}\right) \right) \\
& \;\;+ (1-w_1-w_2)
c^{C}_{\rho^{(C,S)}_i}\left( F_{1,i}\left(\widetilde{\bm{X}}^S_{1,i,obs};\tilde{M}^S_{1,i},\Sigma_{1,1}\right),\ldots, F_{J,i}\left(\widetilde{\bm{X}}^S_{J,i,aux};\tilde{M}^S_{J,i},\Sigma_{J,J}\right) \right), \; S \in \left\{P,I\right\},
\end{split}
\end{equation*}
and such that $w_1 + w_2 + (1-w_1-w_2) = 1$. This specifies a mixture of central, upper and lower tail dependence as denoted by the mixture of Archimedian copula models made up of Frank, Clayton and Gumbel members, such that for the source of data $S$, the copula parameters for each Archimedian family member is given by $\rho^{(G,S)}_i > 0$, $\rho^{(C,S)}_i > 1$ and $\rho^{(F,S)}_i \in \mathbb{R}/\left\{0\right\}$. Therefore the total conditional distribution corresponding to the likelihood model considered is given by,
\begin{equation} \label{EqnFullDataLH}
\begin{split}
f\left(\tilde{X}|M,\Sigma,\Omega,\bm{\rho}\right) &= \underbrace{\prod_{i=0}^J
\tilde{c}^P_{\bm{\rho}^P_{i}}\left(F\left(\widetilde{\bm{X}}^P_{i,obs},\widetilde{\bm{X}}^P_{i,aux};\bm{[M]}^P_{\bullet i},\Sigma\right)\right) \tilde{c}^I_{\bm{\rho}^I_{i}}\left(F\left(\widetilde{\bm{X}}^I_{i,obs},\widetilde{\bm{X}}^I_{i,aux};\bm{[M]}^I_{\bullet i},\Sigma\right)\right)}_{\text{Copula Dependence in Data Augmented PIC Likelihood}} \\
& \; \; \times \underbrace{\frac{\exp\left(-\frac{1}{2}\text{tr}\left[\Omega^{-1}\left(\widetilde{X} - M\right)'\Sigma^{-1}\left(\widetilde{X} - M\right)\right]\right)}{\left(2 \pi\right)^{(2J^2+3J+1)/2}\left|\Omega\right|^{(2J+1)/2}\left|\Sigma\right|^{(J+1)/2}}.
}_{\text{Marginal Distribution in Data Augmented Likelihood PIC Model}}
\end{split}
\end{equation}
}
\item{Assume that the tail dependence features of the Data-Augmented copula PIC model are such that the dependence structure is homogeneous accross accident years, $\bm{\rho}^P = \bm{\rho}^P_{i}$ and $\bm{\rho}^I = \bm{\rho}^I_{i}$ for all $i \in \left\{0,1,2,\ldots,J\right\}$.}
\item{Conditional on $\Sigma$, $\bm{\Phi} = \left[\Phi_0,\Phi_1,\ldots,\Phi_J\right]$ and $\bm{\Psi} = \left[\Psi_0,\Psi_1,\ldots,\Psi_J\right]$ the hierarchical prior distribution on the auxiliary payment data for the $i$-th accident year is given by a normal distribution, centered on the development year mean,
\begin{equation}
\begin{split}
&\widetilde{\bm{X}}^P_{i,aux} \sim \mathcal{N}\left(\left[\Phi_{J-i+1},\Phi_{J-i+2},\ldots,\Phi_J\right],\Sigma^P_2\right).
\end{split}
\end{equation}
The hierarchical prior distribution on the auxiliary incurred loss data for the $i$-th accident year is given by
\begin{equation}
\begin{split}
&\widetilde{\bm{X}}^I_{i,aux} \sim
\mathcal{N}\left(\left[\Psi_{J-i+1},\Psi_{J-i+2},\ldots,\Psi_J\right],\Sigma^I_2\right),
\end{split}
\end{equation}
with $\Sigma_2$ the lower portion of covariance $\Sigma$ corresponding to the lower triangle matrix from $(J-i+1)$ through to $J$ for all $i \in \left\{0,1,2,\ldots,J\right\}$.\\
}
\item{For all accident years, $i \in \left\{0,1,\ldots,J\right\}$, the ultimate payment losses and incurred losses are equal a.s., $P_{i,J} = I_{i,J}, \; \mathbb{P}-\text{a.s.}$}
\item{The matrix $\widetilde{\Sigma}$ is positive definite and components of $\bm{\Theta}$ are independent with prior distributions
\begin{equation}
\Phi_i \sim \mathcal{N}\left(\phi_i,s_i^2\right) \; \text{ and } \; \Psi_j \sim \mathcal{N}\left(\psi_j,t^2_j\right)
\end{equation}
and hyper-prior distributions
\begin{equation}
s^2_i \sim \mathcal{IG}\left(\alpha_i,\beta_i\right) \; \text{ and } \; t^2_j \sim \mathcal{IG}\left(a_j,b_j\right)
\end{equation}
for all $i \in \left\{1,\ldots,J\right\}$ and $j \in \left\{0,\ldots,J\right\}$.
}
\item{The matrix $\Sigma$ is distributed as $\Sigma \sim \mathcal{IW}\left(\Lambda,k\right)$ and the copula parameters are distributed as $\rho^{G,P} \sim \mathcal{IG}\left(\alpha^G,\beta^G\right)$, $\rho^{C,P} \sim \mathcal{IG}\left(\alpha^C,\beta^C\right)$ and $\rho^{F,P} \sim \mathcal{N}\left(0,\sigma^F\right)$}
\end{itemize}
\end{model ass}


Hence, we have made precise the auxilliary data scheme used in formulating the Data-Augmented-PIC model. In particular illustrating the importance of the role of the auxiliary data in evaluation of the model and estimation of the PIC claim development factors. Also we note we get indirectly via the data augmentation the distribution for the predicted payment and incurred Loss reserves.

\begin{remark} The following remarks provide motivation for the Data-Augmentation and resulting incorporation of auxiliary payment and incurred Losses data.
\begin{itemize}
\item{The use of data augmentation in the above model structure is critical in the PIC model formulation, since it allows one to ensure that the dependence structure considered (in this case a HAC-Mixture) is consistent both across accident years and across development years. \\
\textbf{Note:} In the case of a linear dependence structure such as with a covariance / correlation matrix under a Gaussian Copula or Independent Copula model, such as those presented previously under Models I,II, III, we have that conditional distributions and marginal distributions are Gaussian. This means that the evaluation of the likelihood is analytic without the need for auxiliary variables.
}
\item{In order to evaluate the likelihood one has two choices, to evaluate the observed data likelihood (Equation (\ref{EqnObsDataLH})) or to evaluate the full data likelihood (Equation (\ref{EqnFullDataLH})). }
\begin{itemize}
\item{The PIC copula model equivalent of Equation \ref{EqnDAGeneric} is the observed data likelihood is given for the $i$-th accident year by
\begin{equation} \label{EqnObsDataLH}
\begin{split}
&p\left(\widetilde{\bm{X}}^{P}_{i,obs},\widetilde{\bm{X}}^{I}_{i,obs}|\bm{\Theta},\Sigma,\Omega,\bm{\rho}\right) \\ \nonumber
& \; \; = \int\cdots \int p\left(\widetilde{\bm{X}}^{P}_{i,obs},\widetilde{\bm{X}}^{I}_{i,obs}|\bm{\Theta},\Sigma,\Omega,\bm{\rho},\widetilde{\bm{X}}^{P}_{i,aux},\widetilde{\bm{X}}^{I}_{i,aux}\right) p\left(\widetilde{\bm{X}}^{P}_{i,aux},\widetilde{\bm{X}}^{I}_{i,aux}|\bm{\Theta},\Sigma,\Omega,\bm{\rho}\right)
d\widetilde{\bm{X}}^{P}_{i,aux}d\widetilde{\bm{X}}^{I}_{i,aux}\\ \nonumber
& \; = \int\cdots \int \tilde{c}^P_{\bm{\rho}^P_{i}}\left(F\left(\widetilde{\bm{X}}^P_{i,obs},\widetilde{\bm{X}}^P_{i,aux};\bm{[M]}^P_{\bullet i},\Sigma\right)\right) \tilde{c}^I_{\bm{\rho}^I_{i}}\left(F\left(\widetilde{\bm{X}}^I_{i,obs},\widetilde{\bm{X}}^I_{i,aux};\bm{[M]}^I_{\bullet i},\Sigma\right)\right) \\
& \; \; \; \; \; \times
f_{\widetilde{X}_{i,aux}}^{MVN}\left(\widetilde{x}_{i,aux};M_{i,aux},\Sigma^P_2 \oplus \Sigma^I_2\right) f_{\widetilde{X}_{i}}^{MVN}\left(\widetilde{x}_{i};M_{i},\Sigma\right) \;  d\widetilde{\bm{X}}^{P}_{i,aux} \; d\widetilde{\bm{X}}^{I}_{i,aux}
%
%
%
\end{split}
\end{equation}
where matrix-variate Gaussian distributions $f_X^{MVN}()$ and $f_X^{MVN}$ are as defined in Lemma \ref{lemmaMatrixGaussianDist}

with $\widetilde{X}_{i,aux} = Vec\left(\widetilde{\bm{X}}^P_{i,aux},\widetilde{\bm{X}}^I_{i,aux}\right)$, $M_{i,aux} = Vec\left([\Phi_{J-i+1:J}]',[\Psi_{J-i+1:J-1}]'\right)$, \\ $\widetilde{X}_i=[\widetilde{\bm{X}}^P_{i,obs},\widetilde{\bm{X}}^P_{i,aux},\widetilde{\bm{X}}^I_{i,obs},\widetilde{\bm{X}}^I_{i,aux}]$ and $M_i = [\Phi_0,\ldots,\Phi_J,\Psi_0,\ldots,\Psi_J]$ the equivalent mean.}
\item{Clearly, the marginalization required to evaluate the Observed data likelihood involves intractable integration, except in special cases in which the copula models are Gaussian or independence copulas.}
\end{itemize}
\item{The full data likelihood comprised of observed and auxiliary data involves incorporating auxiliary variables to represent the unobserved data in the lower reserve triangle for payment and incurred loss triangles. These become part of the inference procedure and are required to be estimated jointly with the model parameters in the estimation methodology.}
\end{itemize}
\end{remark}
\section{Estimation via Adaptive Data-Augmented MCMC for Claims Reserving PIC Models}
It has been shown for the Independent and Gaussian copula models that we can obtain the observed data likelihood analytically. Therefore the posterior distribution for all the model parameters can be sampled via a MCMC procedure comprised of block Gibbs sampler updates. In the case of a more general copula dependence model in which the observed data likelihood cannot be analytically evaluated pointwise, we must resort to a Data Augmentation scheme. In this case we will be able to perform sampling via a general MCMC Metropolis-Hastings sampler. In particular we will consider automating such a sampler using an adaptive MCMC scheme.

\subsubsection{Adaptive Metropolis within Data-Augmented Copula PIC Models}
This section presents the adaptive proposal we use to sample the parameters and the auxiliary variables. The advantage of an adaptive MCMC mechanism is that it automates the proposal design through consideration of a proposal distribution that learns the regions in which the posterior distribution for the static model parameters and auxiliary data has most mass. As such, the probability of acceptance under such an on-line adaptive proposal is likely to improve as the iterations progress and the generated MCMC samples will ideally have reduced autocorrelation. In such cases the variance of Monte Carlo estimators of integrals of smooth functionals formed from such samples will be reduced.

There are several classes of adaptive MCMC algorithms, see \citet{roberts2009examples}. The distinguishing feature of adaptive MCMC algorithms, compared to standard MCMC, is the generation of the Markov chain via a sequence of transition kernels. Adaptive algorithms utilize a combination of time or state inhomogeneous proposal kernels. Each proposal in the sequence is allowed to depend on the past history of the Markov chain generated, resulting in many possible variants.

\citet{haario2005componentwise} develop an adaptive Metropolis algorithm with proposal covariance adapted to the history of the Markov chain was developed. \citet{andrieu2008tutorial} is presenting a tutorial discussion of the proof of ergodicity of adaptive MCMC under simpler conditions known as \textit{Diminishing Adaptation} and \textit{Bounded Convergence}. We note that when using inhomogeneous Markov kernels it is particularly important to ensure that the generated Markov chain is ergodic, with the appropriate stationary distribution. Two conditions ensuring ergodicity of adaptive MCMC are known as \textit{Diminishing Adaptation} and \textit{Bounded Convergence}. These two conditions are summarised by the following two results for generic Adaptive MCMC strategies on a parameter vector $\bm{\theta}$. As in \citet{roberts2009examples}, we assume that each fixed MCMC kernel $Q_{\gamma}$, in the sequence of adaptions, has stationary distribution $P\left(\cdot\right)$ which corresponds to the marginal posterior of the static parameters. Define the convergence time for kernel $Q_{\gamma}$ when starting from a state $\bm{\theta} \in E$, as $M_{\epsilon}\left(\bm{\theta},\gamma\right) = \text{inf}\{s \geq 1 : \|Q^s_{\gamma}\left(\bm{\theta};\cdot\right) - P\left(\cdot\right)\| \leq \epsilon $. Under these assumptions, they give the following two conditions which are sufficient to guarantee that the sampler produces draws from the posterior distribution as the number of iterates tend to infinity. The two sufficient conditions are:
\begin{itemize}
\item{ \textit{Diminishing Adaptation:} $\text{lim}_{n\to\infty}\text{sup}_{\bm{\theta} \in E}\|Q_{\Gamma_{s+1}}\left(\bm{\theta},\cdot\right) - Q_{\Gamma_{s}}\left(\bm{\theta},\cdot\right)\|_{tv} = 0$ in probability. Note, $\Gamma_s$ are random indices.}
\item{ \textit{Bounded Convergence:} For $\epsilon > 0$, the sequence $\{M_{\epsilon}\left(\bm{\theta},\Gamma_j\right)\}^\infty_{j=0}$ is bounded in probability.}
\end{itemize}
The sampler converges asymptotically in two senses,
\begin{itemize}
\item{\textit{Asymptotic convergence:} $\text{lim}_{j\to\infty}\|\mathcal{L}\text{aw}\left(\bm{\theta}	 \right)-P\left(\bm{\theta}\right)\|_{tv}=0$ in probability.}
\item{\textit{Weak Law of Large Numbers}: $\text{lim}_{j\to\infty}\frac{1}{j}\sum^{j}_{i=1}\phi \left(\bm{\theta}\right)=\int \phi(\bm{\theta})P(d\bm{\theta}) $ for all bounded $\phi : E \to R$.}
\end{itemize}

In general, it is non-trivial to develop adaption schemes which can be verified to satisfy these two conditions. In this paper we use the adaptive MCMC algorithm to learn the proposal distribution for the static parameters in our posterior $\bm{\Phi}$. In particular we work with an adaptive Metropolis algorithm utilizing a mixture proposal kernel known to satisfy these two ergodicity conditions for unbounded state spaces and general classes of target posterior distribution, see \citet{roberts2009examples}
for details.

\subsubsection{Euclidean and Riemann-Manifold Adaptive Metropolis within Data-Augmented Copula PIC Models}
This section presents the specific details of the Adaptive Metropolis algorithm that we combine with Data-Augmentation to obtain an MCMC sampler for the Data Augmented Mixture Copula PIC Model proposed. This involves specifying the details of the proposal distribution in the AdMCMC algorithm which samples a new proposed update vector $\bm{\Upsilon}^*$ and matrix $\widetilde{\Sigma}^*$ from an existing Markov chain state $\bm{\Upsilon}$ with
$$\bm{\Upsilon} = \left[\bm{\Phi},\bm{\Psi}, s^2_{0:J}, t^2_{0:J},\bm{\rho}, \widetilde{\bm{X}}^P_{1,aux},\ldots,\widetilde{\bm{X}}^P_{J,aux},\widetilde{\bm{X}}^I_{1,aux}\ldots, \widetilde{\bm{X}}^I_{J,aux}\right]$$ and matrix $\widetilde{\Sigma}$. At the $j$-th iteration of the Markov chain we have existing state $\bm{\Upsilon}^{(j-1)}$ and $\widetilde{\Sigma}^{(j-1)}$ which is used to construct the proposal distribution  $q\left(\bm{\Upsilon}^{(j-1)},\bm{\Upsilon}^{*}\right)q\left(\widetilde{\Sigma}^{(j-1)},\widetilde{\Sigma}^{*}\right)$. The choices we make for the two proposals will involve a novel development of a new adaptive proposal for positive definite matrices, required for the covariance matrix $\widetilde{\Sigma}$ should we choose not to specify it as diagonal.

\textbf{Euclidean Space Adaptive Metropolis for Static Parameters:}\\
We first detail the proposal for updating $\bm{\Upsilon}$ using a mixture of multivariate Gaussian distributions as specified for an Adaptive Metropolis algorithm which involves sampling from the proposal
\begin{equation} \label{AdapMetroEuclid}
q\left(\bm{\Upsilon}^{(t-1)},\cdot\right)=w_{1}\mathcal{N}\left(\bm{\Upsilon};\bm{\Upsilon}^{(t-1)},\frac{\left(2.38\right)^{2}}{d}\mathbb{C}\text{ov}\left(\left\{\bm{\Upsilon}^{(j)}\right\}_{0 \leq j \leq t-1}\right)\right)+\left(1-w_{1}\right)\mathcal{N}\left(\bm{\Upsilon};\bm{\Upsilon}^{(t-1)},\frac{\left(0.1\right)^{2}}{d}I_{d,d}\right),
\end{equation}
where we define the sample covariance for Markov chain past history by $\mathbb{C}\text{ov}\left(\left\{\bm{\Upsilon}^{(j)}\right\}_{0 \leq j \leq t-1}\right)$ and we note the following recursive evaluation, which significantly aids in algorithmic computational cost reduction
{\small{
\begin{equation}
\begin{split}
\mathbb{E}\left(\left\{\bm{\Upsilon}^{(j)}\right\}_{0 \leq j \leq t}\right) & =\mathbb{E}\left(\left\{\bm{\Upsilon}^{(j)}\right\}_{0 \leq j \leq t-2}\right)+\frac{1}{t}\left(\bm{\Upsilon}^{(t-1)}-\mathbb{E}\left(\left\{\bm{\Upsilon}^{(j)}\right\}_{0 \leq j \leq t-1}\right)\right)\\
\mathbb{C}\text{ov}\left(\left\{\bm{\Upsilon}^{(j)}\right\}_{0 \leq j \leq t}\right) &=\frac{1}{t+1}\left(\left(\bm{\Upsilon}^{(t-1)}-\mathbb{E}\left(\left\{\bm{\Upsilon}^{(j)}\right\}_{0 \leq j \leq t}\right)\right)\left(\bm{\Upsilon}^{(t-1)}-\mathbb{E}\left(\left\{\bm{\Upsilon}^{(j)}\right\}_{0 \leq j \leq t}\right)\right)'-\mathbb{C}\text{ov}\left(\left\{\bm{\Upsilon}^{(j)}\right\}_{0 \leq j \leq t-1}\right)\right)\\
&\;\; + \mathbb{C}\text{ov}\left(\left\{\bm{\Upsilon}^{(j)}\right\}_{0 \leq j \leq t-1}\right).
\end{split}
\end{equation}
}}
The theoretical motivation for the recommended choices of scale factors 2.38, 0.1 and dimension $d$ are provided in \citet{rosenthal2008optimal}.

\textbf{Riemannian Manifold Adaptive Metropolis for Covariance Matrices:}\\
Next we develop a novel proposal distribution for the sampling of the covariance matrix $\widetilde{\Sigma} \in \text{Sym}^+(d)$ in an adaptive MCMC proposal, restricted to the Riemann manifold of symmetric, postive definite $(d \times d)$ matrices, denoted by the space $\text{Sym}^+(d)$.

\begin{remark} First, we note two properties of the marginal posterior $p\left(\left. \widetilde{\Sigma}\right|\left\{ \widetilde{X}^P_{i,obs},\widetilde{X}^I_{i,obs} \right\}_{0\leq i \leq J}\right)$: its distribution is restricted to the Riemann-manifold of symmetric positive definite matrices, but in general will not be Inverse-Wishart; second, the Markov chain samples drawn from this marginal distribution at iteration $t$, $\left\{\widetilde{\Sigma}^{(s)}\right\}_{0\leq s \leq t}$, are not independent. The consequence of this is that we cannot simply apply the property of closure under convolution of independent Wishart distributed random matrices to find a suitable proposal.
\end{remark}

Therefore, we will adopt a strategy to perform adaptive moment matching of a distribution with support $\text{Sym}^+(d)$. We detail one possibility involving an inverse Wishart distribution fitted to the sample mean of the marginal posterior for the covariance. We note that future work could also consider specifying a distribution on the superset of the Riemannian manifold of symmetric positive definite matrices, given by the Riemannian manifold of symmetric matrices $\text{Sym}^+(d) \subset \text{Sym}(d)$. 

\textbf{Adaptive Metropolis inverse Wishart Mixture:}
We note that one way to achieve this is a mixture of inverse Wishart distributions given by
\begin{equation}
q\left(\widetilde{\Sigma}^{(t-1)},\cdot\right)=w_{1}\mathcal{IW}\left(\widetilde{\Sigma};\Lambda^{adap}_t\left(\left\{\widetilde{\Sigma}^{(s)}\right\}_{0 \leq s \leq t-1}\right),p\right)+\left(1-w_{1}\right)\mathcal{IW}\left(\widetilde{\Sigma};\Lambda,p\right).
\end{equation}
Here, the adaptive proposal mixture component is specified through fixing the degrees of freedom $p$ and then selecting $\Lambda^{adap}_t$ with respect to the sample average of the covariance matrices  $\left\{\widetilde{\Sigma}^{(s)}\right\}_{0\leq s \leq t-1}$ which are samples from the matrix-variate marginal posterior in the Markov chain, thereby adapting the proposal to the Markov chain history. To perform the moment matching (Equation (\ref{EqnMomentMatch})), we note that we need to ensure that the sample average considered is restricted to the Riemann-manifold of positive definite matrices.
\begin{equation} \label{EqnMomentMatch}
\Lambda^{adap}_t\left(\left\{\widetilde{\Sigma}^{(s)}\right\}_{0 \leq s \leq t-1}\right) = \widehat{\widetilde{\Sigma}}^{(t-1)} \left(p - dim(\widetilde{\Sigma}) -1 \right).
\end{equation}
This is satisfied through the choice of the estimator 
\begin{equation}
\widehat{\widetilde{\Sigma}}^{(t-1)} = \frac{1}{t-1}\sum_{s=1}^{t-1} \widetilde{\Sigma}^{(s)}.
\end{equation}
To see this we observe that since we only form positive linear combinations of matrices on this manifold, with a scaling, such linear combinations will always remain on the manifold $\text{Sym}^+(d)$.


\FloatBarrier

\section{Real Data Analysis}
To illustrate the proposed models and compare with existing models and estimation methods in the actuarial literature we consider, as in \citet{merz2010paid}, the example presented in \citet{dahms2008loss} and \citet{dahms2009claims} (Tables 10 and 11). As in the second analysis framework in \citet{merz2010paid}, we treat the claim development factors, the likelihood dependence parameters and the hyperparameters on the claim development factor priors as parameters which we incorporate into the posterior inference.

We present two sets of results, the first studies the performance of the adaptive Markov chain Monte Carlo algorithms developed for the estimation and inference of the posterior distributions for the PIC-Copula models for Gaussian Copula (Models III) and the Data-Augmented-Mixture-Copula PIC (Models IV). The second stage of results assesses the estimation of predictive distributions and dependence features of the PIC claims reserving models compared to the independent PIC Model, the payment only model and the incurred only models. In particular, we focuss analysis on the data sets studied in \citet{merz2010paid} for comparison of the influence of dependence features in PIC models versus independence assumptions when performing PIC claims reserving.

\noindent \textbf{Convergence Analysis:} In all the Markov chain Monte Carlo simulations, for each model (payment, payment-incurred Gaussian copula Model III; and Data-Augmented hierarchical Archemdean mixture copula Model IV), we carried out convergence diagnostics. This included the Gelman-Rubin R-statistics (all less than 1.5), the ACF plots for each parameter were checked to ensure all parameters had ACF's which were less than 10\% by lag 20. Then the first 20\% of samples were discarded as burnin and the remaining samples were used in inference results presented below.

\subsection{Results: Euclidean and Riemann-Manifold Adaptive Metropolis for hierarchical Bayesian Copula PIC Models}
In the simualtion results, we consider a block Gibbs sampler with the following three stages performed at each iteration of the adaptive Metropolis-within-Gibbs sampler for the PIC Model III and Model IV:\\
\begin{enumerate}
\item[]{\textbf{Stage 1:} Perform exact sampling of the development factors and their hyperparameters under the conjugacy results developed.\\}
\item[]{\textbf{Stage 2:} Perform Euclidean space Adaptive Metropolis updates of the Augmented Data variables using proposal in Equation (\ref{AdapMetroEuclid}).\\}
\item[]{\textbf{Stage 3:} (Gaussian Copula Model III) - Perform Riemannian space Adaptive Metropolis updates of the covariance matrix in the Gaussian copula. Note, we consider the constrained specifications presented in the ``Dependent Lag Years'' model specification in Section \ref{DevLagYrs}, Equation (\ref{EqnTelescopingDependence}). Under this hierarchical Bayesian model, the joint covariance between all observed payment and incurred loss data under the dependent development years assumption, satisfies a telescoping diagonal block size form covariance matrix structure. Hence, the sampling of this structure can be performed blockwise on each covariance sub-block;\\
(Mixture Clayton-Gumbel Copula Model IV) - Perform Euclidean space Adaptive Metropolis updates of the mixture copula parameters.}
\end{enumerate}

\subsubsection{Hierarchical Bayesian Gaussian Copula (telescoping block covariance) PIC (Model III)}
\begin{figure}[ht]
\includegraphics[height=6cm, width=\textwidth]{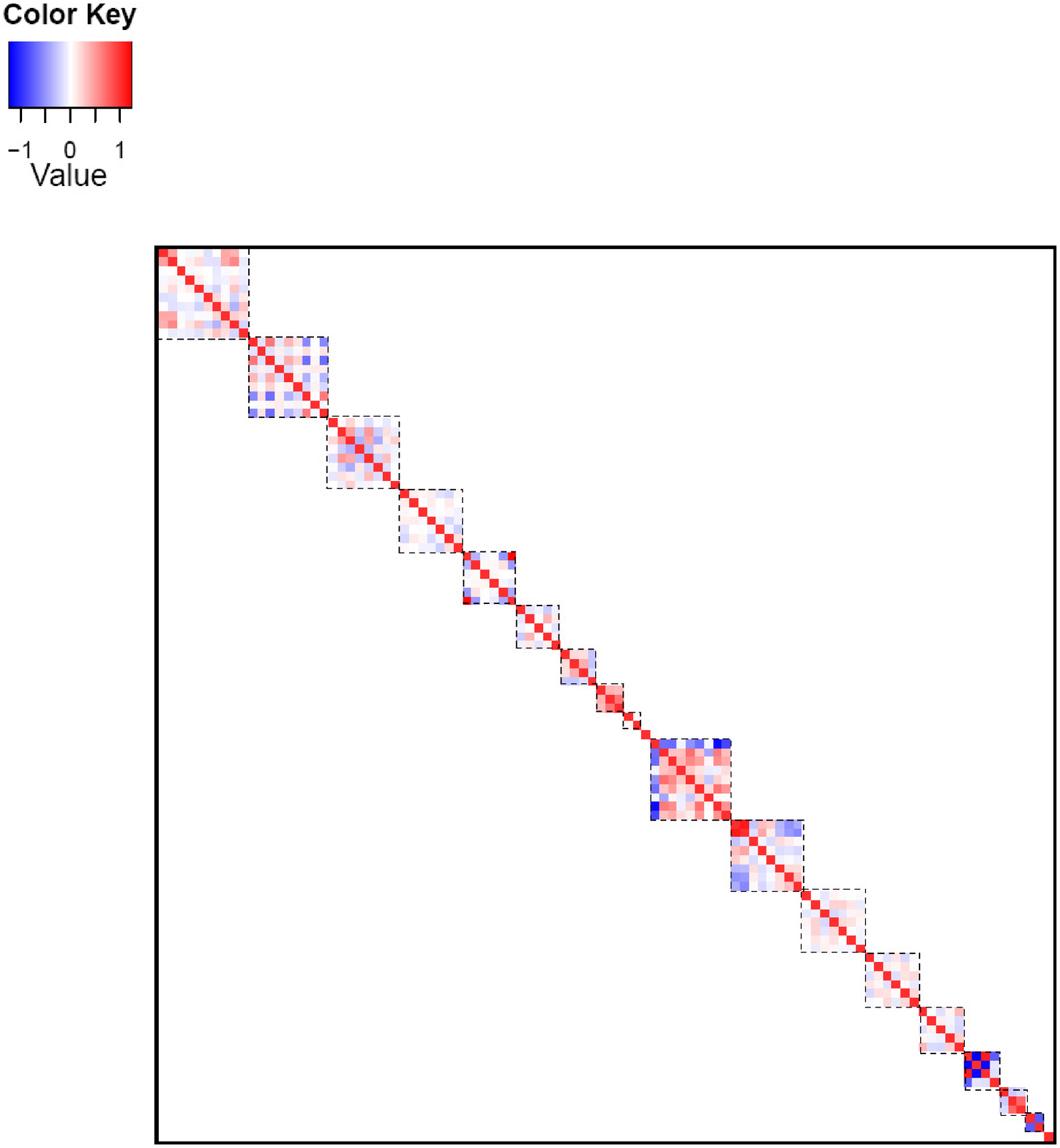}
\includegraphics[height=6cm, width=0.5\textwidth]{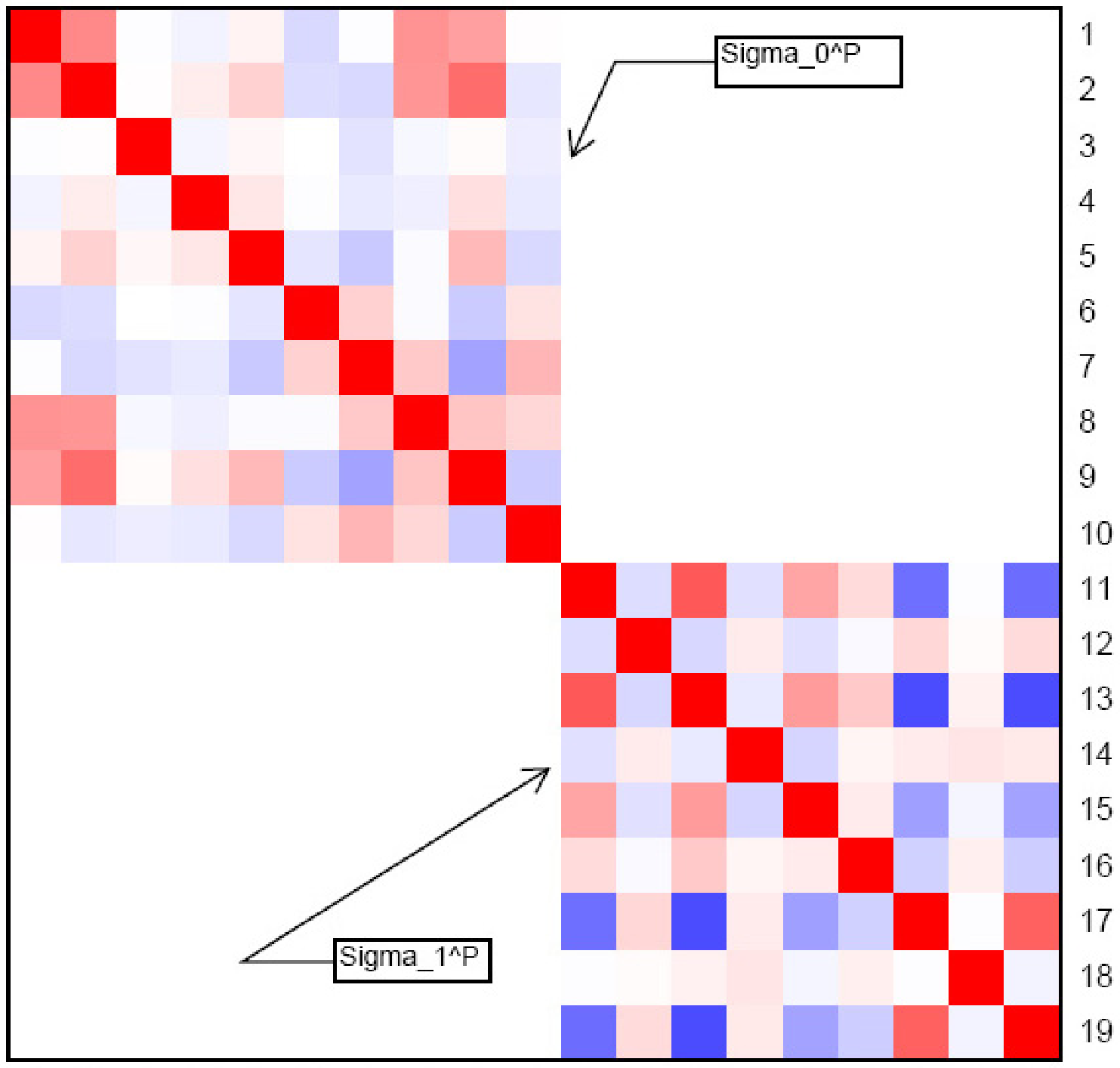}
\includegraphics[height=6cm, width=0.5\textwidth]{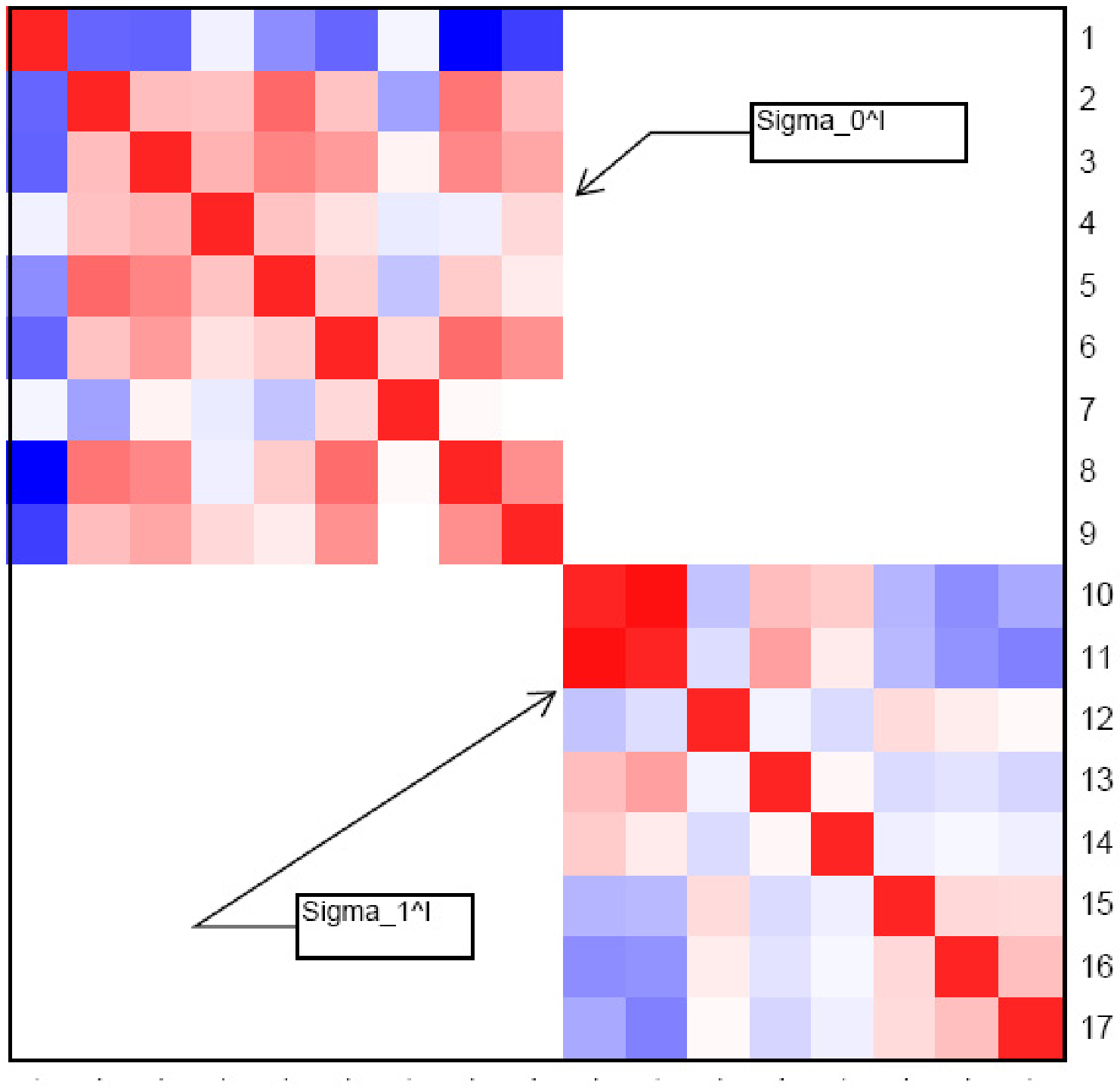}
\caption{\textbf{Top panel:} Heatmap of the posterior distribution for the Gaussian copula covariance matrix $(100 \times 100)$, summarised by the heat map for the mean of correlation structure using samples from the Riemannian Manifold Adaptive Metropolis sampler under restriction to a telescoping diagonal block form. \textbf{Bottom Left Panel:} Heatmap for the posterior distribution sub-block covariance matrices $\Sigma_0^P$ and $\Sigma_1^P$ converted to correlation matrices. \textbf{Bottom Right Panel:} Heatmap for the posterior distribution sub-block covariance matrices $\Sigma_0^I$ and $\Sigma_1^I$ converted to correlation matrices. The color key is given at the top left.
}
\label{PIC_GausCopMod}
\end{figure}

This section presents the estimation results for the Gaussian Copula based PIC models (Model III) on the real data. Figure \ref{PIC_GausCopMod} summarizes the dependence structure by a heatmap for the posterior distribution of the Gaussian copula covariance matrix. As mentioned in the introduction, the telescoping block covariance refers to the fact that the covariance structure is reducing in rank by 1 on each diagonal block for the payment data and then the incurred data. This model has the joint covariance between all observed payment and incurred loss data under the assumption that the development years are dependent, satisfying a telescoping diagonal block size form covariance matrix structure. Summarising the information from such posterior samples for distributions of covariance matrices is non-trivial as discussed in \citet{tokuda2011visualizing}, where they develop a four layer approach. Our article adopts aspects of the ideas proposed in \citet{tokuda2011visualizing} to interpret the features of the posterior distribution samples for the dependence structures.

The posterior mean for estimated PIC covariance structure is obtained by using Monte Carlo samples from the Riemann-Manifold Adaptive Metropolis sampler and given by the estimator,
\begin{equation}
\mathbb{E}\left[\tilde{\Sigma}|\bm{P},\bm{I}\right] = \frac{1}{S}\sum_{s=1}^S \left\{\left(\bigoplus_{i=0}^J \Sigma^{P}_i\right)\oplus\left(\bigoplus_{i=0}^J \Sigma^{I}_i\right)\right\}^{(s)},
\end{equation}
where $\left\{\left(\bigoplus_{i=0}^J \Sigma^{P}_0\right)\oplus\left(\bigoplus_{i=0}^J \Sigma^{I}_0\right)\right\}^{(s)}$ is the $s$-th sample of the $J(J-1) \times J(J-1)$ covariance matrix. The estimated posterior mean covariance matrix is reported in a heatmap for the correlation matrix in Figure \ref{PIC_GausCopMod}. In addition, we present examples based on posterior mean covariance for covariance sub-blocks $p\left(\Sigma_4^P |\bm{P},\bm{I}\right)$ and then for $p\left(\Sigma_4^I |\bm{P},\bm{I}\right)$, where $\Sigma_4^P \in SP^{+}(6)$ and $\Sigma_4^I \in SP^{+}(5)$, again converted to heatmaps of the correlation. We see that although the priors selected for the dependence features in Model III in all cases favoured independence, since the scale matrices were all diagonal i.e. $\Lambda_5^P = \mathbb{I}_{6}$ and $\Lambda^I_4 = \mathbb{I}_{5}$, the resulting summaries of the marginal posteriors of the covariances clearly indicate non-trivial dependence patterns in the development years within the payments data and the incurred loss data. This is observed throughout each sub-block covariance matrix.

Table \ref{Tab:CovPost} provides a second summary of the posterior for the covariance matrix which further demonstrates features of the dependence properties in the payment and incurred data per accident year and involves the estimates of the largest eigenvalue of each block diagonal matrix for the payment and incurred data as summary statistics. These estimates are given by
\begin{equation}
\widehat{\lambda}^{(s)}_i = \arg \max \left(\det(\Sigma_i^{(s)} - \lambda\mathbb{I})=0\right).\\
\end{equation}
The largest eigenvalue provides information on the posterior distribution of the magnitude of the first principal component of each development year, decomposed by accident year. That is, we can quantify in the PIC model, by accident year, the proportion of residual variation in the log payments for accident year $i$ currently unexplained by the development factors $\Phi_{0:J-i}$, which were jointly estimated in the PIC model and assumed constant accross each accident year (i.e. constant per development year) for parsimony. We can also repeat this for the incurred loss data. Suppose that a principal component analysis is performed, decomposing the variation in the payment and incurred data for each accident year $i$ with respect to the variation unexplained by the development factors in the PIC model. Then, up to proportionality, the distribution of the eigenvalues corresponds to the proportion of contribution from the leading eigenvector (principal component). When this is coupled with the fact that we can also easily obtain samples from the marginal posterior distribution of the leading eigenvector of the covariance matrix for the $i$-th accident year's payment of incurred loss data in the PIC model, then we get complete information per accident year on the ability of the development factors in the PIC model to explain variation in the observed loss data. Table \ref{Tab:CovPost} summarises the results for the average PCA weight (largest eigenvalue) and average posterior eigenvector.

\begin{figure}[ht]
\includegraphics[height=6cm, width=\textwidth]{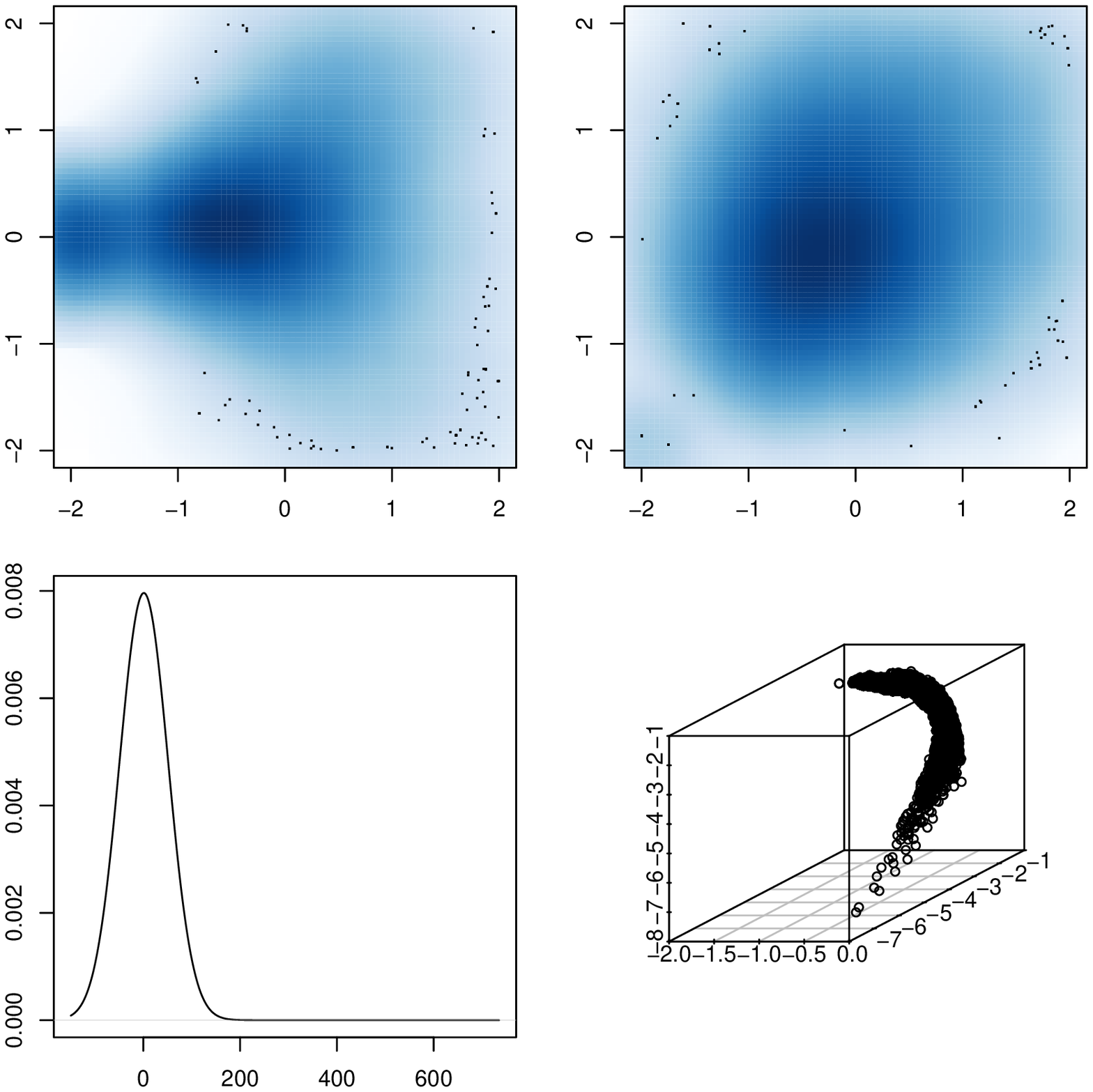}
\includegraphics[height=6cm, width=\textwidth]{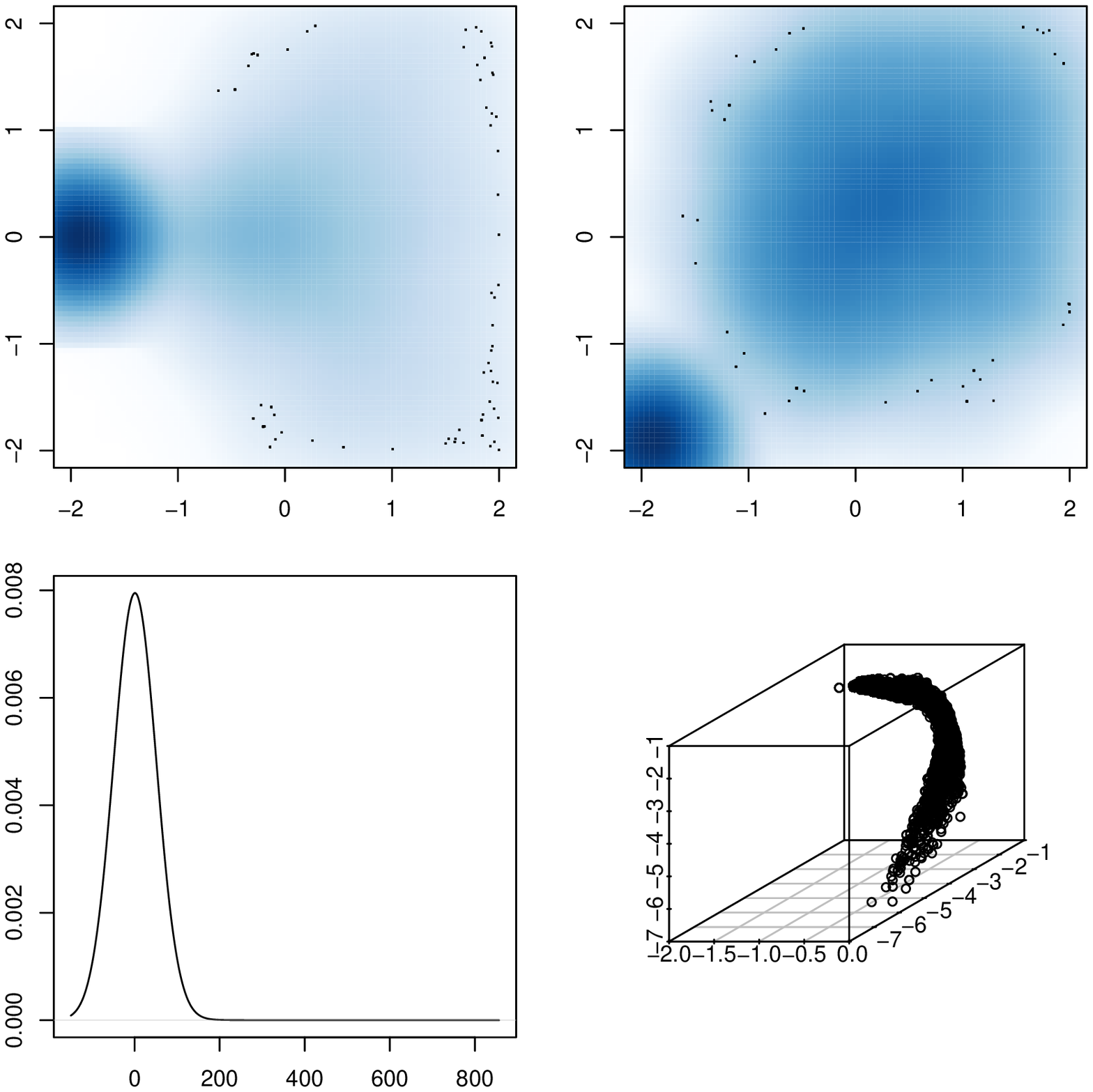}
\caption{Heatmaps for the block diagonal covariance matrices $\Sigma_4^P$ ($2 \times 2$ sub-plot 1) and $\Sigma_4^I$ ($2 \times 2$ sub-plot 2). These are obtained using samples from the Riemannian Manifold Adaptive Metropolis sampler. Samples from the Posterior distribution of the telescoping diagonal block size form covariance matrix structures of the Gaussian copula under the hierarchical Bayesian model which has the joint covariance between all observed payment and incurred loss data under the dependent development years. Each set of $4 \times 4$ panels, starting from the top, summarizes the posterior distributions for the covariance matrices for $s \in \left\{P,I\right\}$ according to:
\textbf{\textit{Top Left Panel}}: contour map of posterior samples $\log\left[\Sigma_4^s\right]_{1,1}$ vs $\log\left[\Sigma_4^s\right]_{5,5}$.
\textbf{\textit{Top Right Panel}}: contour map of posterior samples $\log\left[\Sigma_4^s\right]_{1,1}$ vs $\left[\Sigma_4^s\right]_{1,5}$.
\textbf{\textit{Bottom Left Panel}}: kernel density estimator of the posterior distribution of the trace of the covariance matrix using samples $\left\{\log \text{tr}\left(\Sigma_4^s\right)\right\}$.
\textbf{\textit{Bottom Right Panel}}: scatter plot of posterior samples of the first, second and third largest eigenvalues scaled by total of the eigen valuse - (PCA weights - for linear combinations of the development factors when explaining variation in observed payment and incurred data for a given accident year).
}
\label{PIC_GausCopModP4I4CovPosts}
\end{figure}

\citet{tokuda2011visualizing} develops a framework which formalizes an approach to the summary of dependence structures. For the running example of results that we present for distributions $p\left(\Sigma_4^P |\bm{P},\bm{I}\right)$ and $p\left(\Sigma_4^I |\bm{P},\bm{I}\right)$, under such an approach the third and fourth layers of summary are presented in Figure \ref{PIC_GausCopModP4I4CovPosts}. This involves the presentation of contour maps of these marginal posteriors that are constructed using adaptive MCMC samples of these matrices.

\begin{figure}[ht]
\includegraphics[height=5cm, width=0.45\textwidth]{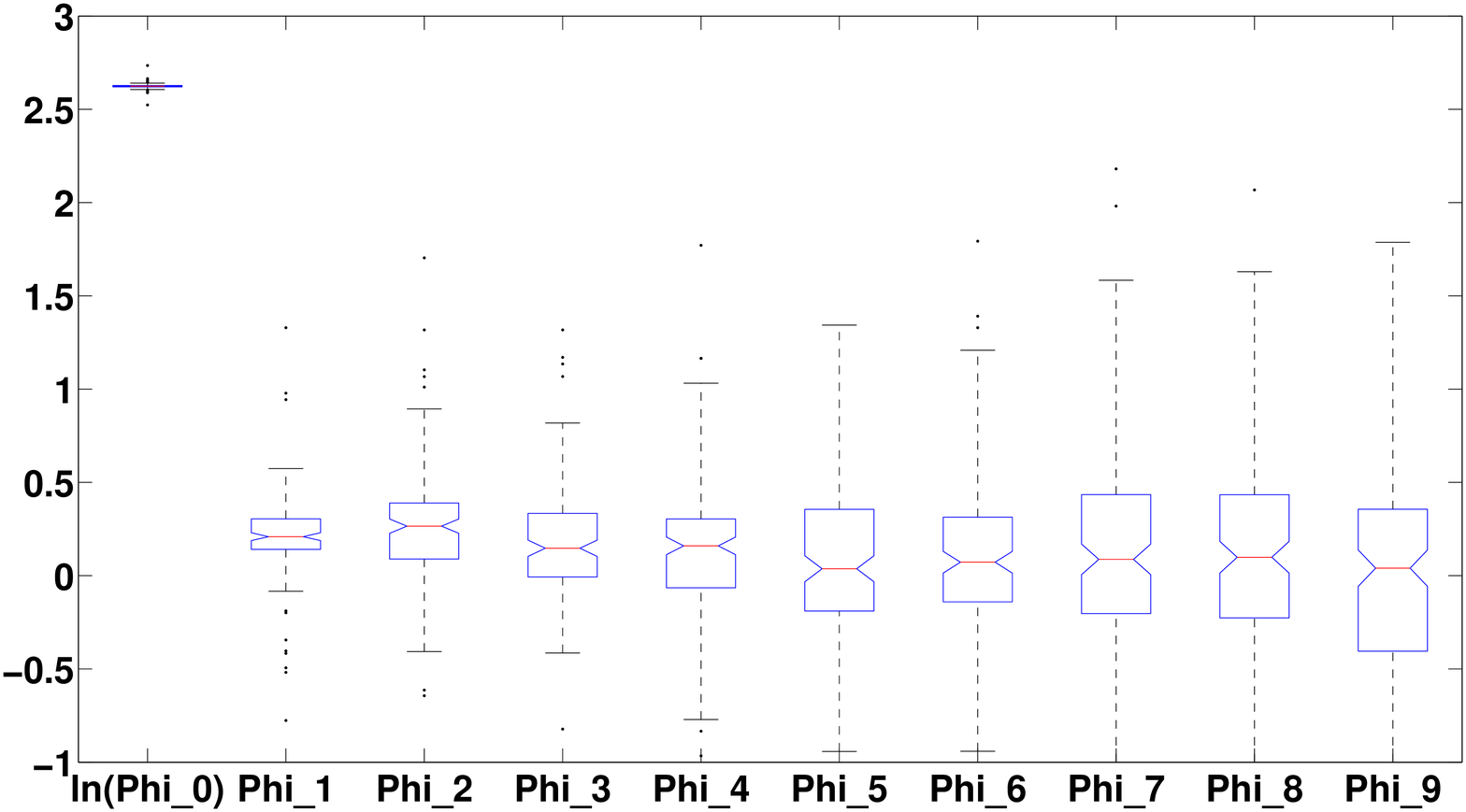}
\includegraphics[height=5cm, width=0.45\textwidth]{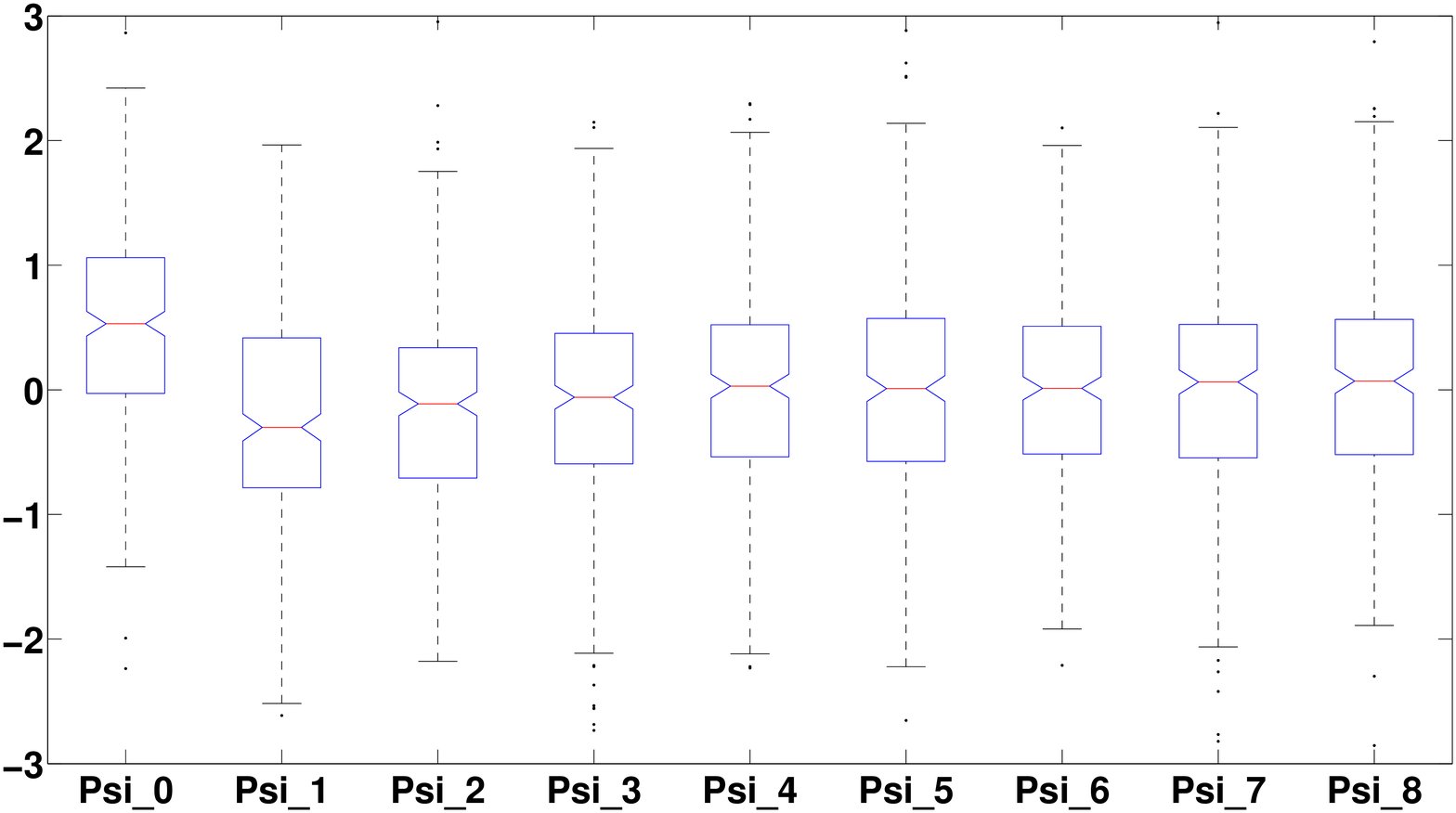} \\
\includegraphics[height=5cm, width=0.45\textwidth]{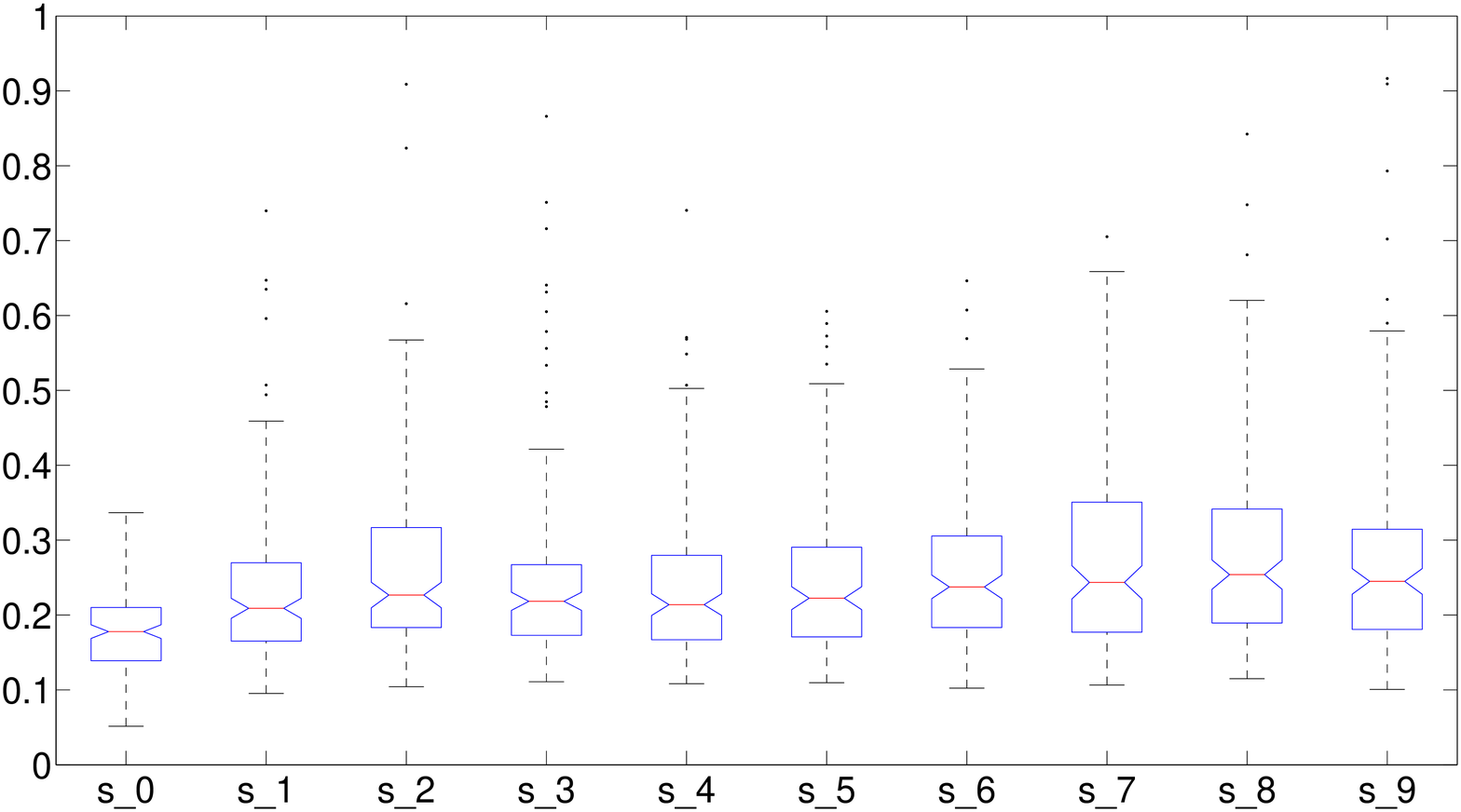}
\includegraphics[height=5cm, width=0.45\textwidth]{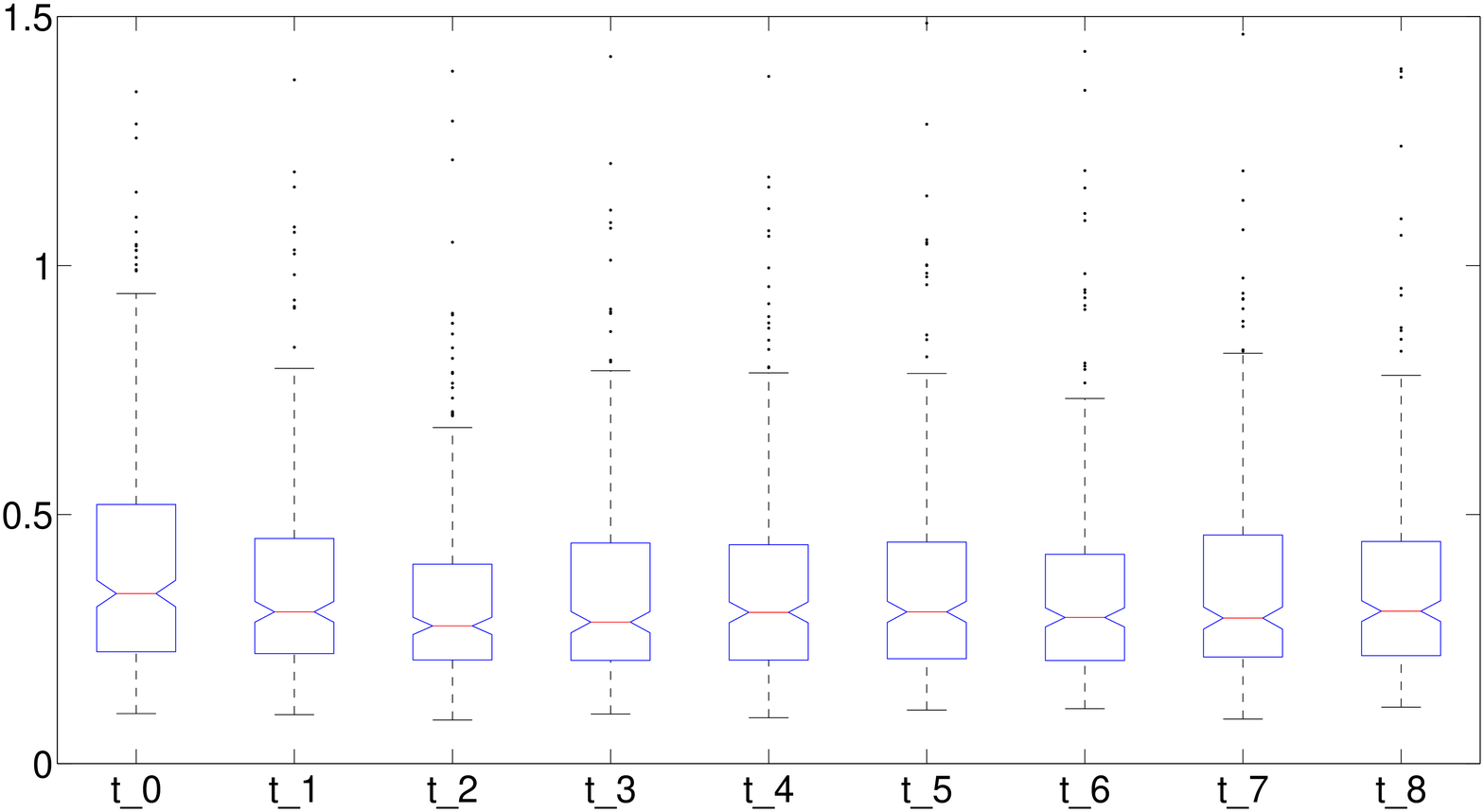}
\caption{Boxplot summaries of the marginal posterior distributions obtained using samples from the Riemannian Manifold Adaptive Metropolis sampler. Samples from the Posterior distribution under a telescoping diagonal block size form covariance matrix structures of the Gaussian copula under the hierarchical Bayesian model which has the joint covariance between all observed payment and incurred loss data under the dependent development years.
\textbf{\textit{Top Left Panel}}: box plots of marginal posterior distributions for $p\left(\Phi_i|\bm{P},\bm{I}\right)$.
\textbf{\textit{Top Right Panel}}: box plots of marginal posterior distributions for $p\left(\Psi_i|\bm{P},\bm{I}\right)$.
\textbf{\textit{Bottom Left Panel}}: box plots of marginal posterior distributions for $p\left(s_i|\bm{P},\bm{I}\right)$.
\textbf{\textit{Bottom Right Panel}}: box plots of marginal posterior distributions for $p\left(t_i|\bm{P},\bm{I}\right)$.
}
\label{PIC_GausCopModDevFactMargPosts}
\end{figure}

In Figure \ref{PIC_GausCopModDevFactMargPosts}, the development factors for payment and incurred data marginal posterior distributions are presented along with the posteriors of the hyperparameters for the Gaussian Copula based PIC models (Model III). Finally, we also compare the estimated posterior marginal distributions of the development factors for the payment and incurred loss triangles for the models: payment only model; the incurred only model; the Gaussian Copula (Model III) dependent model; the PIC [Full] independent model and the PIC [Partial] independent model of \citet{merz2010paid}. The results of this comparison include the posterior mean estimates of $\mathbb{E}\left[\Phi_i|\bm{P},\bm{I}\right]$ and $\mathbb{E}\left[\Psi_i|\bm{P},\bm{I}\right]$, for all $i \in \left\{0,1,\ldots,J\right\}$ and the posterior quantiles for left and right tails as measured by the fifth and ninety-fifth percentiles, given in Table \ref{Tab:MMSEModels}. \textit{We note that the results in this section for the Gaussian copula models are obtained using the log ratio observational data and the restults for the Mixture Archimedian copula model are more conveniently obtained using the log observations (not ratio data).}

It is also worth noting other approaches that can be adopted in the case of the Gaussian copula model. One could also included a data-augmentation stage in the analysis as was utilised in the Mixture Archimedian copula example. In addition, the covariance matrices could have been specified under different structures with more or less parsimony. The examples utilised in this section were those which provided a reasonable trade-off between parsimonious model specification, while allowing a meaningful decomposition of the results.

The results of the comparison between the Gaussian copula PIC model and the independent PIC model illustrated that whilst the posterior marginal mean development factor estimates are not affected by the dependence feature included, the marginal posterior shape is affected. This is reflected by the comparison of the posterior confidence intervals for the Gaussian copula PIC model when compared to the payment or incurred individual models where there is a significant difference present in the shapes of the marginal posterior. It is expected that this will have implications for the estimation of reserves using these different will be quantified in the next section.

\subsubsection{Data-Augmented hierarchical Bayesian Mixture-Archimedian Copula PIC (Model IV)}
\begin{figure}[ht]
\includegraphics[height=8cm, width=\textwidth]{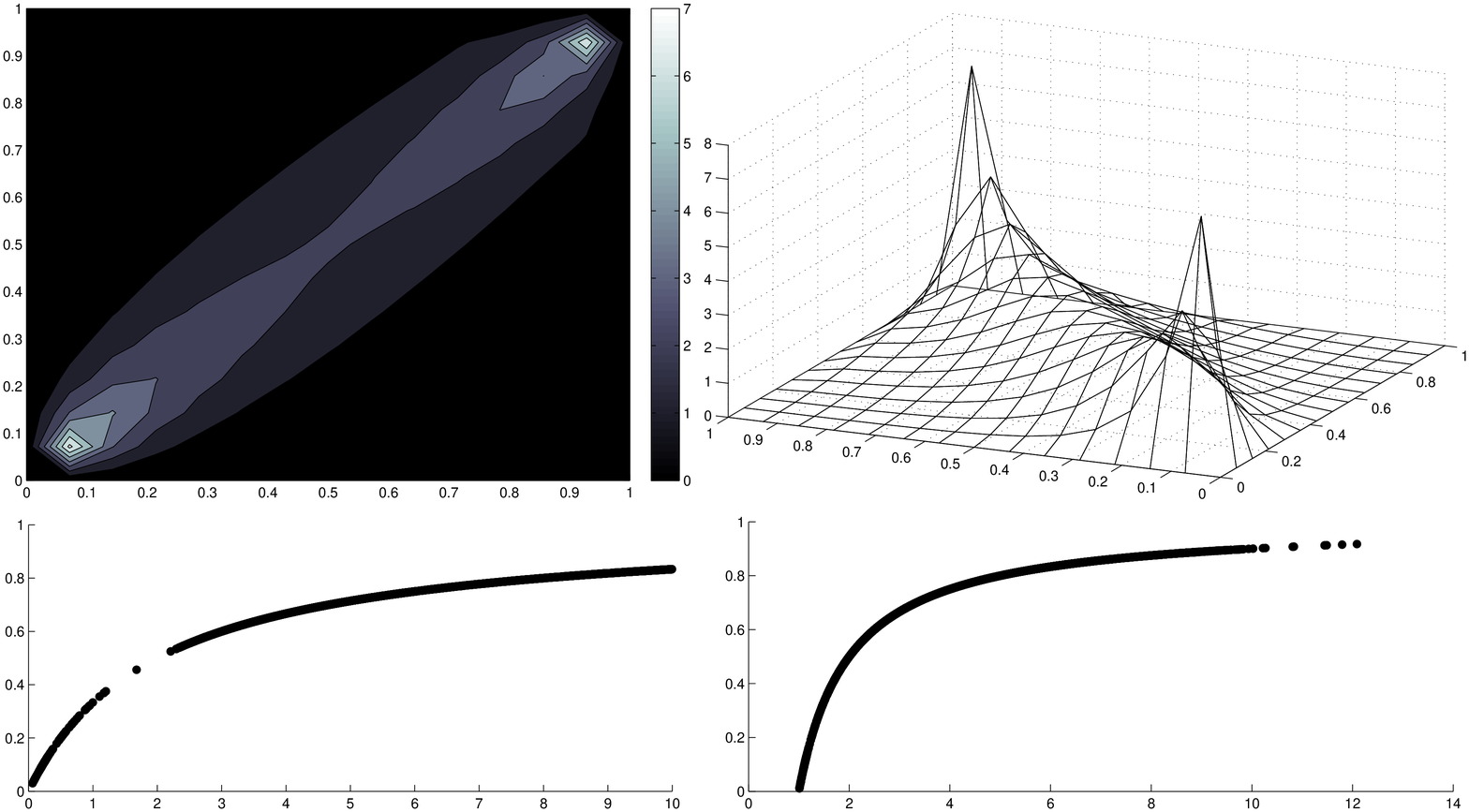} \par \noindent
\caption{Copula Dependence Parameter Posterior distributions estimated under the Data-Augmented Mixture Copula PIC Model IV. A mixture of Archimedean copula models is considered, with Clayton and Gumbel copula choices, allow for possible asymmetry in the tail dependence over development years.We chose uniformative uniform priors $U[0,20]$ for the copula parameters.
\textbf{\textit{Top Left Panel}}: Contour map of posterior estimated mixture copula dependence distribution between development years over paid and incurred loss data, with homogeneous dependence assumptions over accident years (estimated from posterior mean of $\rho_C^{MMSE}$ and $\rho_G^{MMSE}$.
\textbf{\textit{Top Right Panel}}: Surface plot of posterior estimated mixture copula dependence distribution between development years over paid and incurred loss data, with homogeneous dependence assumptions over accident years (estimated from posterior mean of $\rho_C^{MMSE}$ and $\rho_G^{MMSE}$.
\textbf{\textit{Bottom Left Panel}}: Scatter plot of posterior samples used to estimate Kendall's tau rank correlation versus copula parameter for the Clayton mixture component.
\textbf{\textit{Bottom Right Panel}}: Scatter plot of posterior samples used to estimate Kendall's tau rank correlation versus copula parameter for the Gumbel mixture component.
}
\label{PIC_ArchMixCopMod}
\end{figure}

This section presents the estimation results for the mixture of Clayton and Gumbel Copula based PIC models (Model IV) on the real data are presented in this section. Figure \ref{PIC_ArchMixCopMod} presents a summary of the mixture copula dependence structure obtained from posterior samples of the copula parameters under the hierarchical Bayesian model. The results in this section are obtained using the log observational data, not ratio data. The figures summarise succinctly the estimated posterior dependence structure for the hierarchical Bayesian mixture Copula model, through plots of the dependence structure as captured by the estimatd mixture copula distribution, the scatter plots of copula parameter for the lower tail and rank correlation (Kendall's tau) and the upper tail copula parameter versus rank correlation. These results clearly demonstrate posterior evidence for non-trivial tail dependence features in the payment and incurred data, as well as potential for asymmetry in the upper and lower tail dependence. Note, uniformative prior choices were made on the copula parameters with uniform priors over $[0,50]$ and $[1,50]$ respectively, indicating these estimated copula parameters are data driven results.

\begin{figure}[ht]
\includegraphics[height=5cm, width=0.45\textwidth]{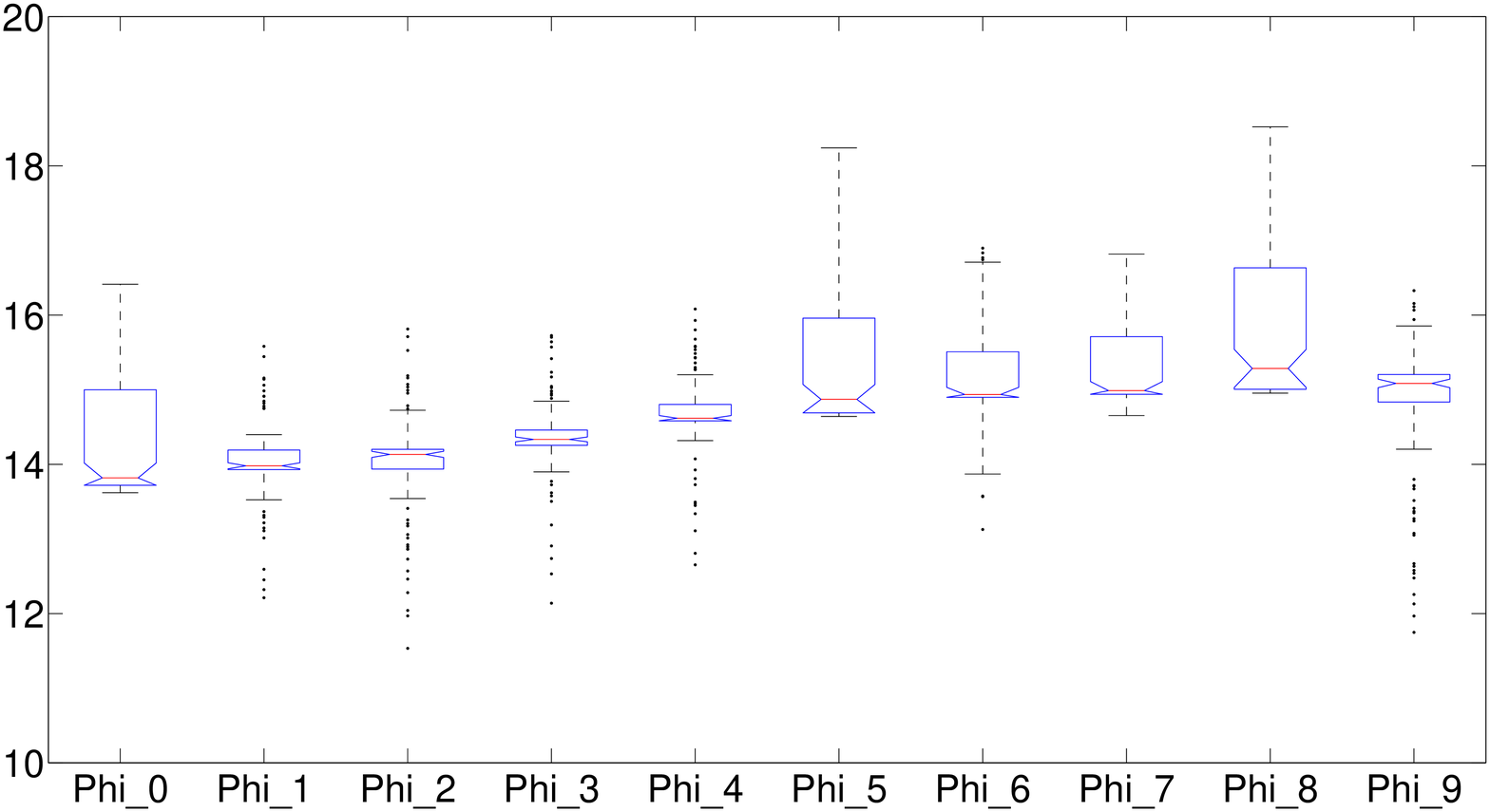}
\includegraphics[height=5cm, width=0.45\textwidth]{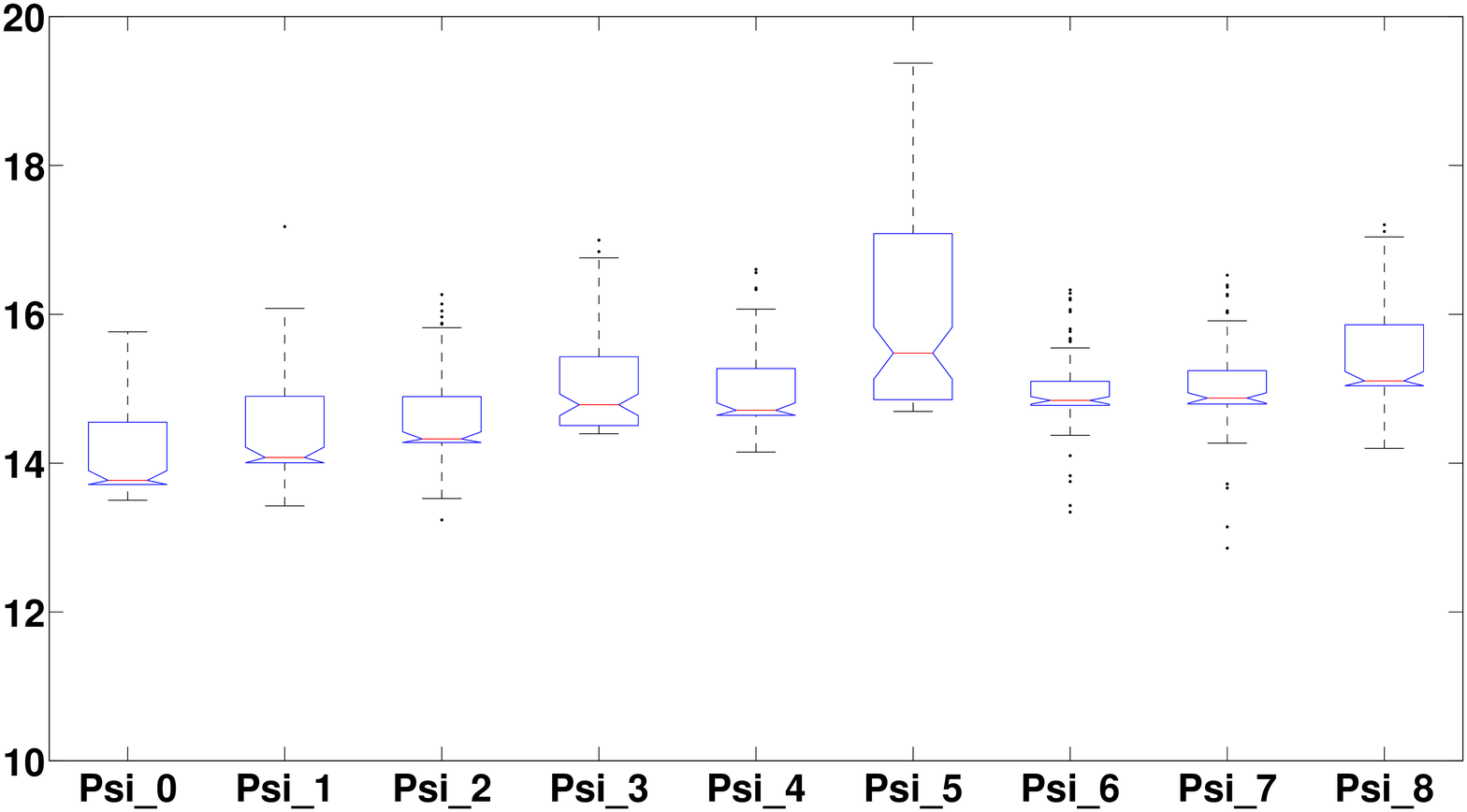} \\
\includegraphics[height=5cm, width=0.45\textwidth]{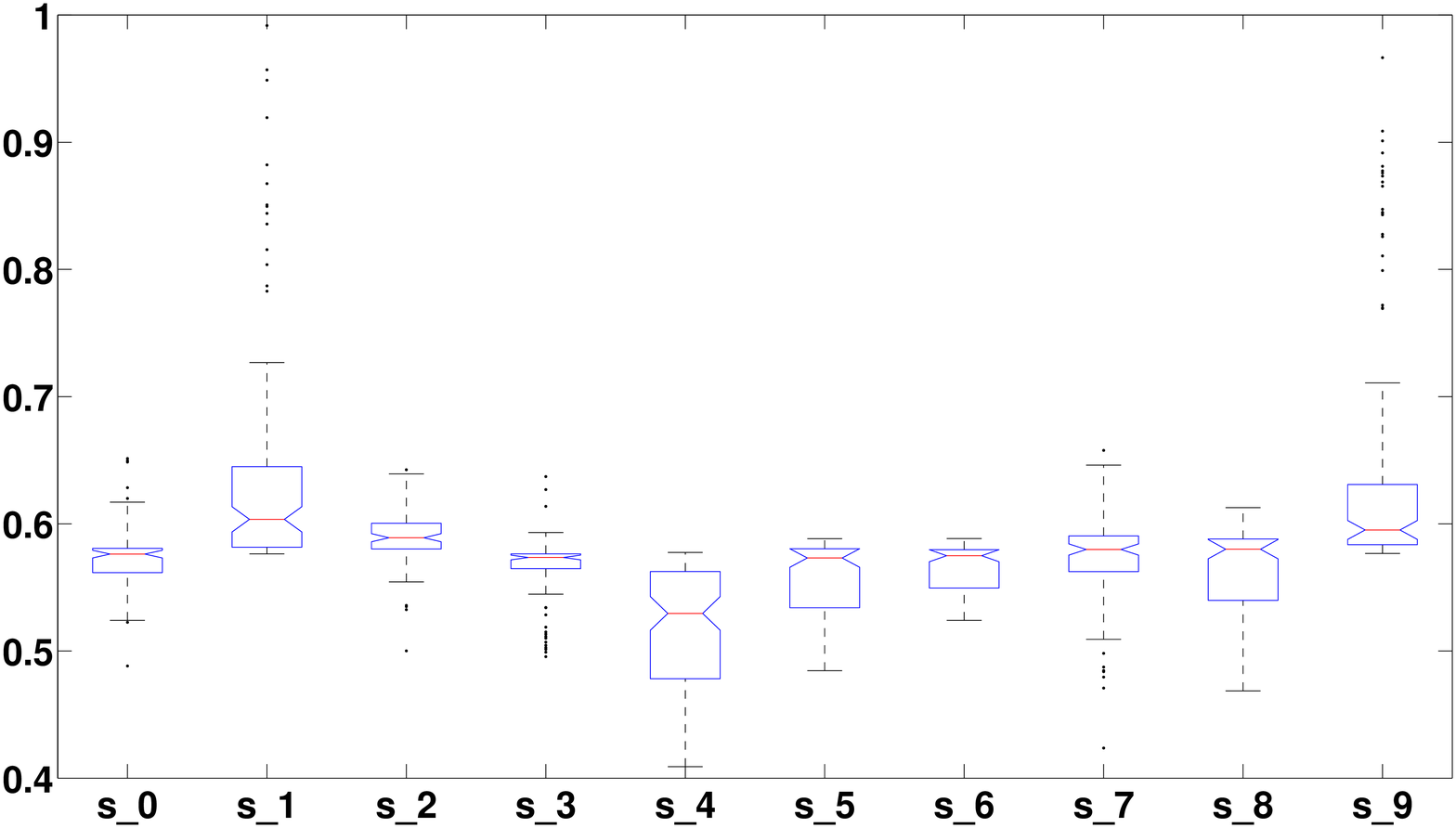}
\includegraphics[height=5cm, width=0.45\textwidth]{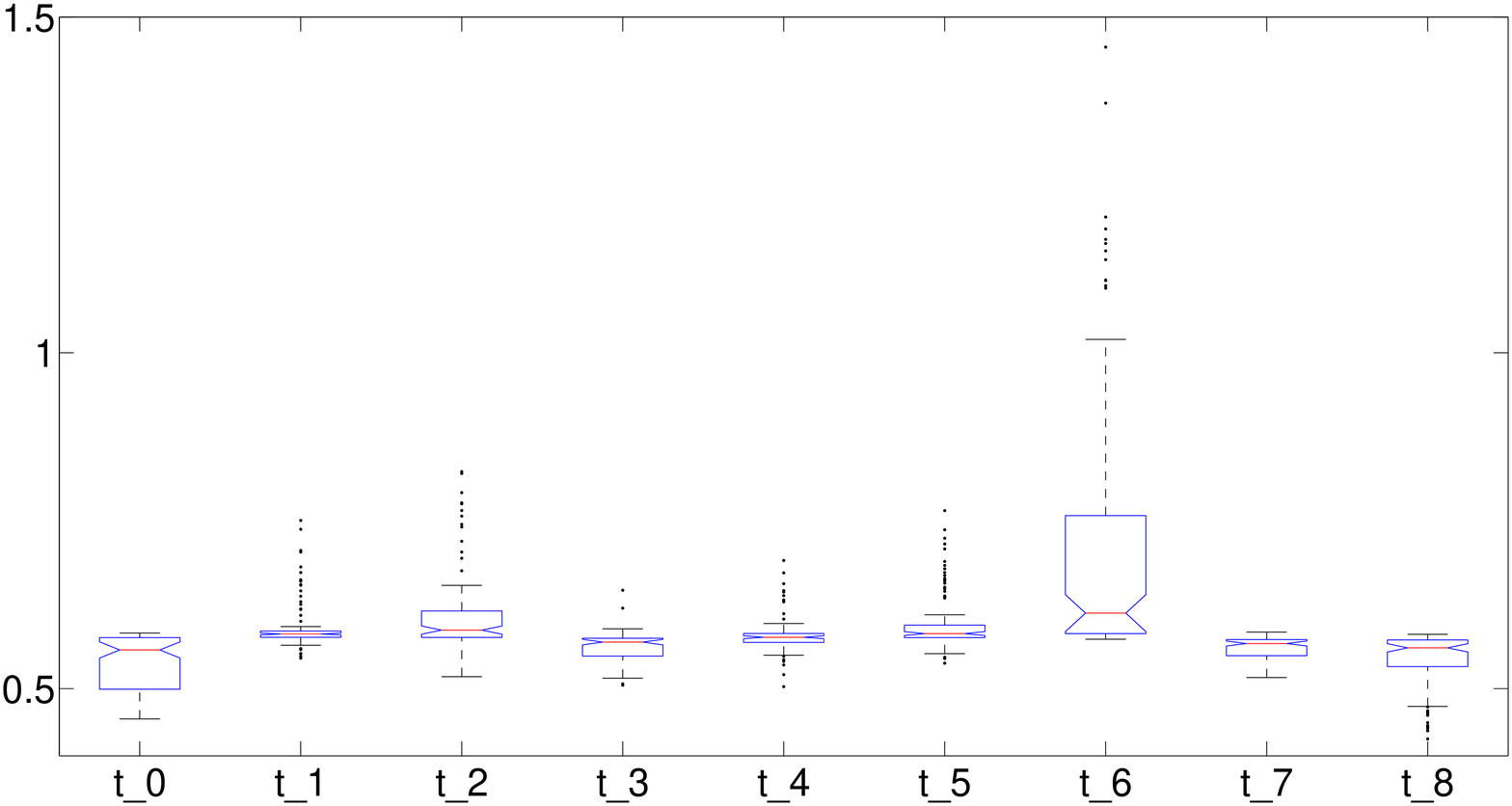}
\caption{Boxplots of the marginal posterior distributions of the development factors and hyperparameters.
\textbf{\textit{Top Left Panel}}: box plots of marginal posterior distributions for $p\left(\Phi_i|\bm{P},\bm{I}\right)$.
\textbf{\textit{Top Right Panel}}: box plots of marginal posterior distributions for $p\left(\Psi_i|\bm{P},\bm{I}\right)$.
\textbf{\textit{Bottom Left Panel}}: box plots of marginal posterior distributions for $p\left(s_i|\bm{P},\bm{I}\right)$.
\textbf{\textit{Bottom Right Panel}}: box plots of marginal posterior distributions for $p\left(t_i|\bm{P},\bm{I}\right)$.
}
\label{PIC_MixCopModDevFactMargPostss}
\end{figure}

Figure \ref{PIC_MixCopModDevFactMargPostss} presents the development factors for payment and incurred data marginal posterior distributions along with the hyperparameter marginal posteriors for the Data-Augmented Mixture Copula based PIC models (Model IV).

\FloatBarrier

\section{Comparison of PIC reserving with Gaussian Copula PIC and Mixture Archimedian Copula PIC Models}

This section discuss the effect of modelling the dependence structures on the reserving estimates. First we note two important details in calculating the reserves. We need to be able to draw samples from the predictive distributions for the payment and incurred data given below, for each accident year $i$, using
\begin{equation*} \label{Eqn:predictive}
p\left(P_{i,J}|\bm{P},\bm{I}\right) = \int p\left(P_{i,J}|P_{i,1:J-i},\bm{\Theta}\right)p\left(\bm{\Theta}|\bm{P},\bm{I}\right)d\bm{\Theta}\;\; \mathrm{and} \;\; p\left(I_{i,J}|\bm{P},\bm{I}\right) = \int p\left(I_{i,J}|I_{i,1:J-i},\bm{\Theta}\right)p\left(\bm{\Theta}|\bm{P},\bm{I}\right)d\bm{\Theta}.\\
\end{equation*}
In general it is not possible to solve these integrals analytically. Howerver, for the Gaussian copula models developed in this paper, under the results in Lemma \ref{LemmaTransformData}, one adopt the results of \citet{merz2010paid}[Theorem 2.4] to obtain analytic Gaussian predictive distributions.  Alternatively, the predicitive distributions can be estimated as described in \citet[Section 3.3]{peters2010chain}. Although the results in Table \ref{Tab:MMSEModels} demonstrate that the incorporation of the dependence structures does not significantly alter the posterior mean of the development factors for the payment and incurred loss data, it is clearly possible for the predictive distribution to be altered, since the shape of the posterior distribution is altered by the dependence features. Second, regarding the hierarchical mixture Archimedian copula model, it does not admit an analytic solution for the predictive distribution. This does not matter if a data augmentation stage is set up in the joint posterior distribution to sample cumulative payments, since then we can use the MCMC sampler output for the ultimate cumulative payment and incurred losses in each accident year.

Finally, we also note that a simple Monte Carlo based approximation for the ultimate claim can be constructed. Take the samples from the MCMC output for the PIC model of interest (sampled from the complete PIC model with dependence features present) and then utilise these samples to construct a Laplace approximation to the predictive observation distribution for example $p\left(P_{i,J}|P_{i,1:J-i},\bm{\Theta}\right)$ which involves a normal approximation around the MAP or locally around each Monte carlo sample for the development factors, with precision given by the sampled observation covariance structure. Though this is not required, as we have shown for the Gaussian copula models independence models, it may be useful for alternative copula based models with simple data-augmentation approaches. In addition a second alternative would be to utilise in the predictive distribution the marginal distributions.

\begin{figure}[ht]
\includegraphics[height=8cm, width=\textwidth]{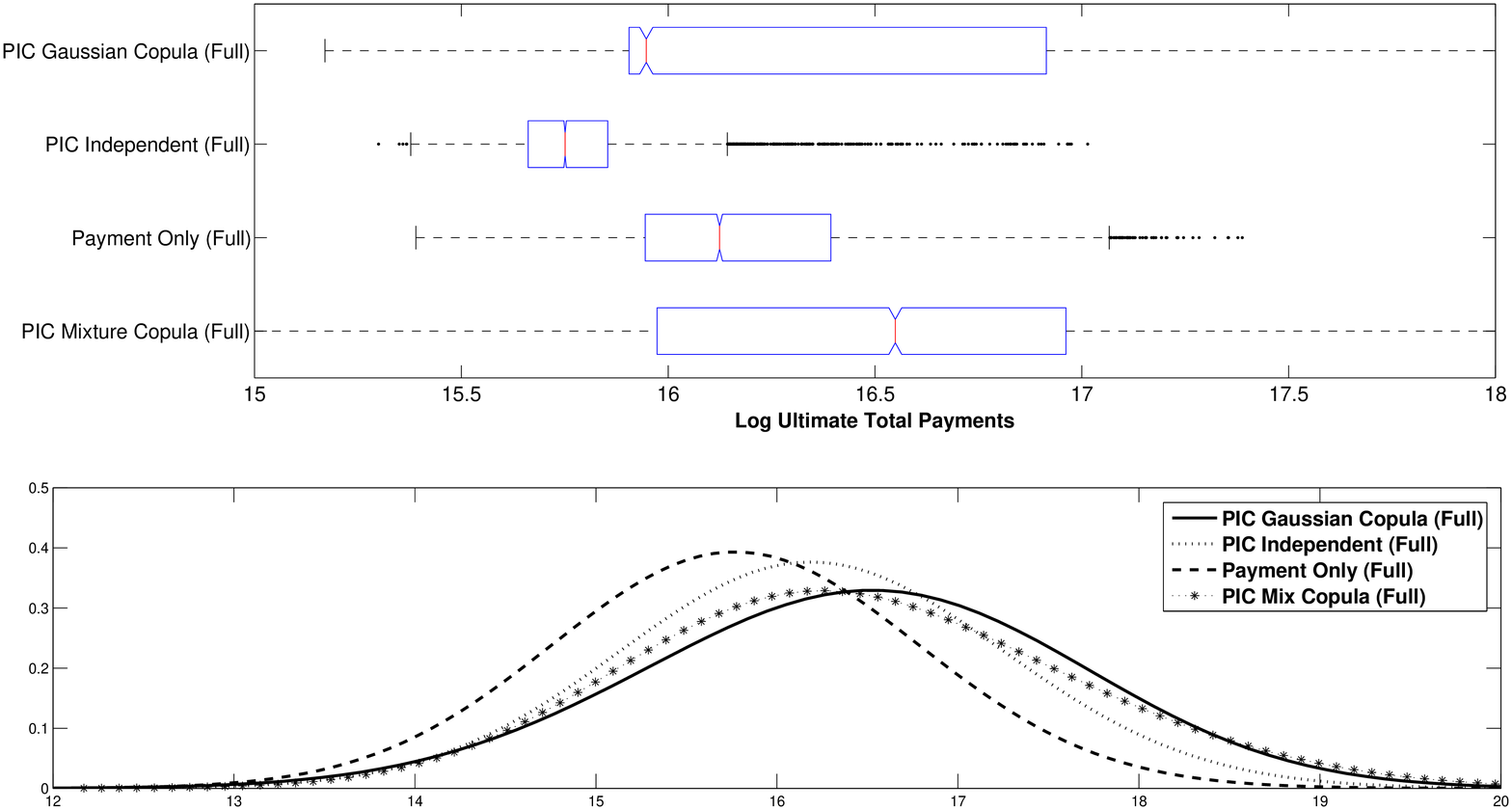} \par \noindent
\caption{Boxplots of the predictive distributions obtained from the MCMC samples. Ultimate Bayesian predictive distributions for log payment data from the payment only predictive distribution, the Full Independent PIC model, and the hierarchical PIC Mixture Copula model via Data Augmentation predictive distribution.
\textbf{\textit{Left Panel}}: Posterior predictive distribution box plots from samples.
\textbf{\textit{Right Panel}}: Kernel density estimates of the predictive distributions.
}
\label{PIC_UltimateRatio}
\end{figure}

Figure \ref{PIC_UltimateRatio} presents the log posterior predictive distribution for the ultimate total claim given by the predictive distribution for the log of the cumulative payment over each accident year $\sum_{i=0}^JP_{i,J}$ for the full Bayesain models which incorporate priors on observation error, development factors and hyperpriors for precision of the development factors. We see that all three models are in good agreement with each other with the dependence parameters affecting the variance and tail behaviour of the distributions.

\begin{figure}[ht]
\includegraphics[height=7cm, width=\textwidth]{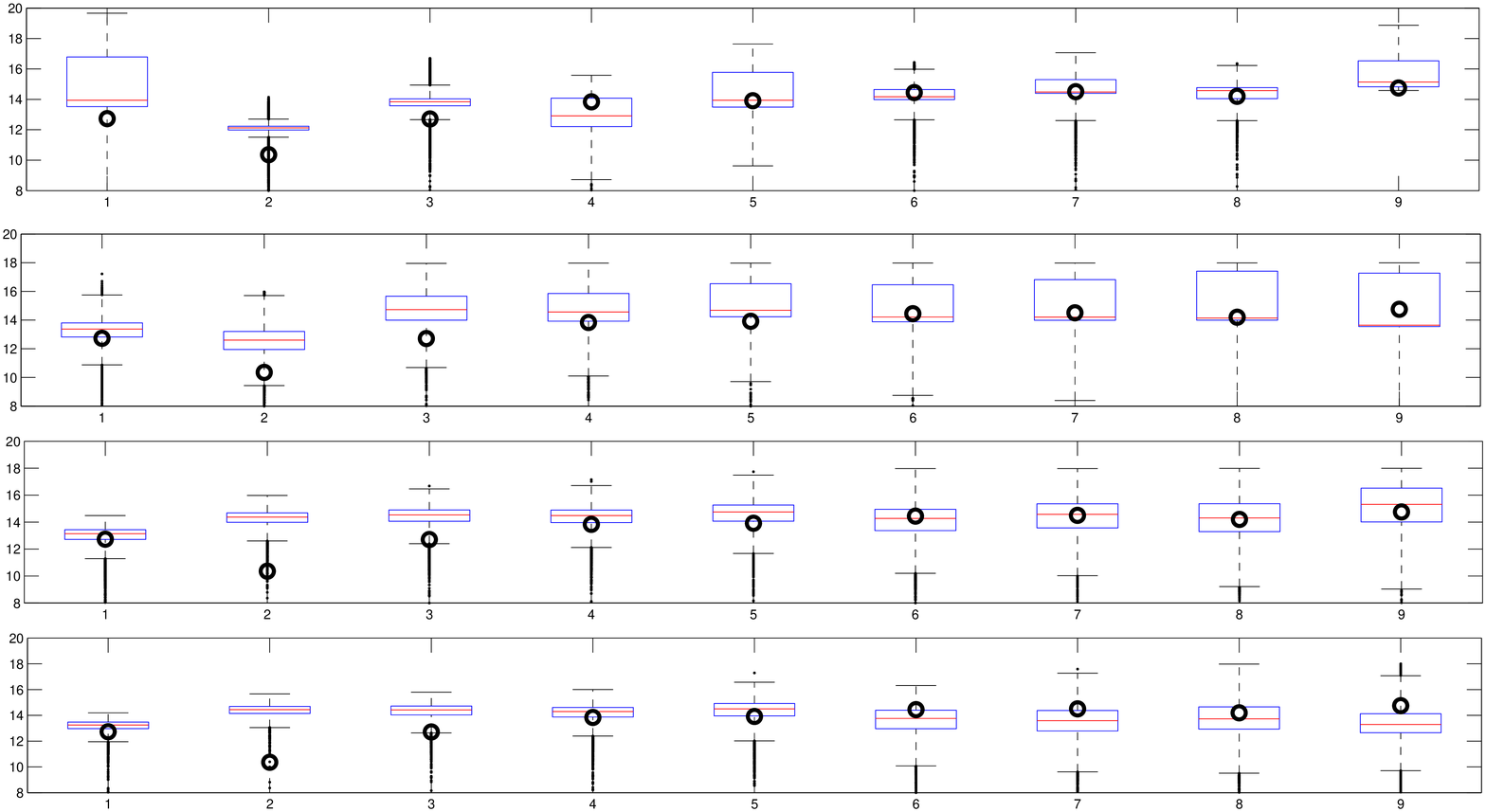} \par \noindent
\caption{Boxplots of log ultimate Bayesian predictive reserve distributions for payment data per accident year, compared to (Partial) PIC Independent posterior mean estimates from \citet{merz2010paid} (karge unfilled black circles). \textbf{Top Row:} the (Full) hierarchical PIC Mixture Copula model via Data Augmentation; \textbf{Second Row from Top: } the (Full) hierarchical PIC Gaussian Copula model; \textbf{Third Row from Top: } the (Full) Independent PIC model; \textbf{Bottom Row: } the (Full) payment Only model.
}
\label{PICReservesDist}
\end{figure}

Next we consider the distributions of the outstanding loss liabilities estimated using the S samples from the MCMC obtained for the posterior PIC model. We denoted these by random variables $\left\{R(\bm{P},\bm{I})^{(s)}\right\}_{s=1:S}$ where $R(\bm{P},\bm{I})^{(s)} = P_{i,J} - P_{i,J-i}$ and depending on whether payment, incurred, or both data is present we denoted by $R(\bm{P})^{(s)}$, $R(\bm{I})^{(s)}$ and $R(\bm{P},\bm{I})^{(s)}$ respectively. Figure \ref{PICReservesDist} presents the MCMC estimated claims reserve marginal posterior predictive distributions for each accident year per model developed.
We compared our results to those obtained in \citet{merz2010paid} and find good agreement between the mean reserve per accident year and each proposed model. In addition, we note the possible differences between the distributions can be attributed to the utilisation of the full versus partial hierarchical Bayesian models in this paper and the different dependence structures considered. Additionally, we note that further analysis on comparisons to existing models in the literature can be obained for the models of \citet{mack1993distribution}, \citet{dahms2008loss} and \citet{quarg2004munich} for this data analysis in \citet{merz2010paid} [Table 4] and in the spreadsheet provided by Professor Mario Wuethrich at URL\begin{footnote}{URL:\url{http://www.math.ethz.ch/~wueth/claims_reserving3.html}}\end{footnote}.

\section{Conclusions}

This paper extends the class of PIC models to combine the two different channels of information as proposed in \citet{merz2010paid} by introducing several novel statistical models for the dependence features present within and between the payment and incurred loss data. This allows us to obtain a unified ultimate loss prediction which incorporates the potential for general dependence features. To achieve this we developed full hierarchical Bayesian models which incorporate several different potential forms of dependence, including generalized covariance matrix structure priors based on inverse Wishart distributions and conditional Bayesian conjugacy in the PIC independent log-normal model. This forms a general class of Gaussian copula models which extends the approach of \citet{happ2011paid}.

Second, we develop a class of hierarchical mixture Archimedian copula models to capture potential for tail dependence in the payment and incurred loss data, again developing and demonstrating how to appropriately construct a full Bayesian model incorporating hyperpriors. In this regard, we also develop a class of models in which data-augmentation is incorporated to both overcome challenging marginal likelihood evaluations required for the MCMC methodology to sample from the PIC Bayesian models. This had the additional feature that it also allowed for joint Bayesian inference of the reserves as part of the posterior inference.

Finally, to perform inference on these approaches we developed an adaptive Markov chain Monte Carlo sampling methodology incorporating novel adaptive Riemann-manifold proposals restricted to manifold spaces (postive definite symmetric matrices) to sample efficiently the covariance matrices in the posterior marginal for the Gaussian copula dependence. We make these advanced MCMC accessible to the actuarial audience to address challenging Bayesian inference problems in Claims Reserving modelling.

The consequence of these models for actuaries is that a new extended suite of flexible dependence structures have been incorporated into the recently proposed PIC models. These can now be extended and compared to existing chain ladder methods. We perform an analysis on real payment and incurred loss data discussed in \citet{merz2010paid} and compare our models with the analysis for the independent PIC model (partial) and the (full) Bayesian PIC model as well as several different dependent models and the payment only model. Furthermore, we provide reference on further comparisons to the alternative models of \citet{mack1993distribution}, \citet{dahms2008loss} and \citet{quarg2004munich} for this data.

\section{Acknowledgement}
We would like to thank Prof. Mario Wuethrich of ETH for his discussions on this topic and for introducing us to the family of models that we have explored. The research of Robert Kohn was partially supported by ARC Discovery Grant DP0667069.
\FloatBarrier

\bibliographystyle{plainnat}	
\bibliography{IME_PIC}		

\begin{table}[htbp] \label{Tab:CovPost}
	\centering
		\begin{tabular}{|p{1cm}|p{2cm}p{2cm}p{2cm}p{8cm}|}
 \multicolumn{5}{p{15cm}}{Posterior Covariance Matrix for payments and incurred Loss Gaussian Copula}\\ \hline
Sub-Block & Ave. $\widehat{\lambda}^{(s)}_i$	& Std.Dev $\widehat{\lambda}^{(s)}_i$ &	 $[Q0.05;Q0.95]$ for $\widehat{\lambda}^{(s)}_i$ & Ave. Principal Eigen Vector $\widehat{\bm{v}}$ \\\hline
$\Sigma_0^P$	&2.52 &13.15&	[0.15;11.09]& [0.10,-0.12,-0.07,-0.03,-0.05,-0.03,0.01,-0.06,-0.03]
\\
$\Sigma_1^P$	&1.97 &13.92&	[0.15;8.30] & [0.05,-0.08,-0.04,-0.02,-0.01,-0.01,-0.02,-0.01,-0.01]\\
$\Sigma_2^P$	&0.94	&8.48 &	[0.14;3.19] & [0.06,-0.10,-0.06,-0.02,-0.02,-0.01,-1.1e-3,-0.01]\\
$\Sigma_3^P$	&0.75	&6.38 &	[0.14;1.92] & [0.08,-0.12,-0.05,-0.03,-0.01,-6.9e-5,-2.3e-3]\\
$\Sigma_4^P$	&0.76 &6.81 &	[0.13;0.25] & [0.12,-0.13,-0.06,-0.03,-0.01,0.01]\\
$\Sigma_5^P$	&0.70	&5.93	& [0.12;0.23] & [0.14,-0.15,-0.08,-0.01,0.01]\\
$\Sigma_6^P$	&1.11 &9.90	& [0.12;0.24] & [0.21,-0.20,-0.07,0.03]\\
$\Sigma_7^P$	&2.16 &18.07&	[0.10;0.26] & [0.27,-0.25,0.07]\\
$\Sigma_8^P$	&5.44	&34.67&	[0.08;20.92]& [-0.47,0.43]\\
$\Sigma_9^P$	&1.95	&10.28&	[0.06;11.56]& Not Applicable \\\hline
$\Sigma_0^I$	&1.69 &4.78 &	[0.13;6.57] & [0.10,-0.12,-0.07,-0.03,-0.05,-0.03,0.01,-0.06,-0.03]\\
$\Sigma_1^I$	&1.08	&3.66 &	[0.13;4.93] & [0.03,-0.12,-0.05,-0.03,-0.01,2.4e-3,0.01,0.01]\\
$\Sigma_2^I$	&0.80	&3.26 &	[0.12;3.76] & [0.09,-0.12,-0.06,-0.02,-0.01,3.2e-3,-0.01]\\
$\Sigma_3^I$	&0.66	&3.15 &	[0.11;3.27] & [0.10,-0.12,-0.07,-0.03,-0.02,1.5e-3]\\
$\Sigma_4^I$	&0.65 &3.97 &	[0.10;3.23] & [0.15,-0.15,-0.07,-0.02,0.02,]\\
$\Sigma_5^I$	&1.00 &5.02 &	[0.09;5.62] & [6.6e-12,-6.6e-12,-2.9e-12,-1.0e-12,8.5e-13]\\
$\Sigma_6^I$	&1.15 &7.52 &	[0.08;6.31] & [0.31,-0.24,0.07]\\
$\Sigma_7^I$	&5.26 &25.29&	[0.06;32.87]& [-0.50,0.42]\\
$\Sigma_8^I$	&1.03	&3.82 &	[0.04;6.30] & Not Applicable  \\\hline
		\end{tabular}
\end{table}

\newpage

\begin{landscape}
\begin{table}[htbp]
	\centering
		\begin{tabular}{|c|cc|cc|cc|c|cc|}
 \multicolumn{6}{c}{Posterior Marginal Distributions for Development Factors}\\ \hline
Factor & \multicolumn{2}{|c|}{PIC Gaussian Copula (Full)} & \multicolumn{2}{|c|}{PIC Independent (Full)} & \multicolumn{2}{|p{4cm}|}{payment or incurred only (Full)} & \multicolumn{1}{|p{4cm}|}{\citet{merz2010paid} PIC Independent (Partial)} & \multicolumn{2}{|p{4cm}|}{PIC Mixture Clayton-Gumbel Copula (Full)} \\ \hline
 & Post.Ave. &	$[Q0.05;Q0.95]$ & Post.Ave. &	$[Q0.05;Q0.95]$ & Post.Ave.&	$[Q0.05;Q0.95]$  & Post.Ave. & Post.Ave. & $[Q0.05;Q0.95]$\\\hline
$\Phi_0$ & 13.79	&[13.55;14.03]&	14.51 & [13.19;15.01]	& 13.77 &[13.68;13.86] & 13.78& 13.72 & [13.63;15.96]\\		
$\Phi_1$ & 0.21		&[-0.16;0.58]	&	0.18	& [0.05;0.29]   & 0.20	&[0.12;0.27]   & 0.22 & 13.98 & [13.64;15.05]\\	
$\Phi_2$ & 0.25		&[-0.25;0.77]	&	0.22	& [0.08;0.34]   & 0.23	&[0.14;0.31]   & 0.24 & 14.10 & [12.80;15.14]\\	
$\Phi_3$ & 0.18		&[-0.44;0.81]	&	0.17	& [0.04;0.30]   & 0.15	&[0.06;0.24]   & 0.17 & 14.32 & [13.92;15.29]\\	
$\Phi_4$ & 0.15		&[-0.55;0.86]	& 0.16	& [0.02;0.30]   & 0.13	&[0.04;0.23]   & 0.16 & 14.61 & [14.32;15.60]\\	
$\Phi_5$ & 0.13		&[-0.63;0.91]	& 0.15	& [1.9e-3;0.30] & 0.12	&[0.01;0.22]   & 0.14 & 14.71 & [14.64;16.85]\\	
$\Phi_6$ & 0.10		&[-0.71;0.92]	& 0.12	& [-0.04;0.30]  & 0.08	&[-0.04;0.20]  & 0.11 & 14.92 & [14.82;16.50]\\	
$\Phi_7$ & 0.07		&[-0.79;0.93]	& 0.13	& [-0.05;0.33]  & 0.05	&[-0.09;0.19]  & 0.07 & 14.96 & [14.90;16.44]\\	
$\Phi_8$ & 0.08		&[-0.81;0.97]	& 0.11	& [-0.09;0.32]  & 0.05	&[-0.12;0.22]  & 0.05 & 15.06 & [14.97;17.59]\\	
$\Phi_9$ & 0.04		&[-0.88;0.98]	& 0.10	& [-0.04;0.52]  & 0.02	&[-0.19;0.24]  & 0.08 & 15.10 & [13.26;15.86]\\ \hline		
$\Psi_0$ & 0.51		&[-0.84;1.85]	& 0.45	& [0.31;0.56]   &	0.52	&[0.38;0.64]   & 0.50 & 13.73 & [13.69;15.33]\\
$\Psi_1$ & -0.15	&[-1.50;1.20]	&	-0.08	& [-0.11;0.12]  &	0.01	&[-0.11;0.12]  & -0.15& 14.01 & [13.94;15.78]\\
$\Psi_2$ & -0.13	&[-1.49;1.23]	&	-0.09	& [-0.15;0.20]  &	0.01	&[-0.12;0.12]  & -0.14& 14.30 & [14.23;15.77]\\
$\Psi_3$ & -3.7e-2&[-1.39;1.34]	&	0.01	& [-0.05;0.21]  &	0.01	&[-0.13;0.13]  & -0.04& 14.54 & [14.43;16.46]\\
$\Psi_4$ & -1.7e-2&[-1.39;1.36]	&	-0.01	& [-0.06;0.23]  &	-0.01	&[-0.15;0.14]  & -0.02& 14.67 & [14.58;16.07]\\
$\Psi_5$ & -7.1e-3&[-1.39;1.38]	&	0.02	& [-0.06;0.21]  &	-0.06	&[-0.17;0.15]  & -0.02& 14.89 & [14.72;18.10]\\
$\Psi_6$ & -7.3e-3&[-1.40;1.39]	&	-0.02	& [-0.05;0.30]  &	-0.01	&[-0.19;0.16]  & -0.01& 14.82 & [14.51;15.87]\\
$\Psi_7$ & -2.4e-3&[-1.40;1.39]	&	0.02	& [-0.05;0.34]  &	-0.06	&[-0.40;0.22]  & -0.01& 14.85 & [14.62;16.05]\\
$\Psi_8$ & -2.0e-4&[-1.40;1.40]	&	-0.01	& [-0.02;0.52]  &	-0.13	&[-0.52;0.25]  & -0.01& 15.07 & [14.98;16.76]\\ \hline
		\end{tabular}
\caption{NOTE: \textbf{(Full)} corresponds to PIC models with results for the FULL hierarchical Bayesian PIC model with priors on development factors, observation variances and hyperpriors on precisions on development factors. The PIC Independent \textbf{(Partial)} of \citet{merz2010paid} are the Bayesian posterior results in which $\sigma_m$ and $\tau_n$ are assumed known. In addition, the PIC Mixture Copula model has posterior development factors on the scale of log cumulative payment data (not ratio data), so the reported posterior mean development factors are for the cumulative payment marginal posterior means (log scale).}	
\label{Tab:MMSEModels}
\end{table}

\end{landscape}

\appendix
\section{Appendix A}

\begin{lemma}[Properties of Matrix-Variate Gaussian Distribution] \label{lemmaMatrixGaussianDist} 
A $p \times n$ random matrix $X$ is said to have a matrix variate Gaussian distribution with $p \times n$ mean matrix $M$ and covariance matrix $\Sigma \otimes \Psi$ where $\Sigma$ and $\Psi$ are in the spaces of symmetric positive definite matrices given by $\Sigma \in \mathbb{S}\mathbb{D}^+\left(\mathbb{R}^p\right)$ and $\Psi \in \mathbb{S}\mathbb{D}^+\left(\mathbb{R}^n\right)$ if the $pn \times 1$ dimensional random vector $Vec\left(X'\right)$ has a multivariate normal distribution $Vec\left(X'\right) \sim N\left( Vec(M'), \Sigma\otimes\Psi \right)$.
Furthermore, if $X$ is distributed according to matrix-variate Gaussian distribution $X \sim N_{p,n}\left(M,\Sigma \otimes \Psi \right)$ then the density is given by
\begin{equation}
f^{MVN}_X\left(x;M,\Sigma,\Psi\right) = \frac{\exp\left(-\frac{1}{2}\text{tr}\left[\Sigma^{-1}\left(X - M\right)'\Psi^{-1}\left(X - M\right)\right]\right)}{\left(2 \pi\right)^{np/2}\left|\Sigma\right|^{n/2}\left|\Psi\right|^{p/2}}
\end{equation}
In addition the following properties are satisfied for a matrix-variate Guassian (see \cite{gupta2000matrix} Chapter 2):
\begin{enumerate}
\item{If $X \sim  N_{p,n}\left(M,\Sigma \otimes \Psi \right)$, then $ X' \sim N_{n,p}\left(M', \Psi \otimes \Sigma \right)$}
\item{If $X \sim  N_{p,n}\left(M,\Sigma \otimes \Psi \right)$, and partition $X$, $M$, $\Sigma$, and $\Psi$ as
\begin{equation}
\begin{split}
X = \begin{bmatrix} X_{1r} \\ X_{2r} \end{bmatrix}, \;\; \mathrm{and} \;\;
X = \begin{bmatrix} X_{1c} & X_{2c} \end{bmatrix}
\end{split}
\end{equation}
with $X_{1r}$ the $(m \times n)$ sub-matrix, $X_{2r}$ the $(p-m \times n)$ sub-matrix,     
$X_{1c}$ the $(p \times t)$ sub-matrix and $X_{2c}$ the $(p \times n-t)$ sub-matrix. With analogous partitions of the mean matrix $M_{1r}$, $M_{2r}$, $M_{1c}$ and $M_{2c}$ and covariance matrices
\begin{equation}
\begin{split}
\Sigma = \begin{bmatrix} \Sigma_{11} & \Sigma_{12} \\ \Sigma_{21} & \Sigma_{22} \end{bmatrix}, \;\; \mathrm{and} \;\;
\Psi = \begin{bmatrix} \Psi_{11} & \Psi_{12} \\ \Psi_{21} & \Psi_{22} \end{bmatrix},
\end{split}
\end{equation}
with $\Sigma_{11}$ the $(m \times m)$ sub-matrix, $\Sigma_{12}$ the $(m \times p-m)$ sub-matrix, $\Sigma_{22}$ the $(p-m \times p-m)$ sub-matrix, $\Psi_{11}$ the $(t \times t)$ sub-matrix, $\Psi_{22}$ the $(n-t \times n-t)$ sub-matrix. Then the following properties are true
\begin{equation}
\begin{split}
X_{1r} &\sim N_{m,n}\left(M_{1r}, \Sigma_{11}\otimes \Psi\right)\;\; \mathrm{and}\;\; X_{1c} \sim N_{p,t}\left(M_{1c}, \Sigma\otimes \Psi_{11}\right)\\
X_{2r}|X_{1r} &\sim N_{p-m,n}\left(M_{2r} + \Sigma_{21}\Sigma_{11}^{-1}\left(X_{1r} - M_{1r}\right), \Sigma_{22.1}\otimes \Psi\right)\\
X_{2c}|X_{1c} &\sim N_{p,n-t}\left(M_{2c} + \left(X_{1c} - M_{1c}\right)\Psi_{11}^{-1}\Psi_{12}, \Sigma\otimes \Psi_{22.1}\right)\\
\end{split}
\end{equation}
with $\Sigma_{22.1} = \Sigma_{22} - \Sigma_{21}\Sigma_{11}^{-1}\Sigma_{12}$ and $\Psi_{22.1} = \Psi_{22} - \Psi_{21}\Psi_{11}^{-1}\Psi_{12}$.
}
\end{enumerate} 
\end{lemma}

\begin{lemma}[Properties of Matrix Variate Wishart Distributions] \label{lemmaWishartDist} Suppose that the $(p \times p)$ positive definite matrix $\Sigma$ has an inverse Wishart distribution, with positive definite $(p \times p)$ scale matrix $\Lambda$, degrees of freedom parameter $k > p-1$, and density
\begin{equation}
f(\Sigma|\Lambda,k)= \frac{\left|\Lambda\right|^{m/2}\left|\Sigma\right|^{-(k+p+1)/2}e^{-\mathrm{trace}(\Lambda \Sigma^{-1})/2} }{ 2^{kp/2}\Gamma_p(k/2)},
\end{equation}
where $\Gamma_p(\cdot)$ is the multivariate gamma function. The mean and mode of this distribution are given respectively by
\begin{equation}
\mathbb{E}\left[\Sigma|\Lambda,k\right] = \frac{1}{k-p-1}\Lambda, \;\; \mathrm{and}\;\; m\left(\Sigma\right) = \frac{1}{k+p+1}\Lambda.
\end{equation}
Furthermore, the following marginal and conditional properties of the inverse Wishart distribution are relevant. Consider a partition of the matrices $\Lambda$ and $\Psi$ as
\begin{equation}
\Lambda = \begin{bmatrix} \Lambda_{11} & \Lambda_{12} \\ \Lambda_{21} & \Lambda_{22} \end{bmatrix}, \; \Sigma = \begin{bmatrix} \Sigma_{11} & \Sigma_{12} \\ \Sigma_{21} & \Sigma_{22} \end{bmatrix}
\end{equation}
with $\Lambda_{ij}$ and $\Sigma_{ij}$ denoting $p_{i}\times p_{j}$ matrices, then the following properties are satisfied (see \citet{gupta2000matrix}[Chapter 3, Section 3.4]):
\begin{enumerate}
\item{The random sub-matrix $\Sigma_{11}$ is independent of $\Sigma_{11}^{-1} \Sigma_{12}$ 
}
\item{The marginal distribution of any sub matrix on the diagonal of the matrix $\Sigma$ is distributed as inverse Wishart. For example, the sub random matrix $\Sigma_{11}$ is as inverse Wishart with $\Sigma_{11} \sim \mathcal{IW}\left(\Lambda_{11} , k-p_{2}\right)$;
}
\item{The marginal distribution of sub random matrix $\Sigma_{22\cdot 1}$ is inverse Wishart $\Sigma_{22\cdot 1} \sim \mathcal{IW}\left(\Lambda_{22\cdot 1}, k\right)$.
}
\end{enumerate}
\end{lemma}

In Lemma \ref{lemmaInvWishartDist} below we present details for the matrix-variate Inverse Wishart distribution, see \cite{gupta2000matrix}[Chapter 3.4, Definition 3.4.1 and Theorem 3.4.1]
\begin{lemma}[Properties of Matrix-Variate Inverse Wishart Distribution] \label{lemmaInvWishartDist} 
A random $p \times p$ matrix $V=\Sigma^{-1}$ is distributed as Inverse Wishart, with degrees of freedom $m$ and $p \times p$ parameter matrix $\Psi$, denoted $V \sim IW_{p}(m,\Psi)$ with density
\begin{equation}
f(\Sigma|\Psi,m)= \frac{2^{-1/2(m-p-1)p}|\Psi|^{1/2(m-p-1)}}{\Gamma_{p}\left[1/2(m-p-1)\right]|V|^{1/2m}}\mathrm{etr}\left(-1/2 V^{-1}\Psi\right), \; V>0, \Psi > 0, m>2p.
\end{equation}
\end{lemma}

\section{Appendix B}

The family of Archimedean copula models has the following useful properties presented in Lemma \ref{LemmaArchCop}.

\begin{lemma} \label{LemmaArchCop} Let $C$ be an Archimedean copula with generator $\varphi$. Then according to \citet[Lemma 4.1.2 and Theorem 4.1.5]{nelsen2006introduction}, the following properties hold:
\begin{enumerate}
\item{$C$ is an Archimedean copula if it can be reprsented by
\begin{equation*}
C(u,v) = \varphi^{[-1]}\left(\varphi(u) + \varphi(v)\right)
\end{equation*}
where $\varphi$ is the generator of this copula and is a continous, strictly decreasing function from $[0,1]$ to $[0,\infty]$ such that $\varphi(1)=0$ and $\varphi^{[-1]}$ is the pseudo inverse of $\varphi$.}
\item{$C$ is symmetric, $C(u,v) = C(v,u)\; \forall (u,v)\in [0,1]\times[0,1]$}
\item{$C$ is associative, $C(C(u,v),w) = C(u,C(v,w)) \; \forall (u,v,w) \in [0,1]^3$.}
\item{If $c>0$ is any constant, then $c\varphi$ is a generator of $C$}
\item{According to \citet[Definition 4.7.6]{denuit2005actuarial}, the extension of the Archimedean copula family to n-dimensions is achieved by considering the strictly monotone generator function $\varphi$ such that $\varphi:(0,1] \rightarrow \mathbb{R}^+$ with $\varphi(1) = 0$, then the resulting Archimedean copula can be expressed as,
\begin{equation*}
C(u_1,u_2,\ldots,u_n) = \varphi^{-1}\left(\sum_{i=1}^n \varphi(u_i)\right).
\end{equation*}
}
\end{enumerate}
\end{lemma}

The members of the Archimedean copula family utilised in this manuscript are given below in Lemma \ref{lemmaArchCop}.

\begin{lemma} \label{lemmaArchCop} From the results in \citet[Section 4.3, Table 4.1]{nelsen2006introduction} the distribution and density functions of the Clayton copula are given respectively as:
\begin{align}
	C^C(u_1,...,u_n) &= \left(1-n + \sum_{i=1}^n u_i^{-\rho^C}\right)^{-1/\rho^C},\\
	c^C(u_1,...,u_n) &= \left(1 - n + \sum_{i=1}^n(u_i)^{-\rho^C}\right)^{-n-\frac{1}{\rho^C}}\prod_{i=1}^n \Bigl((u_i)^{-\rho^C-1}\bigl((i-1)\rho^C+1\bigr)\Bigr),
\end{align}
where $\rho^C \in [0,\infty)$ is the dependence parameter. The Clayton copula does not have upper tail dependence. Its lower tail dependence is $\lambda_L = 2^{-1/\rho^C}$.
The distribution function of the Gumbel copula is 
\begin{align}
	C^G(u_1,...,u_d) &= \exp \left(-\left[\sum_{i=1}^d\bigl(-\log (u_i)\bigr)^{\rho^G}\right]^{\frac{1}{\rho^G}}\right),
\end{align}
where $\rho^G \in [1,\infty)$ is the dependence parameter. The Gumbel copula does not have lower tail dependence. The upper tail dependence of the Gumbel copula is $\lambda_U = 2 - 2^{1/\rho^G}$.
The distribution function of the Frank copula is
\begin{align}
	C^F(u_1,...,u_n) &= \frac{1}{\rho}\ln \left(1 + \frac{\prod_{i=1}^n (e^{\rho^F u_i}-1)}{(e^{\rho^F}-1)^{n-1}}\right),
\end{align}
where $\rho^F \in \mathbb{R}/\{0\}$ is the dependence parameter. The Frank copula does not have upper or lower tail dependence. We note that the density functions for Gumbel and Frank does not admit simple recursive expressions in terms of their density functions, but they can be obtained via partial differentiation
\begin{align}
	c(u_1,...,u_n) &= \frac{\partial^n}{\partial u_1,...\partial u_n}C(u_1,...,u_n).
\end{align}
\end{lemma}

\section{Appendix C}
\label{AppendixProofCopMix}
\noindent \textbf{Proof} {\textsl{ The proof of Lemma \ref{lemmaCopMix} requires one to demonstrate that the resulting distribution function
\begin{equation*}
\begin{split}
\tilde{C}\left(u_1,u_2,\ldots,u_n\right) &=\int_{[0,u_1]\times[0,u_2]\times\cdots\times[0,u_n]}\tilde{c}\left(x_1,x_2,\ldots,x_n\right)dx_{1:n} \\
&=\sum_{i=1^m} w_i \int_{[0,u_1]\times[0,u_2]\times\cdots\times[0,u_n]} c_i\left(x_1,x_2,\ldots,x_n\right)dx_{1:n} \\
&= \sum_{i=1^m} w_i C_i\left(u_1,u_2,\ldots,u_n\right)
\end{split}
\end{equation*}
satisfies the two conditions of a n-variate copula distribution given in [Definition 2.10.6] of \citet{nelsen2006introduction}. The first of these conditions requires that for every $\bm{u} = \left(u_1,u_2,\ldots,u_n\right)\in \left[0,1\right]^n$, one can show that $\tilde{C}\left(\bm{u}\right) = 0$ if at least one coordinate of $\bm{u}$ is 0. Clearly since we have shown that
$\tilde{C}\left(\bm{u}\right) = \sum_{i=1^m} w_i C_i\left(\bm{u}\right)$ and given each member  $C_i\left(u_1,u_2,\ldots,u_n\right) \in \mathcal{A}^n$ is define to be in the family of Archimedean copulas each of which therefore satisfies this condition for all such points $\bm{u}$, then it is trivial to see that the probability weighted sum of such points also satisfies this first condition. Secondly one must show that for every $\bm{a}$ and $\bm{b}$ in $[0,1]^n$, such that $\bm{a} \leq \bm{b}$ (i.e. $a_i < b_i \; \forall i\in\left\{1,2,\ldots,n\right\}$) the following condition on the volume for copula $\tilde{C}$ is satisfied, $V_{\tilde{C}}([\bm{a},\bm{b}]) \geq 0$. As in \citet{nelsen2006introduction} we adopt the notation for the n-box, $[\bm{a},\bm{b}]$, representing $[a_1,b_1]\times[a_2,b_2]\times\cdots\times[a_n,b_n]$ and we define the n-box volume for copula distribution $\tilde{C}$ by [Definition 2.10.1, p.43] of \citet{nelsen2006introduction} giving
\begin{equation*}
\begin{split}
V_{\tilde{C}}([\bm{a},\bm{b}]) &= \sum \text{sgn}(\bm{c})\tilde{C}\left(\bm{c}\right)\\
&=\triangle_{a_1}^{b_1}\triangle_{a_2}^{b_2}\cdots\triangle_{a_n}^{b_n} \tilde{C}\left(\bm{c}\right)
\end{split}
\end{equation*}
where the domain $\text{Dom}\tilde{C}$ of the mixture copula $\tilde{C}$ satisfies $[\bm{a},\bm{b}] \subseteq \text{Dom}\tilde{C}$. In addition we note that this sum is understood to be taken over all vertices $\bm{c}$ of n-box $[\bm{a},\bm{b}]$ and $\text{sgn}(\bm{c}) = 1$ if $c_k = a_k$ for an even number of $k$'s or $\text{sgn}(\bm{c}) = -1$ if $c_k = a_k$ for an odd number of $k$'s. Equivalently, we consider $\triangle_{a_k}^{b_k}\tilde{C}\left(\bm{t}\right) = \tilde{C}\left(t_1,t_2,\ldots,t_{k-1},b_k,t_{k+1},\ldots,t_n\right) - \tilde{C}\left(t_1,t_2,\ldots,t_{k-1},a_k,t_{k+1},\ldots,t_n\right)$.
In the case of the mixture copula, we can expand the volume of the n-box $[\bm{a},\bm{b}]$ as follows
\begin{equation*}
\begin{split}
V_{\tilde{C}}([\bm{a},\bm{b}]) &= \sum \text{sgn}(\bm{c})\tilde{C}\left(\bm{c}\right)= \sum_{i=1}^m \sum w_i \text{sgn}(\bm{c}) C_i\left(\bm{c}\right) = \sum_{i=1}^m \sum w_i V_{C_i}([\bm{a},\bm{b}])
\end{split}
\end{equation*}
hence we see that since each component $C_i\left(u_1,u_2,\ldots,u_n\right)$ is a member of the set of Archimedean copula distributions $\mathcal{A}^n$, therefore for each component we have that $V_{C_i}([\bm{a},\bm{b}])\geq 0$ for all $i\in\left\{1,2,\ldots,m\right\}$.
}}
\EndProof

\end{document}